\documentclass[lettersize,journal]{IEEEtran}
\usepackage{amsmath,amsfonts}
\usepackage{algorithmic}
\usepackage{algorithm}
\usepackage{array}
\usepackage{textcomp}
\usepackage{stfloats}
\usepackage{url}
\usepackage{verbatim}

\usepackage{cite}
\newtheorem{definition}{Definition}
\usepackage{tabularx}
\usepackage{xcolor}
\usepackage[caption=false,font=footnotesize,labelfont=rm,textfont=rm]{subfig}
\usepackage{makecell}

\usepackage{multirow}

\usepackage{longtable}

\usepackage{graphicx}

\newcommand{\yf}[1]{}
\newcommand\blue{}

\newcommand{\etal}{\textit{et al}. }
\newcommand{\ie}{\textit{i}.\textit{e}., }
\newcommand{\eg}{\textit{e}.\textit{g}., }

\begin{document}

\title{A Survey on Federated Analytics: Taxonomy, Enabling Techniques, Applications and Open Issues }

\author{Zibo Wang, Haichao Ji, Yifei Zhu,~\IEEEmembership{Member,~IEEE,} Dan Wang,~\IEEEmembership{Senior Member,~IEEE,} and Zhu Han,~\IEEEmembership{Fellow,~IEEE}
\thanks{This work is supported by the National Key R\&D Program of China (Grant No. 2023YFB2704400). 
The work of D. Wang is supported by RGC GRF 15200321, 15201322, 15230624, ITC ITF-ITS/056/22MX, ITS/052/23MX, and PolyU 1-CDKK, G-SAC8.
The work of Z. Han is partially supported by NSF ECCS-2302469, CMMI-2222810, Toyota, Amazon and Japan Science and Technology Agency (JST) Adopting Sustainable Partnerships for Innovative Research Ecosystem (ASPIRE) JPMJAP2326.}
\thanks{Z. Wang, H. Ji, and Y. Zhu are with UM-SJTU Joint Institute, Shanghai Jiao Tong University, Shanghai, China. Y. Zhu is also with Cooperative Medianet Innovation Center (CMIC), Shanghai Jiao Tong University, China. 
E-mail: \{wangzibo, haichao.ji2000, yifei.zhu\}@sjtu.edu.cn}
\thanks{D. Wang is with Department of Computing, The Hong Kong Polytechnic University, Hong Kong, China.
E-mail: csdwang@comp.polyu.edu.hk}
\thanks{Z. Han is with the Department of Electrical and Computer Engineering at the University of Houston, Houston, USA, and also with the Department of Computer Science and Engineering, Kyung Hee University, Seoul, South Korea.
E-mail: hanzhu22@gmail.com}
\thanks{The corresponding author is Yifei Zhu.}
\thanks{Manuscript received xxx xx, xxxx; revised xxx xx, xxxx.}}

\markboth{Journal of \LaTeX\ Class Files,~Vol.~14, No.~8, August~2021}%
{Shell \MakeLowercase{\textit{et al.}}: A Sample Article Using IEEEtran.cls for IEEE Journals}


\maketitle

\begin{abstract}
The escalating influx of data generated by networked edge devices, coupled with the growing awareness of data privacy, has restricted the traditional data analytics workflow, where the edge data are gathered by a centralized server to be further utilized by data analysts. To continue leveraging vast edge data to support various data-incentive applications, computing paradigms have promoted a transformative shift from centralized data processing to privacy-preserved distributed data processing. The need to perform data analytics on private edge data motivates federated analytics (FA), an emerging technique to support collaborative data analytics among diverse data owners without centralizing the raw data. 
Despite the wide applications of FA in industry and academia, a comprehensive examination of existing research efforts in FA has been notably absent. This survey aims to bridge this gap by first providing an overview of FA, elucidating key concepts, and discussing its relationship with similar concepts. We then thoroughly examine FA, including its key challenges, taxonomy, and enabling techniques. Diverse FA applications, including statistical metrics, frequency-related applications, database query operations, FL-assisting FA tasks, and other wireless network applications are then carefully reviewed. We complete the survey with several open research issues, future directions, and a comprehensive lessons learned part. This survey intends to provide a holistic understanding of the emerging FA techniques and foster the continued evolution of privacy-preserving distributed data processing in the emerging networked society.

\end{abstract}

\begin{IEEEkeywords}
Federated analytics, privacy, data science, security, distributed systems, federated learning, Internet-of-Things
\end{IEEEkeywords}

\section{Introduction}

There's been a surge in data volume generated recently by the exponential growth of the Internet of Things (IoT) devices, where the number of connected devices is forecasted to reach 125 billion by 2030\cite{125billion}. This IoT expansion is expected to generate 79.4 ZB of data by 2025\cite{80ZB}. 
The utilization of these data drives numerous significant and predominant data-driven applications in the fields of science and industry, ranging from chemical design to recommender systems.
As the utilization of these edge data increases, uploading and processing these data centrally presents huge challenges for data communication and computing.
Coincident with the extraordinary increase in the volume of data has also been a growing appreciation of the importance of data privacy, reflected by the enactment of stringent regulations such as the General Data Protection Regulation (GDPR) in EU \cite{EuropeanParliament2016a} and the California Consumer Privacy Act (CCPA) in California \cite{CAdataregulations}. Fueled by the data explosion and privacy concerns, there is a noticeable shift from centralized to distributed data collection, storage, and processing. 
This evolution in data management underscores a new paradigm called federated computation (also known as federated computing) \cite{bharadwaj2022introduction,tong2023federated,bharadwaj2024federated}, where data-oriented tasks are conducted among distributed data owners without uploading the local raw data. 
\blue{In federated computation, distributed data owners convert their local data into informative and privacy-preserving intermediate outcomes (such as gradients in federated learning), and upload them to a centralized server. The intermediate outcome is the transformed result based on certain models specified by the designer from the local dataset (\eg gradient derived from backward propagating ML models specified by the designer). The data analyst, who controls the centralized server, aggregates these intermediate outcomes to derive a global model, or analytics results, which represents a global view of the edge data in all data owners.} 


\blue{While federated computation serves as the foundational infrastructure for distributed computation, federated analytics, and federated learning are distinct but closely related concepts that operate within this framework.
Federated learning (FL) is an instance of this computing paradigm that targets collaborative model training without centralizing the raw data. It has been widely studied, deployed in the real world, and adequately surveyed \cite{nguyen2021federated,lim2020federated,yin2021comprehensive,zhang2021survey,khan2021federated}. Federated analytics (FA) is another emerging instance of the federated computation paradigm, which conducts data analytics tasks on distributed data held by different data owners (clients). The term ``federated analytics'' was first introduced by Google in 2020 \cite{fa20} to describe ``the practice of applying data science methods to the analysis of raw data that is stored locally on users’ devices''.} 
FA extends the application of currently successful FL to broader data analytics tasks other than model training for deep learning. \blue{For instance, the first application of FA is conducted by Google researchers \cite{fa20}, where the FA scheme is designed to evaluate the accuracy of a trained FL model against client data in local devices. The primary distinction between FA and FL lies in their different focused tasks, in that FA focuses on descriptive tasks, and FL focuses on predictive tasks. For instance, while FL focuses on training a machine learning model across multiple devices, such as in predictive text algorithms on smartphones, FA can be applied to tasks like aggregating statistical summaries such as collecting statistics, frequent pattern mining, location heatmap, and healthcare data analysis. These tasks do not involve procedures of machine learning model training, so FL cannot be applied.}
In contrast to classical data analytics workflow, which usually requires the clients to upload their raw data to a centralized server, FA prevents any transmission of raw data leaving the client they originated, with in order to save communication costs and preserve the data privacy of clients.  

In an FA algorithm, the clients utilize their local data and the computation model received from the server to conduct local computation. 
\blue{The output of the local computation procedure is called ``insight'' in this paper and it is one form of  ``intermediate outcome'' in federated computation as we discussed previously. The insights are tailored data structures or information that reflect useful knowledge about the local data for the host data analytics task while preserving the privacy of the raw data.} The clients upload the insight to the server to circumvent the raw data transmission, and the server performs insight aggregation to transform the individually generated insights received from the clients to derive the population-level data analytics results. With FA, valuable analytics results can be derived to serve the pervasive data analytics needs of networked applications. The growing data privacy concerns of the users can also be properly handled.


Specifically, FA offers the following benefits:
\begin{itemize}
    \item \blue{\textbf{Communication reduction:} FA, following the paradigm of federated computation, transmits the insight of client data instead of the raw data. The structure of insight is usually designed to contain only valuable information to resolve the data analytics tasks, and the size of insight is usually independent of the number of data samples held by the clients. Therefore, FA can reduce the communication cost in data analytics, especially when the clients hold large-volume raw data with many data samples.}
    \item \blue{\textbf{Privacy preservation:} The tenet of federated computation lets FA prevent the transmission of raw data. The transmitted insight, which only includes abstractive information of the raw data, is less sensitive than the raw data transmission. Therefore, compared to traditional data analytics that requires raw data transmission, FA enhances privacy preservation in utilizing edge data, with both the prevention of raw data transmission and additional privatization mechanisms. FA can be applied to both large-scale mobile devices, as well as several major data silos.} 
    \item \blue{\textbf{Expansive task coverage:} Different from FL which mostly focuses on neural network-based predictive tasks, FA extends the scope and covers the whole spectrum of descriptive tasks in data science. These tasks range from basic calculations to complex tasks involving sophisticated and dedicated data structures and computation procedures. FA, joining forces with FL, successfully addresses the entirety of the learning and analytics problem space, enabling all data-oriented tasks to be completed in a federated environment. }

\end{itemize}

With these unique advantages, FA has offered a range of applications across various industries and sectors. For example, Google has utilized FA ``in support of FL'' to ``measure the quality of FL models against real-world data when that data is not available in a data center'' \cite{fa20}. FA can also assist FL in terms of client selection \cite{wang2021fedacs} and client clustering \cite{sattler2020clustered}. 
Furthermore, sectors where data sensitivity is paramount can benefit from privacy-preserving FA to cooperate on tasks and research without sharing raw data such as healthcare\cite{froelicher2021truly}, finance, and public services\cite{bagdasaryan2021towards}. In these contexts, even competitive parties can conduct FA collaboratively for common interests. The reductions in communication and computation further enable FA to offer wide applications in sectors where data is collected from remote locations like agriculture and environmental monitoring or where large-scale data collections and processing are required such as tasks in smart cities and in industry 4.0. The wide applications and studies of FA make it the proper time to survey this prominent area.

\subsection{Comparison and our contributions}

While numerous surveys have explored related topics such as FL \cite{zhang2021survey,yin2021comprehensive,lim2020federated,nguyen2021federated,khan2021federated} and privacy computing \cite{mollah2017security,zhang2018data,ranaweera2021survey}, a comprehensive review of the current status of FA is still lacking. Only two preliminary surveys exist. Survey \cite{Dan2022survey} offers an overview of FA, its research positioning, motivations, and applications but lacks in-depth task elaboration, focusing more on challenges and solutions. Survey \cite{elkordy2023federated} categorizes FA queries into statistical, set-based, and matrix transformation tasks, but fails to address the full range of data analytics tasks in contemporary FA applications with significant performance implications. Additionally, it overlooks a multi-dimensional taxonomy and a thorough analysis of key enabling techniques from various perspectives.
\blue{In comparison to these previous surveys \cite{Dan2022survey,elkordy2023federated}, this article provides a more comprehensive summary of FA algorithms and systems. We explore a broader set of data analytics tasks to which FA is applied, clarifying ambiguities and examining the overlap between FA and related fields. Our analysis highlights how research in these adjacent areas utilizes FA concepts to address data analytics challenges while preserving privacy. We also integrate these extensive works into this survey. The goal of this article is to offer readers a well-rounded view of the data analytics problems that FA addresses, the enabling technologies behind FA, and the open research questions that require further exploration. Furthermore, we aim to inspire researchers in related fields by demonstrating the relationship between FA and its applications, showcasing how FA has been successfully implemented, and suggesting how it could be further leveraged in these domains.}

The key contributions of this paper are outlined as follows:

\begin{itemize}
    \item\blue{\textbf{Comprehensive survey}. We provide a state-of-the-art, comprehensive survey on FA, covering key challenges, taxonomies, techniques, and a wide range of FA applications.  It offers a holistic view of FA by addressing not only its core principles but also the critical gaps that have persisted in existing literature. }
    \item\blue{\textbf{Taxonomy development}. We present a robust multi-dimensional taxonomy for FA, encompassing aspects such as data analytics tasks, client scale, iteration count, coordination models, and privatization methodologies. It not only provides a comprehensive framework for analyzing existing FA systems but also serves as a guide for designing future solutions that align with the specific needs of diverse applications.}
    \item\blue{\textbf{Enabling techniques}. We summarize the enabling techniques in FA that address privacy preservation, data analytics, and system optimization. They collectively address the dual challenges of maintaining high data utility and strong privacy guarantees, while ensuring that FA systems are scalable, resource-efficient, and practical for real-world applications.}
    \item\blue{\textbf{Application spectrum}. We compile an extensive range of FA applications, including statistical metrics, frequency-related tasks, database operations, machine learning pipeline tasks, and wireless network applications. These applications demonstrate how FA addresses key challenges in privacy-sensitive, large-scale, and distributed environments, making it a transformative technology for industries ranging from IoT to cloud computing and beyond.}
    \item\blue{\textbf{Open research issues}. We identify several critical open research challenges in FA, spanning application scenarios, algorithm design, system optimization, and cross-layer enhancements. These open research challenges outline a roadmap for future exploration and highlight the potential for FA to evolve into a more versatile and impactful framework for privacy-preserving distributed analytics.}
    
\end{itemize}

\begin{table}
\caption{List of important abbreviations, ordered by their occurrence in this survey.}
\centering
\begin{tabular}{c|c}
\hline

FA & Federated analytics\\
IoT & Internet of things\\
FL & Federated learning \\
DP & Differential privacy \\
MEC & Mobile edge computing \\
MPC & Multi-party computation \\
DDM & Distributed data mining \\
LDP & Local differential privacy \\
CDP & Central differential privacy \\
DDP & Distributed differential privacy \\
HE & Homomorphic encryption \\
PSI & Private set intersection \\
FPM & Frequent pattern mining \\
SVM & Support vector machine \\

\hline
\end{tabular}  
\label{table_abb}
\end{table}

\subsection{Structure of the survey}
\begin{figure}[t]
\includegraphics[width=\linewidth]{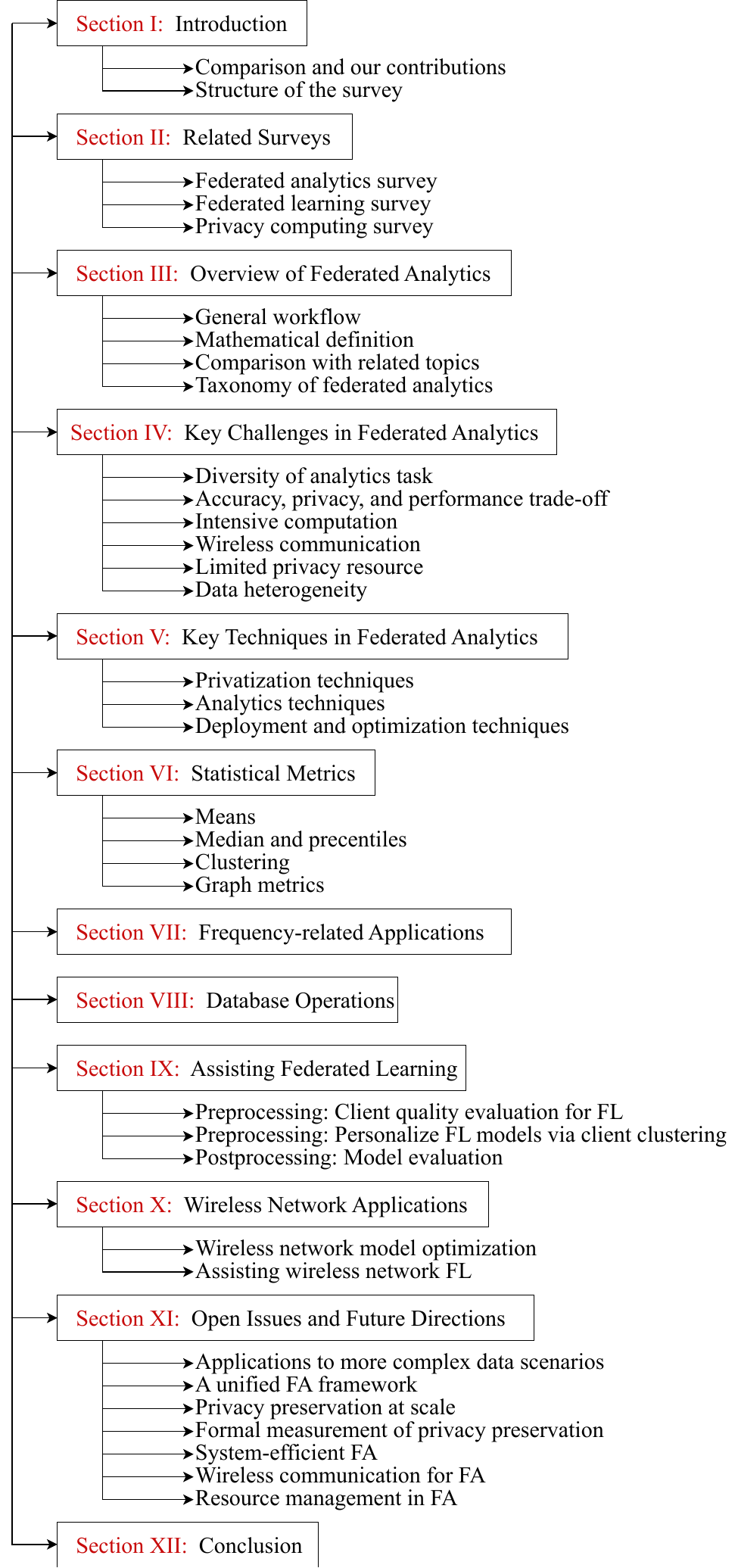}
\caption{Overview of the survey.}
\label{fig_organization}
\end{figure}

The structure of the survey is organized as follows. 
Section \ref{sec_relatedsurvey} reviews surveys of FA and related topics.
Section \ref{sec_overview} provides an overview of FA, including its general workflow, a general mathematical problem formulation, and its comparison with related topics. It also includes our taxonomy of FA works that covers five important dimensions that have significant impacts on its real-world applications and deployments.
In Section \ref{sec_challenge}, key challenges in designing and deploying FA are discussed.
Section \ref{sec_technique}, we introduce key enabling techniques applied in FA, including privatization techniques to realize privacy preservation, analytics techniques to derive data analytics results, and deployment/optimization techniques to enable real-world applications.
In the following five sections, we introduce the vast existing FA solutions in detail, where each section covers FA solutions on a particular class of data analytics tasks: Section \ref{sec_statistical} covers the computation of simple statistical metrics; Section \ref{sec_frequency} covers the tasks related to frequency estimation; Section \ref{sec_database} covers database operations (data analytics tasks formulated as SQL queries);  
Section \ref{sec_assistfl} covers data analytics tasks that are tailor-designed to assist FL systems;
Section \ref{sec_wireless} covers the applications of FA for wireless networks.
Section \ref{sec_openissue} discusses open issues and future research directions in FA, and Section \ref{sec_conclusion} concludes the survey.
An overview of the survey is provided in Fig. \ref{fig_organization}. For reading convenience, we summarize the important abbreviations used in this survey and conclude them in Table \ref{table_abb}.

\begin{figure*}[t]
\centering
\includegraphics[width=1\linewidth]{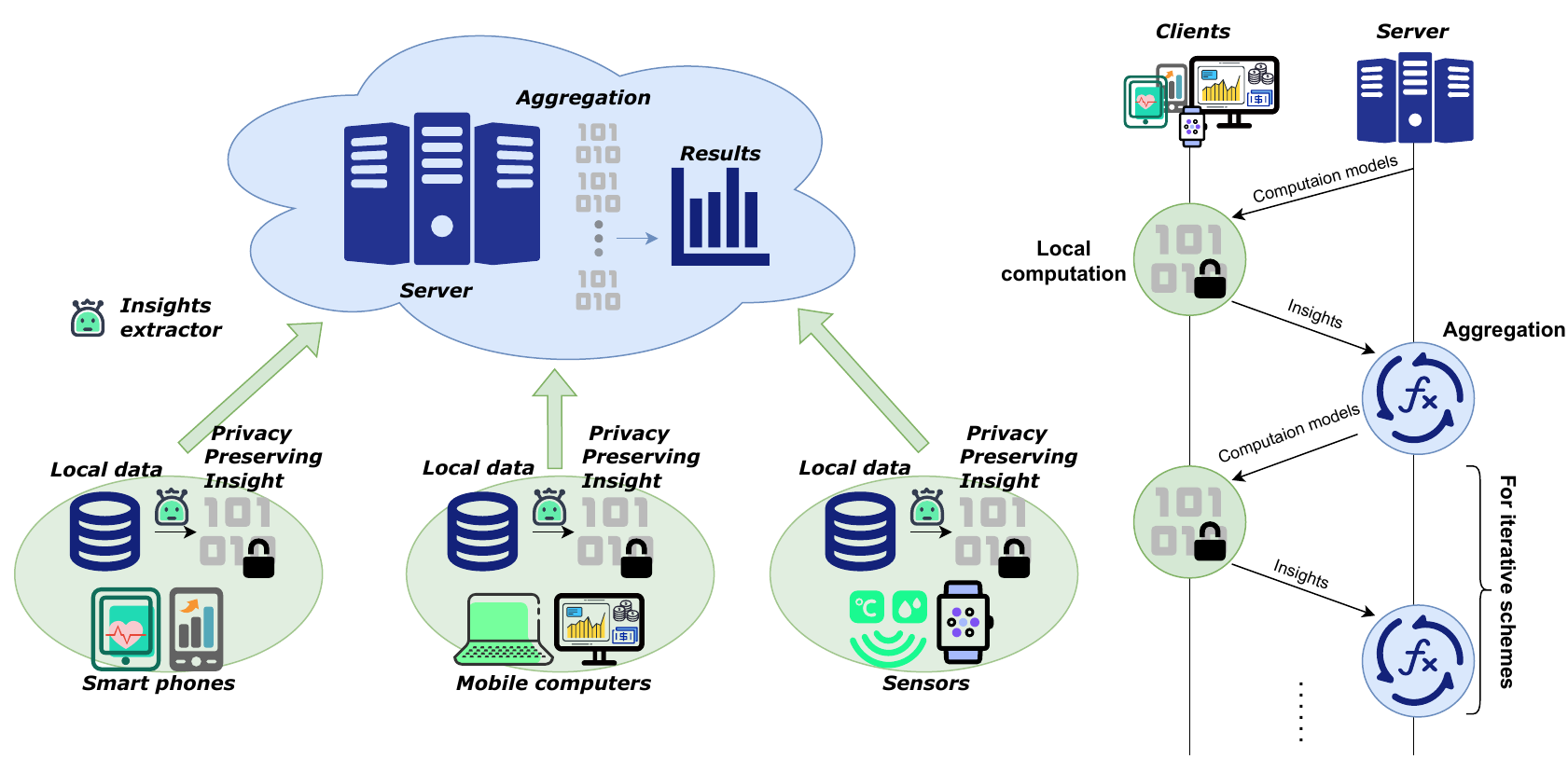}
\caption{Architecture of FA. The left plot demonstrates the interaction between the server and the clients. The clients holding privacy-sensitive local data utilizes the computation procedure of insight extractor to derive privacy-preserving insight, which is aggregated on the server side to derive results. The right plot demonstrates the iterative workflow of FA, where an FA iteration consists of phases of computation model distribution, local computation, insight upload, and insight aggregation. }
\label{fig_FAarchitecture}
\end{figure*}

\section{Related Surveys}\label{sec_relatedsurvey}

\blue{In this section, we begin by first conducting a discussion of the existing survey on FA \cite{Dan2022survey,elkordy2023federated}, systematically reviewing its current research landscape and development trends. Subsequently, we extend our focus to include surveys on two closely related topics: FL \cite{zhang2021survey,yin2021comprehensive,lim2020federated,nguyen2021federated,khan2021federated} and Privacy Computing \cite{mollah2017security,zhang2018data,ranaweera2021survey}. These surveys offer valuable insights into the foundational technologies, methodologies, and challenges closely tied to FA, providing a broader perspective on the research and development of this domain.}

\subsection{Federated analytics survey}

Although the extraordinary advantage of the FA framework in privacy preservation and high-utility data analytics has spurred numerous FA studies, a systematic review of the current status of FA has not been conducted thus far. To date, only two preliminary surveys on FA exist\cite{Dan2022survey,elkordy2023federated}. Survey \cite{Dan2022survey} first reviews the FA, and clarifies its position in the research literature, motivation, and application of it. However, it lacks detailed elaborations on tasks that can be supported by FA, presenting only two cases as examples. They place more emphasis on discussing the challenges, existing and possible solutions of FA in terms of the architecture, privacy problem, computation and communication resource, analytics design, and business models. Authors in \cite{elkordy2023federated} focus on common queries of interest in FA and their corresponding existing solutions, algorithms, and applications. They summarize and divide queries in FA into three categories: statistical queries, set-based queries, and matrix transformation queries. However, this classification is not  comprehensive enough to cover other query types and data analytics tasks in FA, such as database operations, frequency-based queries, and tasks for FL assistance. Additionally, they do not provide a comprehensive summary of common techniques employed in FA, such as privatization, analytics, and system optimization.  

\subsection{Federated learning survey}

FL has obtained significant attention in recent years as a promising approach to distributed machine learning, allowing multiple devices or parties to collaboratively train models without sharing raw data. This paradigm shift addresses critical issues related to data privacy, security, and communication costs, making it particularly relevant in various applications such as healthcare, finance, and IoT. Numerous surveys \cite{zhang2021survey,yin2021comprehensive,lim2020federated,nguyen2021federated,khan2021federated} have been conducted to explore the advancements, challenges, and future directions of FL.

The survey \cite{zhang2021survey} provides an in-depth analysis of recent advances in FL and its future directions. It summarizes the fundamental principles of FL and its advantages over traditional centralized learning approaches and reviews the latest developments in FL, including advancements in communication efficiency, personalization techniques, and secure aggregation methods. They discuss how these innovations have been applied in various sectors, such as finance, healthcare, and autonomous systems, showcasing the versatility and effectiveness of FL. In survey \cite{yin2021comprehensive}, authors comprehensively discuss the challenges, methods, and future directions of FL. It begins by outlining the primary issues in FL, such as data privacy, communication costs, and model accuracy. They review various methods to address these challenges, including differential privacy (DP), secure multiparty computation, and model compression techniques. They also examine the application of FL in various fields like healthcare, finance, and IoT, highlighting the benefits and limitations of FL in these contexts.

There are FL surveys concentrating on specific area. The survey \cite{lim2020federated} focuses on the application of FL in mobile edge networks . It discusses the unique challenges of implementing FL in such environments, including heterogeneous devices, communication costs, and privacy concerns. The authors provide an overview of mobile edge computing (MEC) and its role in facilitating FL, detailing how FL can optimize resource allocation and improve network efficiency. They review existing solutions for integrating FL with MEC, such as edge aggregation and adaptive model updating. The survey also explores various use cases of FL in mobile networks, such as vehicular networks and smart cities. While survey \cite{nguyen2021federated} examines the use of FL in IoT domain. It introduces the key characteristics of IoT, such as vast data generation and the need for real-time processing, which make FL an attractive solution. The authors review the main challenges in applying FL to IoT, including data heterogeneity, device resource constraints, and security issues. They present various strategies to overcome these challenges, such as lightweight models, federated transfer learning, and privacy-preserving techniques. The survey also highlights several applications of FL in IoT, such as smart homes, industrial IoT, and healthcare monitoring. Also in \cite{khan2021federated}, researchers focus on the application of FL in the IoT domain, highlighting recent advancements, a comprehensive taxonomy, and open challenges. They introduce the necessity of FL in IoT due to the massive data generated by distributed devices and the privacy concerns associated with centralized learning and then review recent advancements in FL techniques tailored for IoT, emphasizing metrics such as sparsification, robustness, quantization, scalability, security, and privacy. The survey also presents a detailed taxonomy of FL approaches in IoT, categorizing them based on various parameters such as optimization schemes, incentive mechanisms, security and privacy measures, and aggregation methods.

\subsection{Privacy computing survey}

With the proliferation of data-driven applications and the increasing awareness of privacy concerns, privacy-preserving computing has become a critical area of research. Various surveys have been conducted to explore the different aspects of privacy computing, particularly in the context of edge computing \cite{mollah2017security,zhang2018data,ranaweera2021survey}.

In \cite{mollah2017security}, the authors explore the privacy and security issues inherent in edge computing. It details the architecture of edge computing and how it differs from traditional cloud computing. The authors discuss the primary security challenges such as data confidentiality, integrity, and availability, highlighting the unique threats posed by the distributed nature of edge environments. They provide an overview of existing security solutions, including encryption techniques, authentication protocols, and access control mechanisms. The survey also examines privacy-preserving methods, emphasizing the need for lightweight and efficient solutions to handle the resource-constrained nature of edge devices. Similarly, the survey \cite{zhang2018data} focuses on the security and privacy aspects of multi-access edge computing. It begins by outlining the key components and functionalities and explaining its importance in enhancing computational capabilities at the network edge. The authors review various security challenges specific to multi-access edge computing, such as secure data transmission, user authentication, and intrusion detection. They highlight current solutions and strategies employed to mitigate these issues, including blockchain technology, secure multi-party computation (MPC), and machine learning-based anomaly detection. The survey also addresses privacy concerns, particularly data anonymization and DP techniques, to protect user information in multi-access edge computing environments. In \cite{ranaweera2021survey}, it also delves into data security and privacy-preserving mechanisms within the edge computing paradigm. It discusses the benefits of edge computing over centralized cloud models, particularly in terms of reducing latency and improving data locality. The authors examine the specific security and privacy challenges faced by edge computing, such as secure data storage, secure computation, and user authentication. They provide a comprehensive review of cryptographic techniques like homomorphic encryption, attribute-based encryption, and proxy re-encryption that can be applied to secure data in edge environments. Additionally, the survey covers access control mechanisms and privacy-preserving data aggregation methods.

\section{Overview of Federated Analytics}\label{sec_overview}

\blue{In this section, we provide an overview of the emerging FA. We focus on many important concepts of FA. In Section \ref{subsec_overview_workflow}, we describe the general workflow of FA. In Section \ref{subsec_overview_math}, we propose a general mathematical problem formulation of FA. In Section \ref{subsec_overview_comparison}, we introduce the relationship between FA and related fields. In Section \ref{subsec_taxonomy}, we present taxonomies of FA in five dimensions that help us properly position the diverse FA algorithms and systems.}

\subsection{General workflow}\label{subsec_overview_workflow}
FA aims at conducting data analytics tasks based on the federated data held by multiple clients with privacy preservation. The fundamental principle of FA to realize privacy preservation is that the raw data held by the clients are not transmitted and exposed. In practice, the prevention of raw data exposure is not sufficient to realize a formal privacy guarantee, and extra privatization mechanisms are usually applied in FA for the formal privacy guarantee. 
Since FA and FL are two variations of the federated computation paradigm, they share identical abstractive architecture and workflow, which is defined in the federated computation paradigm.
As is demonstrated in Fig. \ref{fig_FAarchitecture}\footnote{The right plot of the figure only applies for iterative FA schemes according to our taxonomy of iteration pattern (Section \ref{subsec_taxonomy}).}, the typical architecture of FA (server-client model) includes a server, which is hosting data analytics tasks, and multiple clients, that would like to contribute their data to the data analytics task but requires privacy preservation on them.

The right part of Fig. \ref{fig_FAarchitecture} demonstrates the steps of one-shot and iterative FA systems.
In the FA workflow, the clients are in charge of a part of the data analytics task. They receive the computation model from the server and perform local computation based on the received computation model and the client's local data. The result of local computation, which is termed ``insight'' in this survey, is uploaded to the server, where the insights are processed by privatization mechanisms to realize privacy preservation. The server aggregates the received insights with a dedicated aggregation algorithm. The aggregation output may be directly the results of the data analytics task when the FA algorithm is one-shot. When the algorithm is iterative, the aggregation result updates the computation model, and the data analytics results are derived after many rounds of model distribution, local computation, insight upload, and insight aggregation. 

As mentioned above, the same architecture and workflow of FA is also adopted by many classical FL systems. However, it worth to be emphasizing that FA and FL differ a lot when the abstractive architecture and workflow are implemented into specific algorithms and programs. FL and FA perform different computations on the client side, deriving different structures of insights, and are eventually aggregated by the server in different ways. The difference in downstream tasks also affects the abstractive workflow of FA and FL. For example, one-shot schemes (\eg those with only iterations) are much more common in FA than FL, because they are suitable for deriving simple statistics from client data. More discussions and comparisons are present in the following Section \ref{subsec_overview_comparison}.

\subsection{Mathematical definition}\label{subsec_overview_math}
FA considers a system of $n$ data owners, or clients denoted $DO$, and each client $DO_i$ holds local client data $x_i$. These clients would like to collaboratively accomplish a data analytics task, which can be reformulated as deriving the task function
\begin{equation}
    \mathcal{F}(x_1,...,x_n).
\end{equation}
During the procedure of deriving $\mathcal{F}$, the personal raw data $x_i$ should not be exposed to any entity other than $DO_i$.

The aforementioned problem is resolved by the FA paradigm by designing the appropriate insight derivation function $\mathcal{I}_i$ and a global aggregation function $\mathcal{A}$. The clients derive insights from their local data using the insight derivation function $\mathcal{I}_i$, and then aggregate the insights to derive the value of the task function, \ie
\begin{equation}
    \mathcal{F}(x_1,...,x_n) \leftarrow \mathcal{A}(\{\mathcal{I}_i(x_i);i\in DO\}).
\end{equation}
In the majority of existing FA studies, the FA algorithm treats all the data owners equally, and all the clients share the identical insight derivation function (denoted $\mathcal{I}$), and the FA procedure can be written as
\begin{equation}
    \mathcal{F}(x_1,...,x_n) \leftarrow \mathcal{A}(\{\mathcal{I}(x_i);i\in DO\}).
\end{equation}

Based on the aforementioned mathematical formulation, the research problem in designing the FA algorithm is derived: the researchers need to design insight derivation function $\mathcal{I}$ and aggregation function $\mathcal{A}$ so that the FA result $\mathcal{A}(\{\mathcal{I}(x_i);i\in DO\})$ should be identical (or close to) the result of the task function $\mathcal{F}(x_1,...,x_n)$. 

The previous formulation describes the one-shot FA setting. In the iterative FA setting, the insight derivation function should take the computation model as input, and the results of the aggregation function should be the data analytics result in the last iteration and should be updated computation model in other iterations.

\subsection{Comparison with related topics}\label{subsec_overview_comparison}

\begin{table}
\caption{Difference between FL and FA}
\centering
\begin{tabular}{ccc}
\hline
~ & FL & FA \\
\hline
Goal & Training neural networks & Non-training tasks\\
Aggregation & FedAvg & Task dependent \\
Insight & Model weights & Task dependent \\
\hline
\end{tabular}  

\label{table_diff1}

\bigskip

\caption{Difference between non-FA parts of DDM and FA}
\centering
\begin{tabular}{ccc}
\hline
~ & non-FA DDM & FA \\
\hline
Raw data redistribution \& transmission & Allowed & Forbidden\\
Clients and server & Trusted & Untrusted \\
Heterogeneities & Little concerned & Focused \\
\hline
\end{tabular}  

\label{table_diff2}
\end{table}

\begin{figure}[t]
\centering

\includegraphics[width=1.0\linewidth]{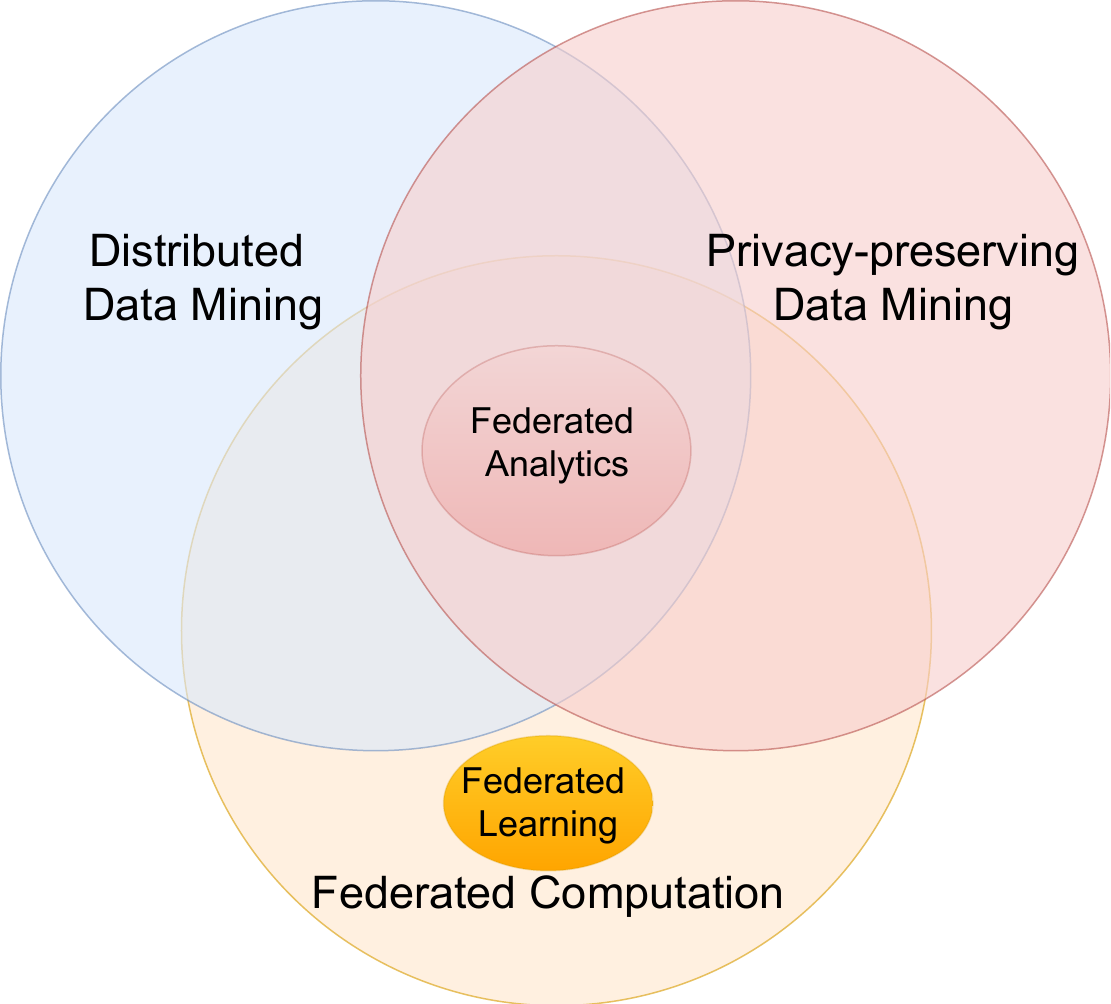}

\caption{Relationship of FA and related topics.}
\label{fig_faflfcRelationship}
\end{figure}

Although the term FA has been introduced in recent years, some core characteristics of FA, such as privacy protection, and distributed computing, can already be found in existing research fields, such as FL, privacy-preserving data mining, and distributed data analytics. 
These topics naturally have certain overlap with FA.
We list these three related topics of FA, introduce the overlapping research, and discuss how FA extends its concepts. \blue{We also provide illustration of their relationships in Fig. \ref{fig_faflfcRelationship}, where FA lies in the overlap region of distributed data mining, privacy-preserving data mining, and federated computation, while FL, which focus on neural network training rather than data mining, is another variation of the federated computation paradigm}.


\subsubsection{Relationship between FL and FA}


Both FL and FA target at conducting intelligent data-oriented tasks without centralizing the raw data. Their major difference can be summarized in Table \ref{table_diff1}. First, FA and FL are differentiated for their goal tasks. In FL, the mechanism is designed to serve a predictive model training task. In most of the FL literature, the ``model training'' in FL exactly refers to deep supervised learning tasks, if no specific context is present. These FL algorithms utilize gradient descent (typically on a neural network).
On the contrary, as its name refers, FA tackles the extensive field of descriptive data analysis tasks, ranging from the simplest task of calculating the average to complex tasks such as graph analytics and video analytics \cite{deng2020fedvision,hu2021feva}. Second, the task range of FA diversifies the forms of local insights and central aggregation. In contrast to the model weights or gradient distilled from the local clients, the insights from FA could be much more diverse and task-dependent, such as privatized hashing results \cite{erlingsson2014rappor}, one-bit response to server query \cite{wang2022fedfpm}, or local clustering centroids \cite{dennis2021heterogeneity}. The procedure of insight aggregation of FA is also diverse and task-dependent compared to FL, such as tree aggregation \cite{zhu2020federated}, Bayesian-based distribution estimation \cite{chen2021digital}, or secret sharing decryption \cite{davidson2022star}.
To be specific, in this survey, if a federated computation scheme is based on an explicit parameterized computation model, and computes by gradient-based methods on the parameterized model, such a scheme will be considered as FL. A federated computation scheme is considered FA rather than FL when it conducts traditional data analytics task or performs computation by a non-gradient descent method.

\subsubsection{Relationship between FA and privacy-preserving data mining}
Privacy-preserving data mining is a comprehensive term including all data analytics schemes that take privacy concerns into consideration. Therefore, FA can be considered as a subfield of privacy-preserving data mining. As the need to perform data analytics with privacy preservation naturally exists, many solutions for privacy-preserving data analytics existed even earlier than the term federated computation (FL was first proposed in 2016 and FA was in 2020). These works (represented by RAPPOR \cite{erlingsson2014rappor}) do not use the term ``federated'', but follow a similar idea and methodology as FA. Therefore, it is reasonable and scientific to re-classify these works as FA.
On the other hand, not all privacy-preserving data mining research can be considered as FA. FA centers around distributed data setting and local privacy preservation. Heterogeneity and the resulting issues born on the client side, such as client dynamics, adversarial attacks, communication overhead, and incentive issues, are all new problems that are not considered in classical privacy-preserving data mining. 

\subsubsection{Relationship between FA and distributed data mining}

Distributed data mining (DDM) refers to the procedure of conducting data analytics tasks while the data is distributed among multiple parties or machines. According to such a comprehensive definition, FA is a subfield of DDM because of its federated data setting. However, the FA studies differ from general distributed data mining in that distributed data mining usually operates in trusted database scenarios with distributed machines supporting computing acceleration. 
In summary, in conventional DDM, the computation nodes are naturally trusted, and the raw data transmission is allowed, which is not allowed in FA. Table \ref{table_diff2} concludes the differences between FA and DDM.

\begin{figure}[t]
\centering
\includegraphics[width=0.5\textwidth]{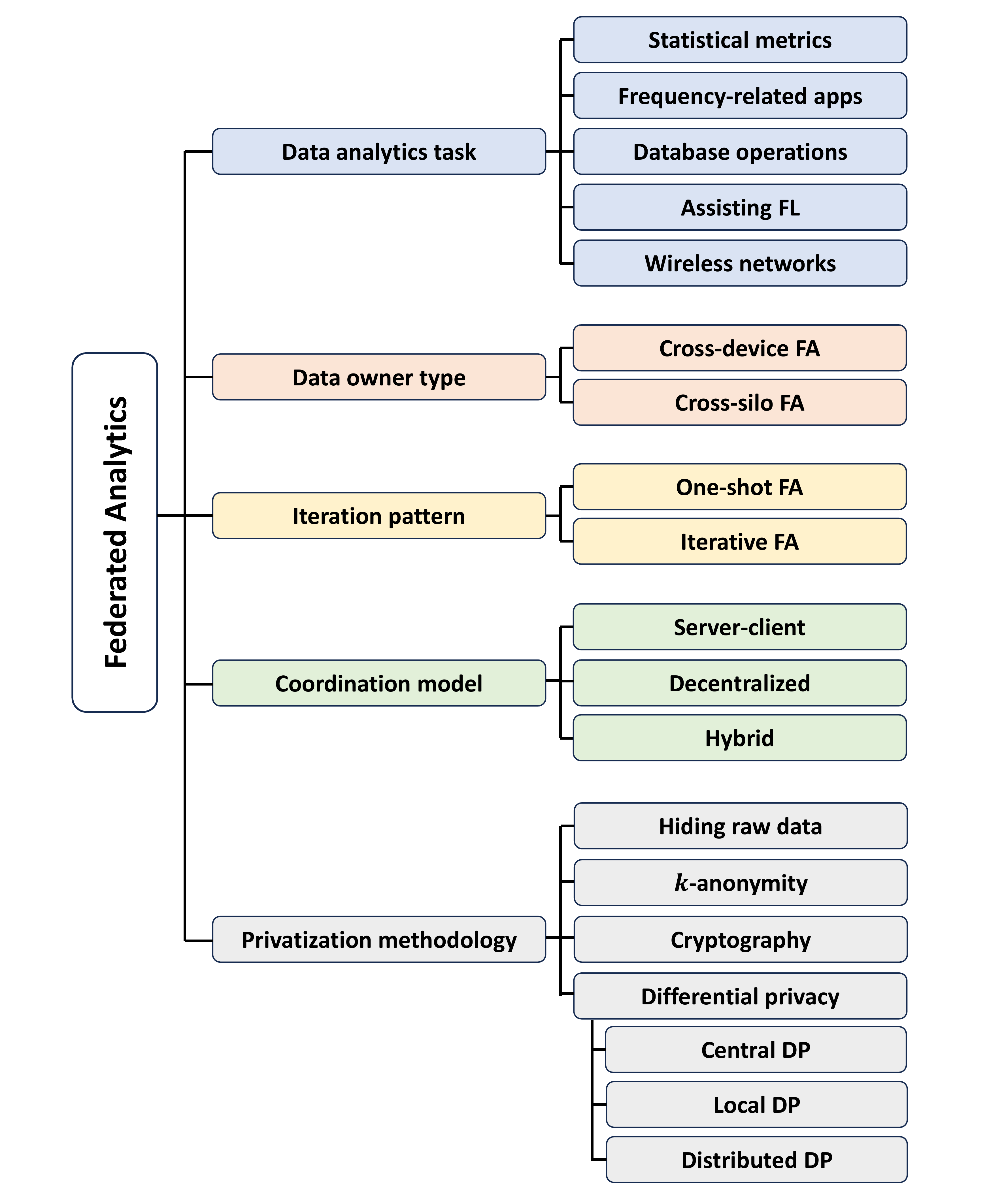}
\caption{An illustration of our taxonomy.}
\label{fig_taxonomy}
\end{figure}

\begin{table*}
\caption{Examples of FA algorithms and systems following the taxonomy of this survey. The ubiquitous privatization methodology of hiding raw data is omitted in the rows if any other methodology is present.}
\label{table_taxonomy}
\centering
\begin{tabularx}{\linewidth}{llllll}
\hline
   \textbf{Reference}& \textbf{Data analytics task} &\textbf{Data owner type} & \textbf{Iter. pattern} & \textbf{Coord. model} & \textbf{Privatization methodology}\\
\hline
\cite{cormode2021bit,ding2017collecting,nguyen2016collecting,ye2019privkv} & Statistical metrics & cross-device & One-shot & Server-client & Local differential privacy\\
\cite{hu2021feva} & Statistical metrics & cross-device & One-shot & Server-client & Hiding raw data\\
\cite{kadhe2020fastsecagg,so2021turbo,roth2019honeycrisp} & Statistical metrics & cross-device & One-shot & Server-client & Cryptography\\
\cite{bonawitz2017practical,bell2020secure} & Statistical metrics & cross-device & One-shot & Hybrid & Cryptography\\
\cite{aggarwal2010secure} & Statistical metrics & cross-silo & Iterative & Decentralized & Cryptography\\
\cite{tueno2019secure} & Statistical metrics & cross-device & One-shot & Server-client & Cryptography\\
\cite{geyer2017differentially,mcmahan2017learning,hu2020personalized}& Statistical metrics & cross-device & Iterative & Server-client & Central differential privacy\\
\cite{wei2021user,kim2021federated,shi2021hfl,pan2022fedwalk,ye2020lf,liu2024edge} & Statistical metrics & cross-device & Iterative & Server-client & Local differential privacy\\
\cite{bohler2020secure} & Statistical metrics & cross-device & Iterative & Decentralized & Cryptography \& Local differential privacy\\
\cite{zhao2022information,so2022lightsecagg} & Statistical metrics & cross-device & Iterative & Server-client & Cryptography\\
\cite{davidson2022star} & Statistical metrics & cross-device & Iterative & Decentralized & Cryptography \& $k$-anonymity\\

\cite{liu2022efficient} & Statistical metrics & cross-device & Iterative & Server-client & Cryptography\\
\cite{li2012sampling} & Statistical metrics &  cross-device & One-shot & Decentralized & Central differential privacy \& $k$-anonymity\\
\cite{iutzeler2017distributed} & Statistical metrics & cross-silo & Iterative & Decentralized & Hiding raw data\\
\cite{dennis2021heterogeneity} & Statistical metrics & cross-device & One-shot & Server-client & Hiding raw data\\
\cite{lubana2022orchestra} & Statistical metrics & cross-device & Iterative & Server-client & Hiding raw data\\
\cite{zhou2022memory,chung2022federated,servetnyk2020unsupervised} & Statistical metrics & cross-silo & Iterative & Server-client & Hiding raw data\\

\cite{erlingsson2014rappor,bassily2015local,qin2016heavy,acharya2019hadamard,acharya2019communication,apple2017learning} & Frequency-based applications & cross-device & One-shot & Server-client & Local differential privacy\\
\cite{bassily2020practical,wang2022fedfpm,chadha2023differentially, wang2018locally,li2022frequent} & Frequency-based applications & cross-device & Iterative & Server-client & Local differential privacy\\
\cite{zhu2020federated,cormode2022sample} & Frequency-based applications & cross-device & Iterative & Server-client & Central differential privacy\\
\cite{bohler2021secure} & Frequency-based applications & cross-device & One-shot & Server-client & Cryptography \& Central differential privacy\\
\cite{wang2024fedweb,bagdasaryan2021towards} & Frequency-based applications & cross-device & Iterative & Hybrid & Distributed differential privacy\\
\cite{boneh2021lightweight} & Frequency-based applications & cross-device & Iterative & Server-client & Cryptography \& Local differential privacy\\

\cite{bater2017smcql,volgushev2019conclave,wang2021secure} & Database operations & cross-silo & One-shot & Decentralized & Cryptography\\
\cite{bater2020saqe} & Database operations & cross-silo & One-shot & Decentralized & Cryptography \& Central differential privacy\\
\cite{poddar2021senate,zhang2022efficient,tong2022hu} & Database operations & cross-silo & One-shot & Server-client & Cryptography\\
\cite{roth2020orchard,roth2021mycelium,margolin2023arboretum} & Database operations & cross-device & Iterative & Server-client & Cryptography \& Central differential privacy\\

\cite{sattler2020clustered,cho2022towards,luping2019cmfl,ruan2022fedsoft,lai2021oort} & Assisting federated learning & cross-device & Iterative & Server-client & Hiding raw data \\

\cite{liang2023efficient} & Assisting federated learning & cross-device & One-shot & Server-client &  Hiding raw data\\
\cite{briggs2020federated,ghosh2020efficient,cho2023communication,wang2020optimizing,wang2021fedacs} & Assisting federated learning & cross-device & Iterative & Server-client & Hiding raw data\\

\cite{wang2020federated,zhao2022semi} & Wireless network application & cross-silo & Iterative & Server-client & Hiding raw data \\
\cite{chen2021federated,mulvey2023cellular,xing2023multi,wang2020federateduav,wang2022content} & Wireless network application & cross-device & Iterative & Server-client & Hiding raw data \\




\hline

\end{tabularx}  
\end{table*}

\subsection{Taxonomy of federated analytics}\label{subsec_taxonomy}
Given the diversity of the existing FA algorithms and systems, it becomes beneficial to provide methods to classify FA problems and solutions. In this section, we provide the taxonomy of FA. It examines the design and application of FA from five important dimensions, including data analytics tasks, scale of clients, number of iterations, coordination models, and threat models. Fig. \ref{fig_taxonomy} illustrates the taxonomy.

\subsubsection{Data analytics task}\label{subsubsec_task}

As the application of FA covers massive and diverse data analytics needs, classifying FA algorithms and systems based on the types of data analytics tasks they solve becomes a natural and useful taxonomy of FA. 
In this survey, we categorize the corresponding data analytics tasks of the FA works we surveyed into five categories: \textit{statistical metrics, frequency-related applications, database operations, assisting FL applications, and wireless network applications}. Following the taxonomy, we respectively introduce FA works focusing on each data analytics task in Sections \ref{sec_statistical}-\ref{sec_wireless}. In some sections, a more detailed classification of the data analytics tasks is present.

The host data analytics task heavily influences the algorithmic design, characteristics, and challenges of the corresponding FA algorithms and systems, and the algorithmic design of FA is expected to be quite different when handling different types of data analytics tasks. One of its reasons is that different data analytics tasks extract and analyze different features from the raw data, which then determine the structure of required insight, the insight generation approach, and the insight aggregation approach. Some researchers would like to break the limit of data analytics task-specific FA design, and propose some unified FA framework \cite{wang2022fedfpm}. However, proposing such an FA framework that can handle all data analytics tasks is still far from reality at this moment. Therefore, the host data analytics task is currently an important component of any FA taxonomy. In Section \ref{subsec_openissue_unified}, we investigate the probability of a unified FA framework as a future research direction.


\subsubsection{Data owner type} 

\blue{FA algorithms and systems can be categorized into \textit{cross-device FA} and \textit{cross-silo FA} based on the essence of their participating clients (data owners). Cross-device FA usually involves massive (millions of) clients with little data and limited computation/communication capacity, while cross-silo FA usually involves a small number of clients, with more data and stronger resources, which have clear objectives for participating in the federation. Cross-device FA and cross-silo FA have significant differences in many aspects, including the utilized privacy mechanisms and application scenarios. For example, cross-device FA is usually deployed in mobile applications with massive edge devices. Therefore, the cross-device FA mechanisms are required to easily scale up, and introduce limited workload to the clients. On the other hand, for cross-silo FA in industrial scenarios (\eg finance, biology), there might be only a few data silos, and each of them possesses a large number of high-quality data. In these cases, the participants are likely to derive highly correctness analytics results, introducing little error compared to the centralized algorithms. At least the performance of FA should be higher than centralized analytics on its local database, \ie the benefit of collaboration should at least cover the degrading in privacy preservation.}


Usually, DP-based mechanisms are better at supporting cross-device FA, because they only require the clients to individually perturb their client uploads in parallel, and the server aggregation function usually has linear complexity \cite{cormode2021bit,ding2017collecting,erlingsson2014rappor,bassily2015local}. In addition, when the privacy model of central differential privacy is considered, increasing the number of participating clients can increase privacy preservation by better hiding the information of individual clients \cite{zhu2020federated,cormode2022sample}.

On the other hand, FA schemes utilizing computation-intensive cryptography tools, though can provide higher analytics utility, are usually hindered by the algorithm, computation, or communication limits. Consequently, they are usually applied to cross-silo FA. 
For example, the garbled circuit \cite{yao1986generate}, a classical and famous cryptography tool employed in some FA studies \cite{bater2017smcql,poddar2021senate}, only supports computation between two parties, and modifications of it usually introduces significant computation and communication overhead to support several more participating clients. To overcome the limit in supported client scale when utilizing cryptography tools, many research efforts are proposed to increase the supported participating clients in cryptography-based FA systems. An example direction of this research is to replace the computation-intensive cryptography tools with lightweight ones, like simple additive masking \cite{bonawitz2017practical,bell2020secure,kadhe2020fastsecagg,so2021turbo}.



\subsubsection{Iteration pattern}\label{subsec_taxonomy_iterative}

\blue{FA algorithm involves steps of computation model distribution, local computation, insight upload, and insight aggregation, where the aforementioned steps compose one FA iteration. FA algorithms can then be classified into \textit{one-shot FA and iterative FA}, based on their patterns in conducting iterations. }

\blue{In one-shot algorithms, the analytics results are derived and the algorithm terminates after one iteration. Since the system only processes one round, there is no memorized ``state'' on the server side, and the computation model of the FA algorithm is fixed. Therefore, the phase of computation model distribution can even conducted along with the deployment of the FA program. Roughly speaking, one-shot schemes are usually applicable in simple FA tasks and incur less communication cost, like federated mean computation \cite{cormode2021bit,ding2017collecting}. }

\blue{On the contrary, other FA algorithms involve multiple iterations until completion. In each iteration, a part of the analytics tasks is conducted, which updates the ``state'' on the server side. Based on the results in previous rounds, the server then provides information to participating clients to guide their local computation. There exist two fundamental ideas (ways) to design an iterative FA algorithm. Firstly, each iteration of FA aims to conduct different parts of the analytics task, where the computation of the later part usually depends on the results of the former parts. For example, many FA solutions for frequent pattern mining \cite{zhu2020federated,cormode2022sample,chadha2023differentially} let different FA iterations reveal frequent sequential patterns with different fixed lengths, where the computation on longer length depends on the shorter ones. In the second way, the computation model is formulated as the optimization of a parameterized model. In each round, the uploaded insight helps the optimization of the model (\eg the insight is essentially the gradients on the model, being consistent with the methodology of FL). Unlike the former category, FA algorithms in the second category can also choose to terminate when the model converges instead of terminating after a predefined fixed number of iterations. Examples of the second category include clustering solutions \cite{chung2022federated} and assisting FL solutions \cite{wang2020optimizing,lai2021oort}.}

    
\subsubsection{Coordination model}

In our previous introduction, we define the entities of FA as one server and multiple clients, where the server hosts the data analytics tasks by distributing the computation model, receiving insights, and aggregating insights, and the clients contribute their local data by conducting local computation and uploading insights. Such systems formulate the vanilla \textit{server-client architecture} of FA, where all the clients only communicate with the centralized server. 
The server-client architecture has become the mainstream form of FA for multiple reasons. First, in the typical server-client architecture, the clients are unaware of other clients. Therefore, a new client can easily register itself in the FA system simply via communicating with the server. Secondly, an FA system under the server-client architecture can easily scale up, because the computation and communication complexity of the client scheme are usually constant, under such architecture. Thirdly, the complexity of the corresponding algorithm design is simplified, since the communication pattern has been defined by the architecture, and the privacy protection model is relatively simple.

\blue{However, as some shortcomings of the server-client architecture are identified by the researchers, optimized alternative architectures, are proposed. We classify them as \textit{hybrid architecture}, where a server-client system is considered, but client-client communication is introduced in addition to the server-client communication. In \cite{bonawitz2017practical,bell2020secure}, the authors propose some FA solutions under hybrid architecture. By leveraging pairwise communication, the clients can enforce encryption on their uploads at a lower cost, and therefore collaborate to enhance privacy preservation against the potentially malicious server. }

In \cite{aggarwal2010secure,bohler2020secure,iutzeler2017distributed}, \textit{fully decentralized architectures} without any centralized server are introduced. Fully decentralized FA algorithms are usually more complex in design but have the potential to obtain advantages including prevention of one-point failure, better privacy preservation, better communication efficiency, and adaptive organization.


\subsubsection{Privatization methodology}

The tenet of FA, as well as more broadly federated computation, is to protect the data privacy of edge data while leveraging these data in data-intensive applications. Therefore, each FA work has to follow one or multiple methodologies in their algorithmic design. As a result, classifying the FA works based on the approaches to enforce privacy preservation becomes an appropriate and beneficial taxonomy in the field of FA. FA works following different privatization methodologies scatter different threat models, and techniques focuses. They can be applied to support diverse use requirements. This taxonomy is highly associated with various privatization techniques leveraged by FA, where Section \ref{DP} provides a thorough demonstration of these techniques. In this survey, we summarize six different privatization methodologies: hiding raw data, local differential privacy, central differential privacy, distributed differential privacy, $k$-anonymity, and cryptography.

\textit{The methodology of hiding raw data} is the fundamental idea of federated computation, and it is applied in all FA works to provide privacy preservation to some extent.
Under the privatization methodology of hiding raw data, the clients will extract insights from the raw data and upload the insights to the server for further data analytics. It prevents the extreme privacy risk in conventional data analytics where the server gathers all raw data from the clients. Since the extracted insight is an indirect transformation of the raw data where the reverse computation is usually not available, and the extracted insight usually possesses a much smaller dimension size than the original data, the server or other attackers cannot easily infer the raw data from the insights; The data privacy of the raw data is thus protected. However, there remain privacy risks as the attacker can still obtain valuable and sensitive information if the insights are na\"ively extracted. Therefore, many FA solutions apply additional privatization methodologies using modern privacy preservation technologies (Section \ref{DP}) to further improve its privacy preservation, which we will discuss next.

\textit{DP-based privacy preservation methodologies}, including local differential privacy (\textit{LDP}), central differential privacy (\textit{CDP}), and distributed differential privacy (\textit{DDP}), are the most widely used additional privatization methodologies in the FA literature, since they provide rigid theoretical guarantees from the statistical perspective. Namely, the attacker cannot infer the existence of any raw data with sufficient confidence from the DP-possessed insights. LDP is the most popular one among DP methodologies because it assumes a local model that fits the settings of the majority of FA scenarios. In the methodology of LDP, the clients individually possess their local insights by adding perturbations to obtain sufficient privacy preservation. CDP considers protecting data from each client in the aggregated results. CDP requires a much smaller magnitude of perturbations, which results in higher data analytics quality but assumes the existence of a trusted aggregator who gathers raw data from the clients and is therefore not desirable in most FA scenarios. DDP is a novel idea of privatization methodology that combines the advantages of CDP and LDP by providing local privacy preservation with CDP-level perturbation magnitude, owing to its ingenious involvement of cryptography tools. DDP is gaining increasing research interest from FA researchers.

\textit{$k$-anonymity} is another privatization methodology from the statistical perspective. It protects the privacy of client data (or its transformed form) by removing all the records with less than $k$ records. Therefore, a record can be seen by the server only when it has at least $k$ clients holding that record. 

\textit{Cryptography} is the last privatization methodology we focus on in this survey. Utilizing various tools from the field of cryptography, it provides privacy preservation from a different perspective from statistics. It focuses on encrypting the whole upload from all clients so that an adversary can only learn the aggregated results, but cannot learn any upload from individual clients. As the statistics perspective and cryptography perspective consider the elimination of different privacy risks, neither of them can thoroughly outperform the other. Some recent FA research even utilizes multiple privatization methodologies to enhance privacy preservation from multiple perspectives. The merge of cryptography and CDP also yields the emerging privatization methodology of DDP, which combines advantages and enhanced privacy preservation.


The previous taxonomy with five dimensions provides us with a powerful tool for grouping and classifying FA algorithms and systems. In Table \ref{table_taxonomy}, we list all FA solutions covered in this survey and show their positions according to our taxonomy.\footnote{If an FA system does not explicitly clarify its cross-silo FA or cross-device FA, we classify it based on its scale of clients.}

\section{Key challenges in Federated Analytics}\label{sec_challenge}

\blue{As FA is expected to handle various data analytics tasks over federated data with strong privacy preservation, there exist many challenges for researchers to realize the ambition of FA. In this section, we investigate some dimensions of FA challenges and discuss how these challenges could be handled by novel FA research attempts.}

\blue{Some of these challenges are unique to FA, such as diverse tasks in terms of lack of a unified framework, defining proper insight forms and aggregation methods, and handling scalability at an unprecedented level. While others are shared with FL, primarily due to the common principles of data decentralization and privacy preservation. However, these common challenges take on distinct forms in FA. Unlike FL, where aggregation primarily deals with model updates, FA requires task-specific aggregation strategies across diverse analytics tasks, making it nontrivial to standardize the framework. Moreover, FA systems often scale to millions or even billions of clients, where every device can contribute to the final analytics results—introducing new challenges in selecting representative clients to ensure statistical reliability. This contrasts with FL, where training a general model can rely on virtual representative datasets obtained via client sampling. Additionally, while avoiding raw data transmission is already a strong privacy guarantee in FL, FA imposes stricter privacy requirements, as descriptive analytics often involve direct statistical summaries. This calls for a more careful trade-off between privacy and utility, further complicated by the scalability of FA systems, making privacy-preserving data aggregation substantially more challenging than in FL.}

\subsection{Diversity of analytics task}

A significant challenge unique to FA lies in the diversity of analytics tasks, which necessitates handling a wide array of data structures, insight extraction methods, and result formats—elements not typically encountered in FL. In FA, tasks vary not only in their objectives but also in the types of input data they handle. Also, this variability requires that FA systems be highly versatile in processing and integrating such diverse data forms to maintain accuracy and consistency in insight extraction. Besides, each analytics task requires specialized aggregation methods designed for the specific type of insights. Furthermore, the output from these tasks also presents varied data structures, from simple numeric aggregates to intricate data visualizations. These aspects significantly complicate the development of universal FA solutions, necessitating tailored approaches that can adapt to the specific requirements of each analytics task while upholding stringent standards of data privacy and system efficiency. \blue{Since FA schemes tackling different data analytics tasks behave quite differently regarding key characteristics. The host data analytics task becomes a good indicator in the taxonomy of FA, as we state in Section \ref{subsec_taxonomy}.}

\subsection{Accuracy, privacy, and performance trade-off}

FA aims to extract and analyze insights from decentralized data sources in a privacy-preserving way. There are trade-offs as regards accuracy, privacy, and performance, which are three of the most important aspects of FA. Firstly, FA schemes utilizing the widely acknowledged privacy-preserving criterion -- DP mechanisms, naturally introduce the trade-off between accuracy and privacy. The details of DP are described in Section \ref{DP} herein. DP mechanisms usually introduce random noise into analysis results. An increase in noise magnitude corresponds to heightened privacy protection since it better obfuscates the contribution of each individual. However, the guarantee of privacy comes at the cost of a reduction in accuracy, and the result deviates from the ground truth due to the injection of noise. Many researchers evaluate, estimate, visualize, and optimize the significant accuracy-privacy trade-off in DP \cite{hay2016exploring,zhao2020not,pannekoek2021investigating,nanayakkara2022visualizing}. Secondly, the trade-off including performance is usually related to another sort of privatization technique related to cryptography. Although it guarantees both privacy and accuracy by encrypting the data of each individual before sharing it and obtaining the real result by corresponding decryption, there is an enormous amount of computation and communication overhead when performing such operations in FA. The trade-offs among privacy, communication, and computation are adequately investigated by researchers in the cryptography society \cite{boyle2017group,ali2021communication,cong2021labeled}. Lastly, there are approximating methods applied in FA to increase the performance of the system which sacrifices the accuracy of results. For example, private sampling algorithms (See Section \ref{subsec_techniques_deployment} for details) utilize less data in the computation and communication process while introducing variance from sampling into the final result. To conclude, balancing these trade-offs is a key challenge in FA. Researchers in \cite{bater2020saqe} treat the multi-dimensional trade-offs as a major design focus in federated query, and such trade-offs are also taken into great account in many FA solutions. How to properly handle the trade-offs when utilizing the diverse privatization tools in FA that fit the characteristics of data analytics scenarios and host task requirements, becomes a critical challenge in FA design.

\subsection{Intensive computation}
In FA, cryptography is commonly used to protect the privacy of each client's data while it is processed and analyzed by multiple parties including the aggregator and other clients. However, cryptographic methods are considerably resource-intensive, both in terms of computation and communication. Operations in cryptography like encryption, decryption, and hashing are computationally intensive. For some advanced cryptographic techniques used in FA like homomorphic encryption and MPC which allow various computations on the encrypted data, the required computation resource is even higher \cite{wang2022protect}. Regarding communication cost, encrypted data tends to be larger than their plaintext counterparts. Furthermore, most cryptographic protocols require multiple rounds of communication between parties \cite{mcmahan2017communication}, exchanging messages like intermediate results. This attribute of cryptography extremely increases the bandwidth required for tasks in FA. 
In addition, managing cryptographic keys is also a complex task in federated settings. They are required to be private and secure in their generation, distribution, rotation, and destruction \cite{yang2018death}. This process will be more complex and computationally intensive if the number of clients increases \cite{roth2019honeycrisp}. These attributes together contribute to the large resource demands for utilizing cryptography in FA, which is a challenge to preserve privacy efficiently in terms of computation costs. \blue{In conclusion, intensive computation is a challenge in many FA solutions, especially those utilizing privatization methodology based on cryptography, according to our taxonomy. Meanwhile, DP-based solutions usually do not introduce a significant computation burden.}

\subsection{Wireless communication}
Wireless communication becomes a major challenge in FA primarily due to its distributed and dynamic nature. 
The slowest participant in this networked system can significantly hinder overall performance, known as the ``straggler effect". This is often exacerbated by varying data transmission rates and unreliable communication channels, leading to delays and inconsistencies in data aggregation. The paper \cite{zang2023general} addresses it in FL by clustering user equipment based on upload times, thereby reducing time divergence among participants. 
In addition, the over-the-air \cite{over-the-air} technique brings efficiency as well as new challenges to FA and FL, such as signal interference and the requirement for precise synchronization. 
Furthermore, the mobility of devices in wireless networks, such as in vehicular networks and aerial networks, introduces another layer of complexity, requiring adaptive strategies for data transmission and model training that can cope with changing network typologies and varying channel conditions. 
The integration of technologies like Reconfigurable Intelligent Surfaces shows promise in enhancing communication efficiency and reliability in these scenarios \cite{mao2023roar}, while they also require sophisticated optimization of resources and careful handling of estimated Channel State Information. 

\subsection{Limited privacy resource}

Privacy resource in the context of FA usually refers to the ``privacy budget'' in the DP mechanisms\cite{liu2024dpbalance}, which intuitively means the amount of allowable privacy loss for each individual within an analysis of the whole data set. Although the information of clients is protected by various privacy-preserving techniques, there is still a risk that repeated analyses such as multiple or iterative queries can expose it. Typically, a portion of the privacy budget is consumed and a specific amount of noise will be added to the results for a query. Managing the privacy budget is important since multiple analyses may be performed on the same group of clients. Every analysis incrementally raises the risk of privacy loss so this cumulative effect ought to be monitored and controlled. Effectively scheduling the privacy budget for the multi-query analysis is a challenge in FA concerning the accuracy of results for each query and total privacy loss for each client. The idea of treating privacy as an allocatable resource is widely accepted, and many prior works present privacy budgeting mechanisms in federated computation systems with a focus on efficiency or fairness \cite{ghodsi2011dominant,li2022dplanner,yuan2022Privacyas,kuchler2023cohere,luo2021privacy,liu2024dpbalance}. \blue{In contrast to the challenge of intensive computation which usually happens in cryptography-based FA solutions, the challenge of limited privacy resource are usually faced by FA solutions with DP-based privatization methodology according to our taxonomy. }



\subsection{Data heterogeneity}

Data heterogeneity, \ie the phenomenon that data possessed by different clients follows heterogeneous distributions, is common in federated systems. As the tenet of privacy preservation restricts raw data from being transmitted, data reallocation or adjustment is usually unavailable for federated systems, making data heterogeneity unavoidable. The challenge of data heterogeneity has become one of the most popular research focuses in the field of FL \cite{flatscale,wang2020optimizing,zhao2018federated,li2019convergence,sattler2019robust,karimireddy2020scaffold}, and also widely exists in FA systems. Since FA systems tackle various data analytics tasks, data heterogeneity introduces diverse influences for different heterogeneous tasks. 
In FL applications, data heterogeneity is usually considered harmful to the convergence and accuracy of FL models.
In \cite{dennis2021heterogeneity}, data heterogeneity is studied in a federated clustering problem, where the data points held by one client are biased to one or several clusters. The authors of \cite{dennis2021heterogeneity} found that data heterogeneity is beneficial for the FA scheme, where a higher analytics performance can be achieved in the heterogeneous data environment. In \cite{wang2021fedacs}, an FA scheme is proposed to measure the severity of data heterogeneity directly. The proposed scheme can figure out clients with low/high data heterogeneity, which provides significant information to other federated tasks in handling the data heterogeneity challenges.

\section{Key Techniques in Federated Analytics}\label{sec_technique}


In this section, we investigate the key enabling techniques that are frequently utilized by FA. These key techniques are categorized by their utilization in FA: In Section \ref{DP}, the privatization techniques, \ie those enhance the privacy preservation of FA, are discussed; in Section \ref{subsec_techniques_analytics}, the analytics techniques, \ie that help derive the analytics results, are presented; in Section \ref{subsec_techniques_deployment}, the deployment and optimization techniques, which improves the system availability, are presented.

\subsection{Privatization techniques} \label{DP}

The tenet of FA prevents the exposure of raw data to any entity other than the data owner (client), which naturally provides privacy preservation to some extent. However, such a ``hiding raw data'' idea fails to provide any formal privacy guarantee. Therefore, existing FA research prefers to enhance the privacy preservation of their schemes by providing some formal and rigid privacy guarantees, by leveraging various privatization techniques. In this part, we analyze the privacy guarantees and the corresponding techniques to realize these techniques.
Since these privatization techniques can usually be applied in arbitrary computation procedures, the following techniques have gained success in various distributed/centralized computation schemes before FA. However, we would like to emphasize that these techniques are uniquely applied in the novel mechanisms and algorithms in FA, compared to other applications. In the following, we will emphasize the novel usage of these techniques in the field of FA.
\blue{We would like to emphasize that the privatization techniques of FA are highly related to our taxonomy of \textit{privatization methodology}. When FA designer decides to apply a particular privatization methodology (like DP or cryptography tools), it becomes necessary to apply the corresponding techniques to enforce the desired privatization methodology.}

\begin{table*}
\caption{Comparison and drawbacks of privatization techniques. The multidimensional performance of privatization techniques is marked as good (+++), medium (++), and bad (+). A privatization technique is considered as good computation/communication efficiency when the execution of such technique introduces small computation/communication overhead.  }
\label{table_drawback_privatization_technique_new}
\centering
\begin{tabularx}{\linewidth}{l|llllll}
\hline
   \textbf{Technique}& \textbf{Privacy}&\makecell{\textbf{Comput.}\\\textbf{efficiency}}&\makecell{\textbf{Commun.}\\\textbf{efficiency}}&\makecell{\textbf{Data utility}}& \textbf{Scalability} & \textbf{Major drawback}\\
\hline
\hspace{2mm}
 CDP & + & +++ & +++ & ++ & +++ & Assumption of trusted aggregator, limitation in local privacy model\\
\hspace{2mm}
 LDP & +++ & +++ & +++ & + & +++ & High magnitude noise, loss in data utility\\
\hspace{2mm}
 DDP & +++ & ++ & ++ & ++ & ++ & Extra overheads from cryptography tools\\
\hspace{2mm}
 $k$-anon. & ++ & +++ & ++ & ++ & ++ & Loss in data utility, limitation in local privacy\\
\hspace{2mm}
 HE & +++ & ++ & +++ & +++ & ++ &  High computation overhead, limited operation support\\

\hspace{2mm}
 MPC & +++ & +++ & ++ & +++ & + &  High communication overhead, limited scalability\\

 \hline


\end{tabularx}  
\end{table*}






\begin{table}
\caption{Summery of applications of key techniques.}
\label{table_privatization_technique}
\centering
\begin{tabularx}{\linewidth}{lll}
\hline
   \textbf{Technique}& \textbf{Task} & \textbf{Reference} \\
\hline
\multirow{4}{*}{LDP} & Mean computation &\cite{cormode2021bit,ding2017collecting,nguyen2016collecting}  \\
 & Medians and percentiles & \cite{bohler2020secure} \\
 & Graph metrics & \cite{pan2022fedwalk,ye2020lf,liu2024edge} \\
 & Frequency-related apps & \cite{erlingsson2014rappor,wang2022fedfpm,apple2017learning} \\
 \hline
 \multirow{2}{*}{CDP} & Mean computation & \cite{roth2019honeycrisp,geyer2017differentially,mcmahan2017learning} \\
  & Frequency-related apps & \cite{zhu2020federated,cormode2022sample} \\
\hline
\multirow{2}{*}{DDP} & Frequency-related apps & \cite{bohler2021secure,bagdasaryan2021towards,wang2024fedweb} \\
 & Database operations & \cite{bater2020saqe,roth2020orchard,roth2021mycelium}\\
 \hline
 \multirow{2}{*}{$k$-anonymity} &  Database operations & \cite{bater2017smcql,davidson2022star} \\
 & Assisting federated learning & \cite{li2020federated} \\
\hline
 \multirow{3}{*}{HE} &  Mean computation & \cite{roth2019honeycrisp} \\
 & Graph metrics & \cite{roth2021mycelium} \\
 & Database operations & \cite{roth2020orchard,margolin2023arboretum} \\

\hline
 \multirow{4}{*}{MPC} &  Mean computation & \cite{roth2019honeycrisp} \\
 & Medians and percentiles & \cite{aggarwal2010secure,tueno2019secure} \\
 & Frequency-related apps & \cite{bohler2021secure}\\
  & Database operations & \cite{volgushev2019conclave,poddar2021senate,zhang2022efficient,tong2022hu,roth2020orchard,margolin2023arboretum} \\


 \hline
 \multirow{2}{*}{Sketching} &  Mean computation & \cite{apple2017learning} \\
 & Frequency-related apps &\cite{erlingsson2014rappor,bonawitz2017practical,bohler2021secure} \\
 

 \hline
 \multirow{2}{*}{\makecell[l]{Specialized \\data structure}} &  Mean computation & \cite{roth2019honeycrisp} \\
 & Frequency-related apps &\cite{zhu2020federated,cormode2022sample,chadha2023differentially} \\


 \hline
 \multirow{2}{*}{\makecell[l]{Optimization \\theory}} &  Database operations & \cite{toka20235g} \\
 & Assisting federated learning &\cite{deng2020fedvision,pandey2021edge,siping23dro} \\


 \hline
 \multirow{3}{*}{\makecell[l]{Game theory and \\incentive design}} & Frequecy-related apps &\cite{shi2023federated} \\
 &  Database operations & \cite{zhao2023crowdfa} \\
 & Assisting federated learning &\cite{yu2022incentive} \\

 \hline
 \multirow{2}{*}{Sampling} & Frequency-related apps &\cite{qin2016heavy,wang2018locally} \\
 &  Database operations & \cite{bater2020saqe,margolin2023arboretum} \\

\hline
\end{tabularx}  
\end{table}

\subsubsection{Local differential privacy}
DP, introduced by \cite{dwork2006calibrating}, is a privatization technique enforced on any data publication mechanism. It has become the most popular criterion for privacy preservation in data analytics, as well as FA. It considers the threat models where the data publication mechanism publishes some outputs based on the raw data, and an adversary tries to infer sensitive information based on the output, such as membership inference attacks \cite{shokri2017membership} and data reconstruction attacks \cite{carlini2019secret}. DP handles such threat models by introducing randomness to the output of the data publication mechanism and restricting the probabilistic distribution of outputs.

LDP, first introduced in \cite{evfimievski2003limiting}\footnote{This paper is published even earlier than DP, and it gives a mathematical definition equivalent to LDP.}, is the most widely used DP variation in FA, as it provides the strong local privacy preservation required by the majority of FA applications. LDP-based FA schemes consider the threat model that the clients upload their insights to an untrusted server, where the untrusted server directly leverages the received insights to perform inferences. LDP is enforced by applying randomness in client uploads so that the probability of any output would not change intensely for clients with arbitrarily different local data. The definition of LDP in an FA setting is given as follows.
\begin{definition}[Local differential privacy]
    Consider a data publication scheme $\mathcal{M}$ executed by the clients in an FA system. Denote $\mathcal{D}$ as the set of all possible client data. $\mathcal{M}$ satisfies $(\epsilon,\delta)$-LDP when 
    \begin{equation}\label{eq_ldpdef}
        \mathbb{P}\big(\mathcal{M}(d_1) = x\big) \leq e^\epsilon  \mathbb{P}\big(\mathcal{M}(d_1) = x\big) + \delta,
    \end{equation}
    for any $d_1, d_2\in \mathcal{D}$, and any possible output $x$ of $\mathcal{M}$.
\end{definition}

$\epsilon$-LDP, a stronger and stricter LDP criterion, can be achieved by setting $\delta=0$ in \eqref{eq_ldpdef}. In the setting of LDP, the data held by a client is considered as a data sample $d\in\mathcal{D}$, and LDP ensures that the probability of deriving any upload will not change intensely for clients possessing different data samples.
An FA scheme reinforced by LDP can guarantee that the insight uploaded to the server cannot be leveraged by an adversary to efficiently infer the raw client data, from the probability theory perspective.

The satisfaction of LDP requires perturbation (randomness) on the client uploads. 
For the binary data, LDP is typically realized by the randomized response, \ie flipping each bit with some probability. The randomized response-based LDP mechanisms usually work in FA applications where the local computation results are naturally in the binary form \cite{cormode2021bit,wang2022fedfpm,qin2016heavy,acharya2019communication} or the client data are encoded into a binary form with sketching techniques \cite{erlingsson2014rappor,bassily2020practical,apple2017learning}.
For the continuous scaler data, LDP is typically realized by introducing continuous scaler noise on the data. Gaussian mechanism and Laplace mechanism are the most widely used schemes to realize LDP for continuous scaler data. The Gaussian mechanism is applied by adding a random noise following the Gaussian distribution into the uploaded scaler, which can guarantee the corresponding $(\epsilon,\delta)$-LDP \cite{wei2021user,kim2021federated}.
Laplace mechanism, introducing a random noise following the Laplace distribution, can satisfy the stronger $\epsilon$-LDP, which have been utilized in \cite{nguyen2016collecting,zhou2021local,margolin2023arboretum}.

LDP has been applied in numerous privacy-preserving algorithms other than FA. When applied in FA, many novel utilizations are present. In conventional non-FA applications, LDP noise is usually directly applied to raw data. In FA applications, LDP is in contrary applied on the insight uploaded to the server. In many FA designs, the insights are designed to have a much smaller dimension compared to the raw data, which decreases the magnitude of LDP noise. These examples include the bloom filter in \cite{erlingsson2014rappor}, count mean sketch \cite{apple2017learning}, and even one-bit insight \cite{wang2022fedfpm}.

\blue{\textit{\textbf{Lessons learned.} DP mechanisms provide statistical privacy preservation by adding noise to the client uploads so that the original values cannot be accurately recognized by an adversary. LDP is the DP variation that provides the strongest statistical guarantee. FA considers an adversarial environment without any trusted aggregator, where each client adds significant noise to their uploads to protect their record against any potential aggregator, which fits most of the FA settings. LDP is realized by requiring clients to individually add sufficient noise so that the noise upload cannot reveal sensitive information about the original values. While LDP provides an advantage in strong privacy preservation, its major drawback is the significant magnitude of noise required to satisfy LDP, which may ruin the data utility and limit the design of the insight structure.}}

\subsubsection{Central differential privacy} CDP, also known as global differential privacy, is the original form of DP. It considers the scenario that a data analytics mechanism processes a database containing many data samples, and the published analytics results can protect the raw data of any single data sample from being inferred by an adversary. The definition of CDP is as follows.
\begin{definition}[Central differential privacy]
    Consider a data publication mechanism $\mathcal{M}$. Denote a pair of neighboring databases $\mathcal{D}$ and $\mathcal{D}'$ that differs in at most one data sample. $\mathcal{M}$ satisfies $(\epsilon,\delta)$-CDP when
    \begin{equation}\label{eq_cdpdef}
        \mathbb{P}\big(\mathcal{M}(\mathcal{D}) = x\big) \leq e^\epsilon  \mathbb{P}\big(\mathcal{M}(\mathcal{D}') = x\big) + \delta,
    \end{equation}
    for any $D, D'$, and any possible output $x$ of $\mathcal{M}$. 
\end{definition}

Similarly, $\epsilon$-CDP can be achieved by setting $\delta=0$ in \eqref{eq_cdpdef}. When enforcing the same privacy parameters, CDP usually requires much smaller noise than LDP, which can improve the correctness of the final analytics results.

Compared to LDP, where the client data are considered data samples and directly protected, CDP cannot be directly applied to many FA scenarios because it considers enforcing protection on the data already aggregated. In the FA literature, there are mainly two settings to enforce CDP. In the first setting, the raw data of each client is considered as a database consisting of multiple data samples. The clients enforce CDP on the database, so that information about each sample is protected. FA research considering this setting includes \cite{hu2020personalized,bater2020saqe}. In the second setting, similar to the setting of LDP, the data within a single client is considered as a data sample in $\mathcal{D}$. The CDP scheme guarantees that an adversary cannot infer the sensitive information of a client by analyzing the final FA result. However, such privacy attacks can succeed when the adversary has access to the uploads of the clients (\eg when the server acts as the adversary). Therefore, the second setting usually assumes a trusted aggregator (server) and sacrifices the formal privacy guarantee. The second setting has been considered in \cite{geyer2017differentially,mcmahan2017learning,zhu2020federated,cormode2022sample}.

These exist diverse techniques to realize CDP. Similar to those used for LDP, the Gaussian mechanism can still be applied for CDP to protect continuous scaler output, which can satisfy $(\epsilon,\delta)$-CDP \cite{hu2020personalized,bater2020saqe,geyer2017differentially,mcmahan2017learning}.\footnote{Laplace mechanism also has theoretically applicable for CDP, but is not selected for researchers due to the high variance of Laplace noise.} Compared to the LDP case, where Gaussian noise is added to each single data sample, applying the Gaussian mechanism for CDP only requires adding noise to the aggregated result. Therefore, the overall noise on the final analytics results is significantly reduced. 
The exponential mechanism tackles the analytics tasks where a single sample in the database serves as the output (\eg the federated median computation task). It realizes $\epsilon$-CDP by calculating a utility score (preference of serving as the output) for each data sample, and then selecting the output data sample based on the probabilistic distribution characterized by the utility scores. The exponential mechanism has been applied to realize CDP and also DDP, which will be introduced later. \cite{bohler2020secure}
Conducting sampling on the data samples in the database is another effective tool to realize CDP. When only a random portion of data samples are selected for the data analytics task, randomness is introduced to the final analytics results, which can satisfy the criterion of CDP in \eqref{eq_cdpdef} after proper mathematical proof. Two approaches are considered to utilize the sampling technique. In \cite{zhu2020federated,cormode2022sample}, sampling is applied to directly realize CDP without any noise injection procedure; in \cite{bater2020saqe,margolin2023arboretum}, sampling can reinforce a scheme that already satisfies DP via noise injection, by deriving the stronger privacy preservation (decreasing the $\epsilon$ parameter). The latter approach can be applied for both reinforcing CDP and LDP.
This approach to realizing CDP is novel as an FA mechanism because it can be realized only when the sampling of clients is available. It is not appropriate for conventional centralized and distributed algorithms where the sampling of clients is not available.

\blue{\textit{\textbf{Lessons learned.}  CDP assumes that the raw data are already gathered by an aggregator, and the noise helps protect individual data records when the aggregated statistics are released. Since conventional FA settings usually do not assume a trusted aggregator, and the clients should preserve data privacy against the server, FA solutions utilizing CDP essentially sacrifice privacy preservation to some extent. Compared to LDP, CDP requires less noise to the uploads and even can be realized via sampling without adding noise to the raw value. CDP can gain higher data utility under weaker adversarial settings.}}

\subsubsection{Distributed differential privacy}
As mentioned above, LDP can achieve a stronger local privacy guarantee, without the assumption of a trusted aggregator; CDP introduces randomness with a smaller magnitude, which derives a higher correctness in final analytics results. DDP is an attempt the combine the advantages of LDP and CDP, or an attempt to apply CDP in the local privacy model. DDP statistically satisfies CDP to privatize the aggregated result. In addition, DDP requires clients to utilize cryptography tools to encrypt their uploads. With the encryption, the aforementioned local privacy risk existing in CDP, that an adversary can infer sensitive information when observing the individual client upload, is then eliminated. \blue{The idea of DDP is inspired by \cite{dwork2006our}, and the primary formal DDP designs is proposed in \cite{shi2011privacy,mcsherry2009privacy}, which are much earlier than the introduction of the term ``DDP''. Essentially, DDP describes the idea of extending the privacy preservation of CDP to the local model in the distributed setting with the help of cryptography tools.}

The deployment of DDP consists of two phases: distributed noise generation and encrypted aggregation. In the distributed noise generation phase, each client injects noise into their local uploads. The noise is deliberately designed so that the aggregated noise of the clients can satisfy the criterion of CDP. In the encrypted aggregation phase, the client upload is possessed by distributed noise generation and uploaded to the server with encryption. The server cannot observe the individual upload of any client, but can only observe the CDP-enforced aggregated uploads.

DDP algorithm designers can freely choose and combine the techniques for distributed noise generation and encrypted aggregation. Many options for these techniques are surveyed in \cite{goryczka2015comprehensive}. DDP has been applied in FA systems, like \cite{bagdasaryan2021towards,bohler2020secure,bohler2021secure,wang2024fedweb}.

\blue{\textit{\textbf{Lessons learned.} DDP utilizes cryptography tools to force an aggregator unable to observe raw uploads during aggregation so that the magnitude of required noise can be reduced, which provides an option of privacy preservation with higher data utility (equivalent to CDP) in the typical FA setting without a trusted aggregator. However, CDP introduces cryptography tools into its design realm, making DDP mechanisms have much higher computation and communication overheads, compared to conventional DP mechanisms.}}

\begin{figure}[t]
\centering
\begin{picture}(300,180)
\put(0,0){
\includegraphics[width=0.8\linewidth]{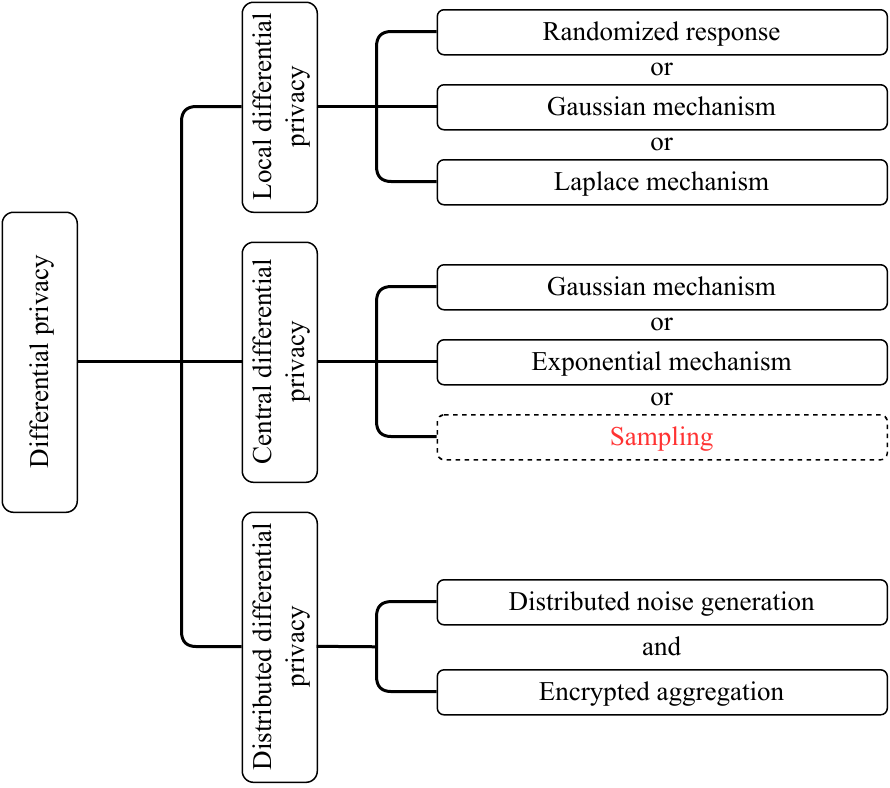}
}
\put(208,174){\tiny\cite{erlingsson2014rappor,cormode2021bit,wang2022fedfpm,qin2016heavy}}
\put(208,167){\tiny\cite{bassily2020practical,acharya2019communication,apple2017learning}}
\put(208,153){\tiny\cite{wei2021user,kim2021federated}}
\put(208,136){\tiny\cite{nguyen2016collecting,zhou2021local,margolin2023arboretum}}
\put(208,112){\tiny\cite{hu2020personalized,bater2020saqe,geyer2017differentially,mcmahan2017learning}}
\put(208,95){\tiny\cite{bohler2020secure}}
\put(208,82){\tiny\cite{zhu2020federated,cormode2022sample}}
\put(208,75){\color{red}\tiny\cite{bater2020saqe,margolin2023arboretum}}
\put(208,30){\tiny\cite{bagdasaryan2021towards,bohler2020secure,bohler2021secure,wang2024fedweb}}
\end{picture}
\caption{Summery of DP variations, their enabling techniques, and existing works applying them. Sampling is not only able to solely satisfy CDP, but also can enhance the privacy preservation of any existing CDP/LDP scheme. Applications following the latter approach are marked {\color{red} red} in the figure.}
\label{fig_dp_variations}
\end{figure}

In Fig. \ref{fig_dp_variations}, we summarize the three kinds of DP variations utilized by FA, and the corresponding techniques to realize these DP variations.

\blue{DP has proven its effectiveness in privacy preservation from the statistical perspective. However, it is sometimes hard to design mechanisms to realize the DP criteria defined in \eqref{eq_ldpdef} or \eqref{eq_cdpdef}, especially in cases when the privatized data is in high dimensionality or complex structure. Therefore, researchers design relaxations of DP that are easier to satisfy and meet particular privacy needs. Examples of these DP criteria include R\'enyi DP \cite{mironov2017renyi}, concentrated DP \cite{dwork2016concentrated}, and dependent DP \cite{liu2016dependence}. Variations of DP are also introduced to handle specific data types with specific definitions of adjacent datasets or units of privacy preservation, such as sample label \cite{chaudhuri2011sample}, streaming data \cite{bao2021cgm}, graph data \cite{kasiviswanathan2013analyzing}, location data \cite{andres2013geo}, and natural language data \cite{feyisetan2020privacy}. Generalizations of DP, like pufferfish privacy \cite{kifer2012rigorous} and blowfish privacy \cite{he2014blowfish}, are also introduced to handle the limitations of DP.}


\newcommand{\privatizationfigwidth}{0.4\textwidth}
\begin{figure*}[!tbp]
    \centering
    \subfloat[Cryptography-based aggregation]
    {
        \includegraphics[width=\privatizationfigwidth]{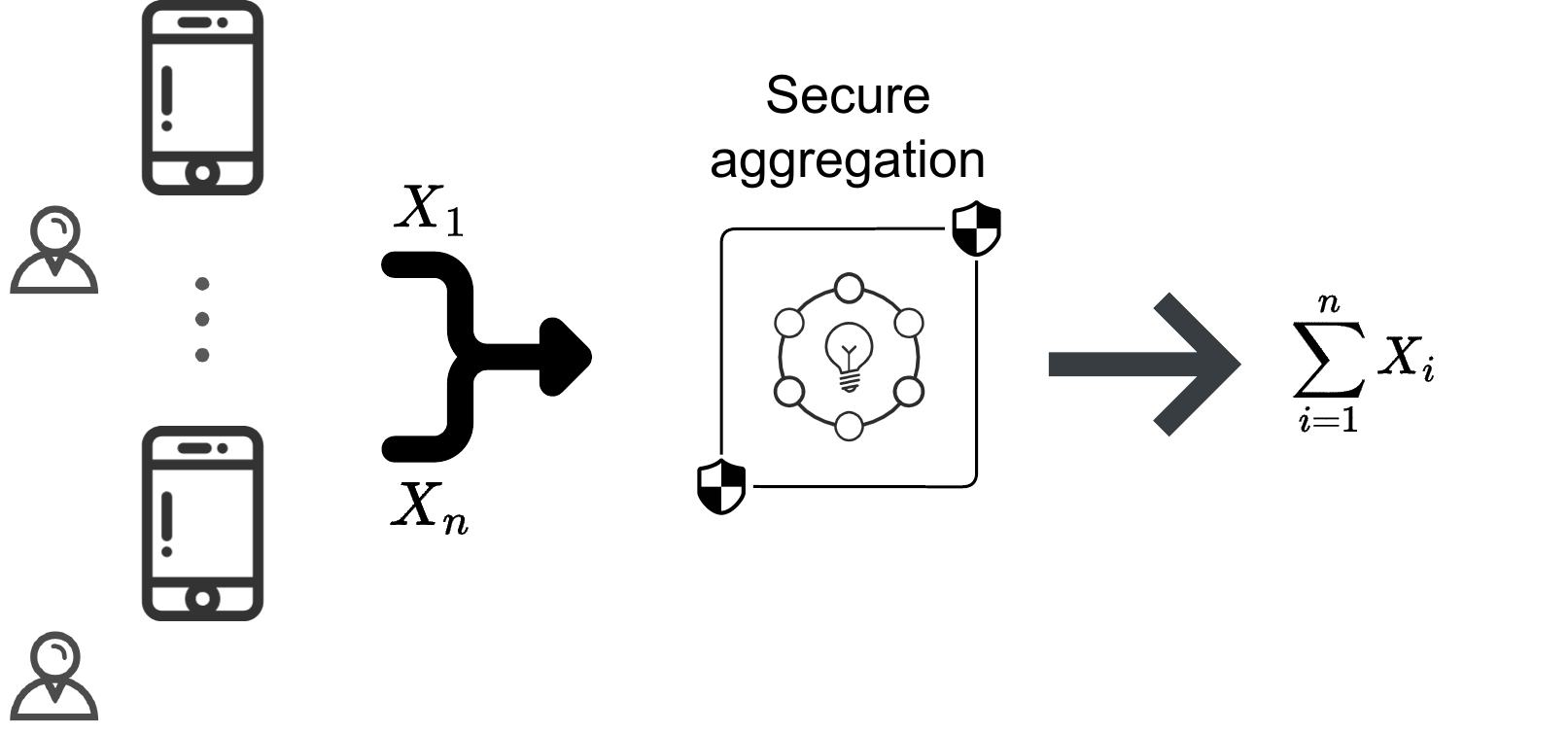}
        \label{subfig_privatization_1}
    }
    \hspace{10mm}
    \subfloat[$k$-anonymity]
    {
        \includegraphics[width=\privatizationfigwidth]{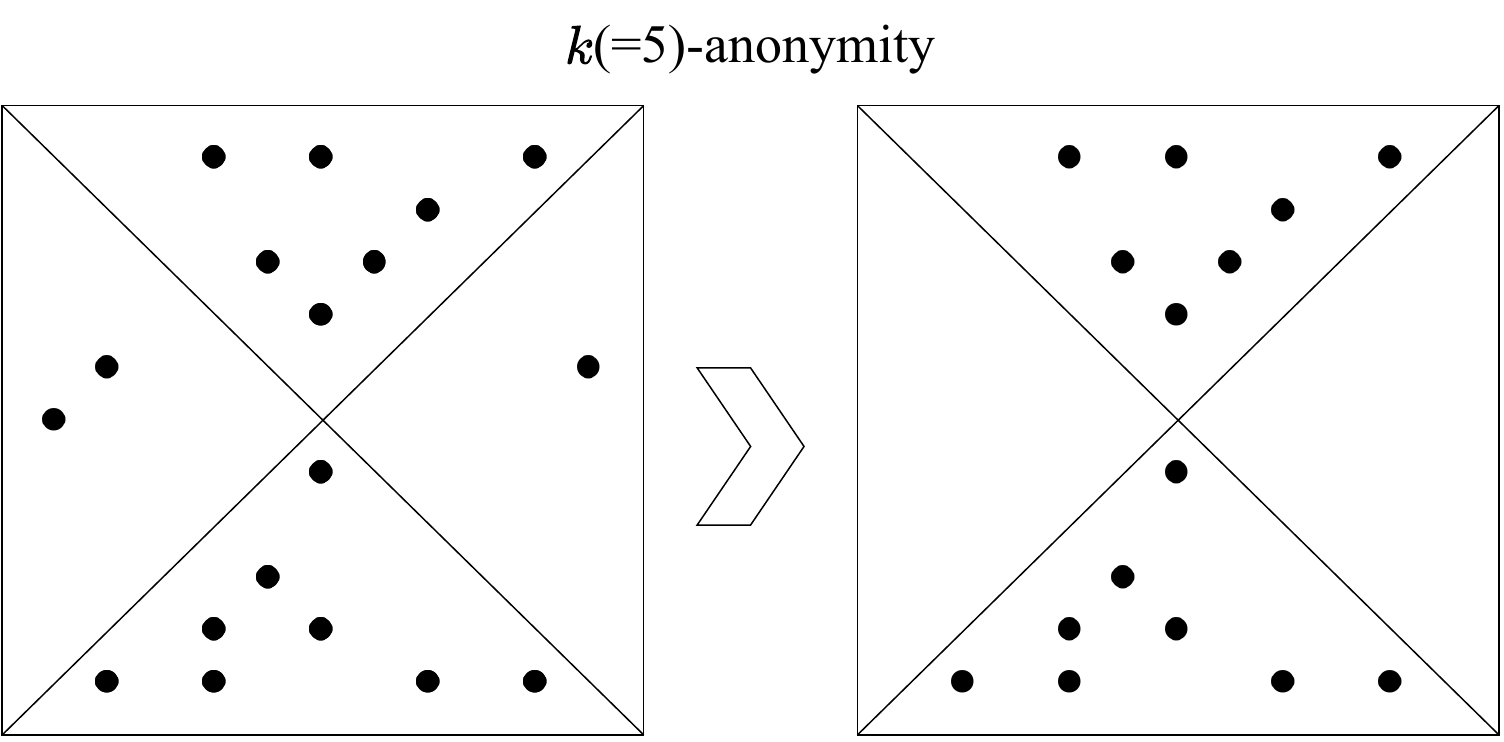}
        \label{subfig_privatization_2}
    }
    \\
    \centering
    \subfloat[LDP]
    {
        \includegraphics[width=\privatizationfigwidth]{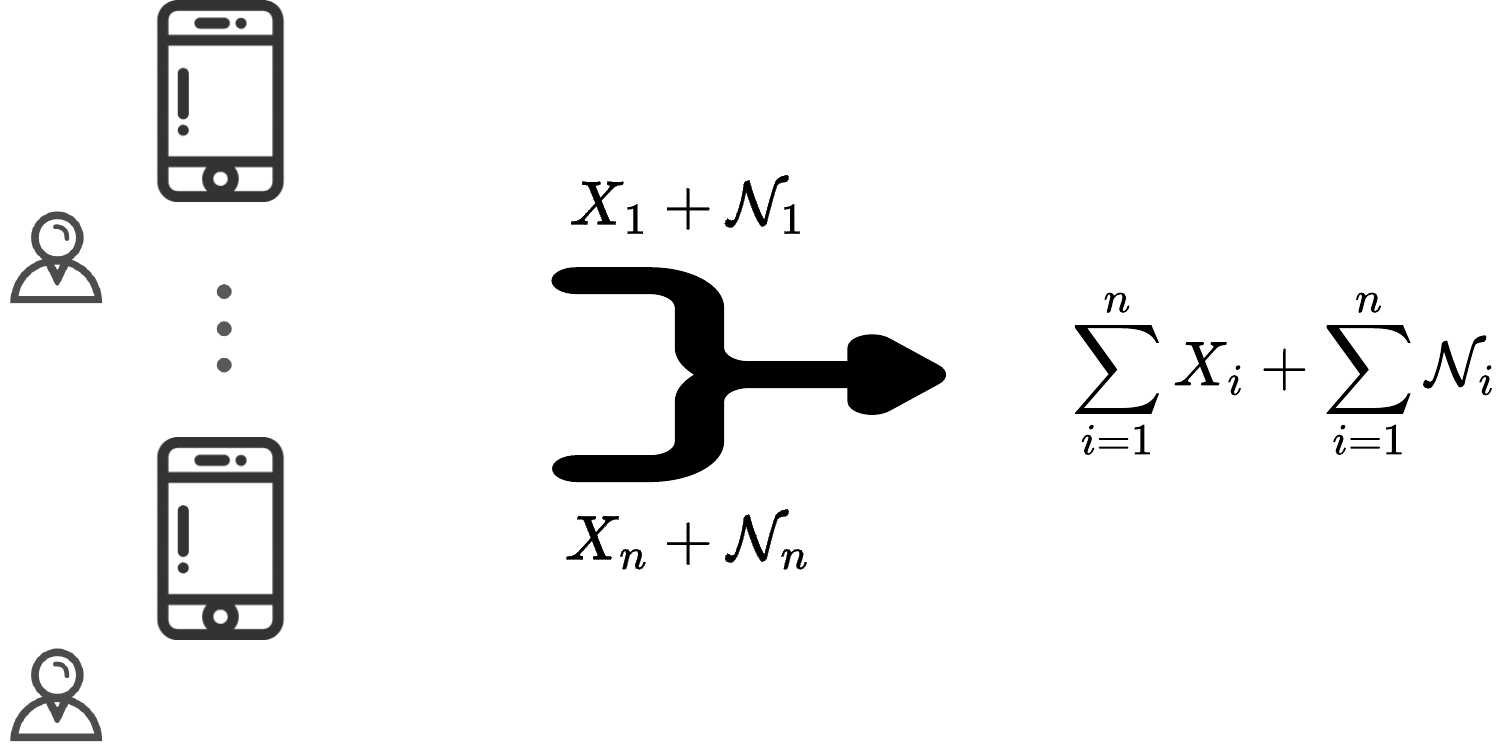}
        \label{subfig_privatization_3}
    }
    \hspace{10mm}
    \subfloat[DDP]
    {
        \includegraphics[width=\privatizationfigwidth]{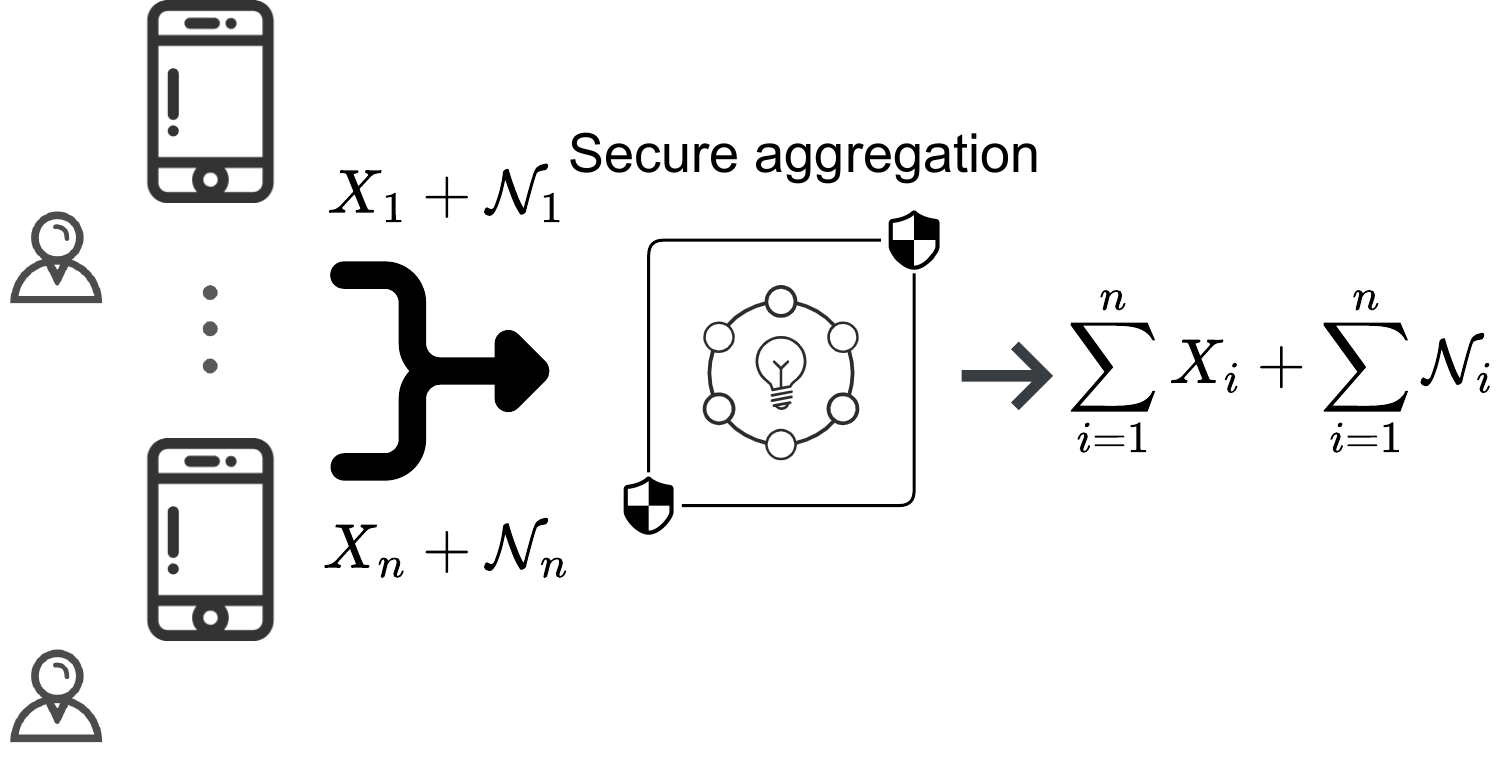}
        \label{subfig_privatization_4}
    }
    \newline
    \caption{Summery of some privatization techniques in FA}
    \label{fig_privatization}
\end{figure*}

\subsubsection{$k$-anonymity} The concept of $k$-anonymity emerged as a privacy-preserving technique, specifically crafted to safeguard individual data in publicly accessible datasets, addressing the growing concerns over personal data security and privacy breaches \cite{samarati1998protecting,de2023real,mahanan2021data}. The main idea is to ensure that any specific individual cannot be linked given any record in the dataset by guaranteeing that any combination of attribute values will indicate at least $k$ individuals. In other words, for every individual in the dataset, there are at least $k-1$ other members with the same information thus it is rarely possible to search out an individual based solely on certain attribute values. To achieve $k$-anonymous, some attributes are either removed or suppressed. There are three types of attributes in datasets: identifiers, non-identifiers, and quasi-identifiers. Quasi-identifiers are attributes that can potentially distinguish individuals together such as age, zip code, and gender, none of which can identify an entity alone. Generally, identifiers are suppressed and quasi-identifiers are either removed or generalized such as replacing the ages with age ranges to achieve $k$-anonymous.

It is a suitable tool in FA in terms of preventing the leakage of sensitive information and the identification of individuals. Usually, $k$-anonymity is utilized in phases before sharing information. The risk of privacy breaches is reduced by guaranteeing that the dataset of each client follows $k$-anonymity before sharing with the server or each other. In \cite{bater2017smcql}, they designed the Private Data Network to enable querying in databases of different parties without disclosure of their raw data by applying $k$-anonymity before data sharing. With the trusted third party as the query planner and results collector, the $k$-anonymous data of clients are then queried and aggregated to solve the questions from users. This setting is particularly relevant to collaboration in clinical research of healthcare clients. There are other methods to achieve $k$-anonymous. Researchers in \cite{li2020federated} utilized sampling and tree pruning to fulfill $k$-anonymity for obtaining DP without noise injection. In \cite{davidson2022star}, authors applied a more straightforward method that only the data sent by more than $k$ clients are chosen in the aggregation phase.

\blue{\textit{\textbf{Lessons learned.} While k-anonymity is a versatile privacy-preserving technique, its application in FA is uniquely suited to address the challenges of data aggregation and querying without compromising individual privacy. The specific methods and practical implementations of k-anonymity are uniquely applied in FA, such as pre-sharing anonymization and the use of trusted third parties. However, k-anonymity has notable limitations that can affect its efficacy. It is vulnerable to attribute linkage attacks, where external knowledge can be used to re-identify individuals, and to homogeneity attacks, where all records in a group have the same sensitive attribute, making them easily identifiable. Achieving k-anonymity often leads to significant data utility loss due to generalization and suppression techniques, which is further exacerbated in datasets with many quasi-identifiers because of the curse of dimensionality.}}

\subsubsection{Homomorphic encryption} Homomorphic encryption (HE) is a powerful privacy-preserving cryptographic method allowing clients to perform mathematical computations such as addition and multiplication on encrypted data \cite{naehrig2011can}. \blue{The output of HE computation is still in the encrypted form, and when the output is decrypted, the result will be identical to the corresponding computation results on encrypted raw data.} HE plays an important role in FA in that it enables the aggregation of encrypted data without access to clients' plaintext data. The collected encrypted data are then performed with mathematical operations to obtain a result which is also encrypted with support of its homomorphic attribute. The outcome is usually sent to a trusted third party holding the private key for decryption and is finally revealed to the querier. In this way, clients cooperate on data analysis tasks but raw data never leaves the local position of their owners. There are several types of HE, including fully \cite{gentry2009fully,gentry2011implementing,van2010fully} and partially \cite{shafagh2017secure,cominetti2020fast,liu2019secure} HE. Specific implementations in FA, such as private set intersection and large-scale data aggregation without a trusted core, showcase its crucial role in privacy-preserving analytics. In contrast, in FL, HE is used more for securing model updates during training rather than for data aggregation, highlighting its distinct and central role in FA.

HE was early applied in private set intersection tasks\cite{chen2017fast,freedman2016efficient,freedman2004efficient,huang2012private,zhang2022efficient}. Generally, in the two-party setting, the sender and receiver first agree on an HE scheme and the receiver generates a public-private key pair. The receiver then sends the encrypted set by public key to the sender who compares it with her encrypted set and sends the intersection back. Finally, the receiver revealed the encrypted intersection entities by the private key. The homomorphic attribute plays a crucial role in the faster computation of the phase in which the sender identifies common encrypted elements. HE is also a suitable tool when there’s a requirement for allowed mathematical operations such as total summation in aggregation parts. In \cite{roth2019honeycrisp}, researchers proposed Honeycrisp to solve the count mean sketch task at a large scale without a trusted core. A committee randomly selected from users generates keys and each client sends their additively homomorphically encrypted data by “two-element” Ring-LWE-based encryption to the server which obtains the total encrypted sum and sends it to the committee for decryption. 

\blue{\textit{\textbf{Lessons learned.} HE is a powerful cryptographic technique that enables mathematical operations on encrypted data without decryption, ensuring privacy in data analysis. In FA, HE allows secure aggregation of encrypted data from multiple clients without exposing their raw data, with the final result sent to a trusted third party for decryption. HE plays a crucial role in privacy-preserving tasks such as private set intersection and large-scale data aggregation, offering strong security guarantees. While its use in FL focuses on securing model updates, in FA, HE primarily supports secure data aggregation, highlighting its distinct roles in different privacy-preserving contexts. However, HE has notable drawbacks that can hinder its practical adoption. Operations on encrypted data are computationally intensive, leading to high processing times, especially for complex computations, and the encrypted data often requires significantly more storage and communication bandwidth due to large ciphertext sizes. Additionally, while Fully Homomorphic Encryption supports arbitrary computations, it is resource-intensive and less practical for real-time applications, unlike Partial Homomorphic Encryption, which has limited functionality. Furthermore, managing and securely distributing encryption keys adds complexity, making implementation and maintenance challenging.}}





\subsubsection{Secure multi-party computation} Secure multi-party computation (MPC) aims to jointly compute functions with inputs from several parties while guaranteeing the privacy of these inputs \cite{ben2008fairplaymp}. \blue{MPC allows the computation with input held by multiple parties, where the result of the computation result can be derived while the input data are held private. MPC ensures privacy and security in FA by allowing multiple parties to collaborate on tasks without compromising the confidentiality of individual data sets of each party, utilizing the technologies of oblivious transfer and secret sharing.} MPC focuses less on the requirement for a large volume of data to maintain privacy and accuracy when compared to DP. Nevertheless, it imposes a heavier burden in terms of computational and communication costs \cite{asharov2012multiparty,wang2017global}. While it can excel in preserving privacy and ensuring precise results with any data size, the trade-off lies in the heightened demand for computational resources and communication overhead among participated parties, rendering it a more resource-intensive approach. Some cryptographic techniques, like HE, garbled circuits, or secret sharing, are usually utilized to enable secure computation among parties. The choice of technique depends on the specific requirements and constraints of various FA tasks. In the computation phase, only intermediate results are shared instead of users’ raw data, which are also usually encrypted to guarantee the throughout the whole process. 

Some researchers directly use MPC as a privatization tool to encrypt the computation results and provide privacy preservation for certain analytics algorithms. MPC is used in tasks like secure communication of the median \cite{aggarwal2010secure,tueno2019secure}, multi-party set intersection \cite{kolesnikov2017practical}, and searching heavy hitters \cite{bohler2021secure}. In \cite{aggarwal2010secure}, it focuses on securely computing the k-th-ranked element (such as the median) of combined confidential datasets from multiple parties. The protocols employ techniques like binary search and consistency checks to ensure privacy while computing the median. While \cite{tueno2019secure} addresses distributed private learning and focuses on computing the median in a way that is both differentially private and efficient in terms of computation and scalability. It utilizes an MPC approach to compute the exponential mechanism for the median, which is also adaptable for other rank-based statistics and machine learning optimizations. In \cite{kolesnikov2017practical}, the authors introduce an efficient method for multi-party private set intersection (PSI), allowing multiple parties to compute the intersection of their datasets without revealing extra information. It uses a new paradigm based on oblivious programmable pseudorandom functions (OPPRF) and avoids computationally expensive public-key operations. Also, the protocol is secure against colluding semi-honest parties and has been demonstrated to be practical for up to 15 parties, each with datasets of a million items. Researchers in \cite{bohler2021secure} present efficient MPC protocols, HH and PEM, to compute differentially private heavy hitters. These protocols use sketches for approximate counts, offering better accuracy than LDP, and are practical for different data sizes: HH is suitable for small datasets (running time linear in data size), while PEM is more efficient for larger datasets (running time sublinear in data domain). The protocols are more accurate than LDP approaches and have an accessible performance on computation and communication costs. 

In addition, some researchers design MPC protocols and platforms for FA applications, rather than directly using MPC to complete certain FA tasks \cite{volgushev2019conclave,zheng2021cerebro,poddar2021senate,zhang2022efficient,tong2022hu}.
The protocol \& platform-based approach provides the users the capacity to apply the MPC scheme to customized tasks and raises the possibility of multipurpose FA systems.
Conclave \cite{volgushev2019conclave} is a query compiler designed to accelerate relational analytics queries by combining data-parallel local cleartext processing with smaller MPC steps. It offers a hybrid MPC-cleartext protocol for cases where parties trust others with specific subsets of data. Cerebro \cite{zheng2021cerebro} is an end-to-end collaborative learning platform that enables multiple parties to compute learning tasks without sharing plaintext data. It addresses the need of organizations to collaboratively use sensitive data, comply with policy regulations, and deal with business competition. Cerebro balances system design for safe collaboration with release policies and auditing, aiming to simplify the complex performance trade-offs between different MPC protocols. Authors in \cite{poddar2021senate} introduce Senate, a system for MPC that allows multiple parties to collaboratively execute analytical SQL queries without exposing individual data, even with malicious adversaries. The main idea of Senate is MPC decomposition which enhances computational efficiency by breaking down cryptographic computations into smaller, parallel units. In \cite{zhang2022efficient}, researchers focus on advancing the field of skyline queries, which are crucial for multi-criteria decision-making systems. It proposes a local dominance-based framework to enhance the efficiency of skyline queries in a vertical data federation setup. The framework decomposes skyline queries into more manageable units, improving the overall query process’s efficiency and security. Hu-Fu \cite{tong2022hu} is the first system dedicated to efficient and secure spatial query processing in a data federation context by optimizing the balance between plaintext and secure operations, minimizing the use of secure operators while maintaining the process's overall security. It’s a significant advancement in data federation that Hu-Fu parses federated spatial queries written in SQL, decomposes them into secure and plaintext components, and securely collects query results, offering a practical solution to the challenges of processing spatial queries securely and efficiently. 
Besides, authors in \cite{roth2019honeycrisp} designed a protocol to distribute the secret key of HE to members of a committee randomly composed of several clients. After the aggregator obtains the total encrypted summation by public key from clients, it sends the result to the committee who then applies MPC to ``fix" the secret key and decrypt the result before revealing it. Functional encryption is another technique related to MPC. It expands upon public-key encryption by allowing an individual with a secret key to decipher a particular function derived from the encrypted content in the ciphertext. In \cite{siping2022federated}, authors proposed the FAA-DL where clients and the server can collaboratively and proactively analyze anomalies based on functional encryption.

\blue{\textit{\textbf{Lessons learned.} MPC plays a pivotal role in ensuring privacy and security in FA by enabling multiple parties to collaborate without revealing their datasets. MPC ensures that only intermediate results are shared, and the raw data remains private throughout the process. However, it comes with a significant trade-off in terms of computational and communication costs, which can be resource-intensive. Despite this, MPC excels in preserving privacy and delivering accurate results, regardless of data size, making it suitable for tasks such as secure set intersection, median computation, and heavy hitters search. Furthermore, the development of MPC protocols and platforms has paved the way for more efficient, flexible, and scalable FA systems, allowing users to apply MPC to various customized tasks. These systems balance the complexities of privacy preservation with system performance, offering practical solutions for secure collaborative data analysis across different domains.}}

In Fig. \ref{fig_privatization}, we demonstrate the mechanism of some privatization techniques. For the cryptography-based aggregation, LDP, and DDP, we take the sum computation task as an example.

\subsection{Analytics techniques}\label{subsec_techniques_analytics}

With the development of FA, there's a paramount requirement for effective and accurate techniques to aggregate insights across decentralized clients. It is communication-intensive to upload encrypted datasets and it is insecure to expose data even with noise while some transformation of data structure may effectively address these challenges. In this section, we discuss the data sketching and some interesting specialized data structures that artfully assist FA.




\subsubsection{Sketching} Data sketching is a technique for creating compact summaries (sketches) of large datasets that approximate key properties, such as cardinality, frequency, or quantiles \cite{wei2015persistent,solar2008sketching,cormode2007sketching}, enabling efficient computation with reduced memory and processing requirements. In FA, the pursuit of insight aggregation across decentralized data nodes demands techniques that are both efficient and scalable. Sketching stands out because it is a method designed with specific attributes tailored for the challenges posed by FA. These attributes \cite{cormode2017data} are query-specific (designed for specific insight extraction tasks), mergeable (allowing for easy aggregation from multiple clients), and extremely compact (ensuring minimal communication cost). Furthermore, various sketching methodologies have been developed to cater to specific data types and tasks. For instance, for set data, the Bloom filter is suitably employed for set intersection tasks. Similarly, the count sketch method is adept at identifying heavy hitters. To address the paramount privacy requirements in FA, sketching often collaborates with privatization methods such as noise addition, randomized responses, and hash functions to ensure the privacy and security of the aggregated insights. Together, these techniques present a robust and private solution for insight aggregation in FA. Unlike FL, which primarily focuses on model training, FA leverages sketching to handle specific data aggregation tasks such as set intersections and heavy hitter identification. Moreover, sketching in FA often integrates privacy-preserving techniques to ensure the confidentiality of the aggregated insights. These attributes highlight the unique and critical role of sketching in FA, distinguishing its application from other areas.

Bloom filters \cite{bloom1970space} is a space-efficient probabilistic data structure used to test whether an element belongs to a set with possible false positives and no false negatives. Bloom filter and its variants are first introduced to FA in Private Set Intersection tasks \cite{kerschbaum2012outsourced,many2012fast,dong2013private,pinkas2014faster}. Researchers in \cite{kerschbaum2012outsourced} combined Bloom filters with Goldwasser Micali HE to compute the intersection which is secure in a malicious model. In \cite{many2012fast}, authors proposed a PSI method using Bloom filters with a secure multiplication protocol SEPIA to obtain an intersection of Bloom filters for each party to find their intersections. Authors in \cite{dong2013private} reduced the hash operations to propose a faster protocol for the PSI task by integrating oblivious transfer with a garbled Bloom filter (GBF). The essential difference between the GBF and Bloom filters is that the GBF uses an array of $\lambda$-bit string where $\lambda$ is a security parameter while Bloom filters use an array of bits. Specifically, the client computes its sets to a Bloom filter and the server also computes its set to a GBF. Then an oblivious transfer protocol is conducted to give the client a GBF of intersection which is finally queried by the client to obtain the result. There were continuing works on Dong \etal\cite{dong2013private} such as optimization of performance based on random OT extension\cite{pinkas2014faster}, enhancement in its malicious-secure variant \cite{rindal2017improved}, and multi-parties intersection\cite{kolesnikov2017practical,inbar2018efficient}. RAPPOR \cite{erlingsson2014rappor} introduced by Google is a widely applicable and practical data collection mechanism providing strong privacy guarantees with high utility. The basic idea of RAPPOR is applying randomized responses to Bloom filters so that each client reports each bit of her Bloom filter with possible untruthful responses to the server for privacy guarantees.

The counter, a common data sketching method, is a dictionary data structure storing the number of occurrences of elements. Apple developed two LDP algorithms based on it: Count Mean Sketch (CMS) and Hadamard Count Mean Sketch (HCMS) to learn dictionary-related tasks such as learning particular words and discovering popular emojis \cite{apple2017learning}. The main idea is to index counters sketched by raw data at the client side by hash functions and Hadamard transform with randomized responses. Researchers in \cite{bonawitz2017practical} proposed TreeHist protocol to solve heavy hitters problem. The protocol utilized a local randomizer which could be regarded as a sampled and noisy version of the count sketch. The sketch is then Hadamard transformed to bits and one bit is sampled for subsequent submission to the server and aggregation to binary prefix tree. In \cite{bohler2021secure}, authors proposed methods that encode counters by Laplace noise addition without costly reconstruction like hash-based techniques under a central DP setting. They saved the computation resources of the aggregator by avoiding reconstructing results from perturbed messages. 

\blue{\textit{\textbf{Lessons learned.} Sketching in FA highlights its critical role in addressing scalability, efficiency, and privacy challenges. Sketching techniques, such as Bloom filters and Count Sketch, stand out for their compactness, mergeability, and task-specific designs, making them well-suited for FA tasks like heavy hitter identification and set intersections. These methods ensure low communication costs while enabling accurate insight aggregation from distributed nodes. To preserve privacy, sketching is often integrated with techniques like noise addition, randomized responses, and hash functions. For example, Bloom filters are enhanced with differential privacy mechanisms like randomized responses in tools such as RAPPOR, while Count Sketch is combined with Hadamard transforms for tasks like frequency estimation. These innovations underscore the adaptability of sketching in balancing privacy and utility in decentralized environments. Sketching proves indispensable for scalable, privacy-preserving insight aggregation in FA.}}
 




\subsubsection{Specialized data structures}
In the context of FA where data analysis is performed across federated devices and servers, some specialized data structures are proposed to help the server to derive the analytics results on particular tasks. These specialized data structures are uniquely tailored to the requirements of FA, facilitating accurate and secure data aggregation.
In \cite{zhu2020federated}, authors proposed the TrieHH algorithm discovering frequently typed words in edge devices, which is an iterative algorithm to find heavy hitters with the Prefix Tree. The Prefix tree, or the Trie, is a tree-like data structure used to store strings like words. Each node in the Trie represents a single character of a string and all end nodes represent an end identifier so that each end node stands for a string and all sibling nodes own a common prefix. In their TrieHH algorithm, a subset of clients is sampled in each iteration and the Trie is updated based on their data points after filtering by a threshold until convergence. This sampling-and-threshold algorithm is used to provide 
CDP instead of the usual noise addition. There’s a following work of it, TrieHH++\cite{cormode2022sample}, which answered more general queries that not only identified heavy hitters but also provided their estimated frequency. 
In \cite{chadha2023differentially}, researchers proposed OptPrefixTree based on previous work. They suggest a flexible algorithm to increase efficacy by utilizing adaptive segmentation, intelligent data selection, and deny lists.

There is also a tree-based method utilized in \cite{roth2019honeycrisp} to guarantee robustness and prevent the aggregator from cheating, which is a summation tree. Authors proposed the Honeycrisp system which uses sparse vector theorem to schedule the DP budget. The aggregator only can conduct summation on collected encrypted data. The committee composed of some clients can decrypt the result and compare it with the guess from the aggregator to decide whether to reveal it. When conducting the total summation in the aggregation phase, the aggregator is asked to generate a summation tree in which each parent node owns two children nodes and is the sum of them under HE. The tree is then checked by clients with each client verifying one pair of a parent and its children to prevent the aggregator from manipulating the aggregated result.

\blue{\textit{\textbf{Lessons learned.} Specialized data structures in FA, such as prefix trees (e.g., TrieHH and TrieHH++) and summation trees, enable efficient, accurate, and privacy-preserving data aggregation. Trie-based methods effectively identify heavy hitters with CDP using iterative updates, while optimized versions like OptPrefixTree enhance flexibility and efficiency through adaptive segmentation and intelligent data selection. Summation trees, as seen in Honeycrisp, ensure robust aggregation with HE and client-side verification, preventing manipulation. These tailored structures highlight the importance of combining privacy, robustness, and task-specific adaptability in FA systems.}}

\subsection{Deployment and optimization techniques}\label{subsec_techniques_deployment}

Beyond the basic privacy afforded by restricting raw data within the data owner, there exists the complex challenge of deploying effective techniques for optimization in some scenarios. These techniques are essential to navigate the distinct problem FA presents in terms of computation, communication, data heterogeneity, incentive mechanism, utility, and privacy. In this section, we list some techniques concentrating on the deployment and optimization problems in FA.

\subsubsection{Optimization theory} The optimization challenges in FA differ significantly from those in a centralized setting. As mentioned in \cite{wang2021field}, federated optimization mainly focuses on communication efficiency, data heterogeneity, computation, and privacy constraints. 

In \cite{deng2020fedvision}, FedVision is introduced, optimizing resource usage in video analytics. This system minimizes network resource consumption and maximizes computational efficiency by integrating black-box optimization with Neural Processes, tailored for dynamic network conditions. The study in \cite{pandey2021edge} presents Edge-DemLearn, an approach that maximizes learning performance and system efficiency. By leveraging distributed computing infrastructure, Edge-DemLearn is a sub-optimal two-sided many-to-one matching algorithm optimizing both the allocation of resources and the generalization of models. This method effectively overcomes the inherent limitations of traditional FL by enhancing model generalization and optimizing the management of user equipment associations within a distributed learning framework. \cite{toka20235g} represents a method that combines FA with 5G technology to optimize data collection and processing in vehicular networks. This approach aims to minimize latency by employing an FA-based solution and a D/M/1 queuing systems-based mathematical model to analyze end-to-end latency. They utilize numerical optimization methods like linear approximation (COBYLA), the Trust Region Constrained Algorithm (TRCA), and the Sequential Least SQuares Programming (SLSQP) to optimize system parameters, focusing on reducing waiting times and maximizing information value. Shi \etal \cite{siping23dro} introduced a Distributionally Robust Optimization (DRO) paradigm to optimize robustness in network traffic classifier learning under noisy labels. DRO is a modeling approach that makes decisions under uncertainty by minimizing the worst-case cost across all possible distributions in a constructed uncertainty set. This strategy maximizes classifier accuracy despite the variability in data quality.

\subsubsection{Game theory and incentive design} In federated settings, incentive mechanisms are crucial for motivating participants to contribute their private data, communicational resources, and computational resources. Incentive design typically balances individual rewards with collective goals to guarantee that participants are fairly compensated and to maintain the efficiency of the system. By employing game theory \cite{he2021game} approaches such as contract theory and multi-leader-follower games, these incentives aim to align participant actions with the overall objectives of the federated tasks. The structure of incentive mechanisms in FA is considerably similar to those in FL \cite{zhan2021survey}, where collaborative data processing and performing tasks require nearly the same strategies. Game theory in FA emerges as a critical tool for resolving complex interactions and optimizing cooperative strategies across various domains. It efficiently addresses challenges in privacy and incentivizes participation, scalability, and effectiveness in decentralized systems.

In \cite{zhao2023crowdfa}, researchers introduce CROWDFA which employs FA in mobile crowdsensing and balances data aggregation, incentive design, and privacy preservation. It uses additive secret sharing for privacy and a novel incentive mechanism, PRAED, to encourage participants' involvement with privacy protection.
Authors in \cite{kang2023blockchain} integrate blockchain with FL in healthcare metaverses, focusing on a user-centric incentive mechanism. It introduces AoI as a metric for data freshness and employs contract theory to incentivize data sharing. The framework uses Prospect Theory to address the service provider's decision-making under uncertainty, aiming to enhance immersive healthcare experiences by ensuring data freshness and privacy.
Another incentive mechanism is developed in \cite{yu2022incentive} for FL/FA systems with multiple tasks. It uses a multi-leader-follower game where rewards are set to motivate data owners, who in turn decide their participation level. The paper provides algorithms for optimal strategy formulation, ensuring effective resource allocation and addressing the challenges of multi-task federated systems.
Shi \etal \cite{shi2023federated} focuses on updating HD maps for autonomous driving using a game-theoretical model. It proposes an overlapping coalition formation game where vehicles collaborate in coalitions to update map data for maximizing utility and map quality. It addresses privacy and incentivizes participation by allowing vehicles to join multiple coalitions, which benefits the scalability and accuracy of map updates.
In \cite{jung2021establishing}, researchers cope with privacy and pricing in data markets. It proposes a federation model where data providers form coalitions with DP. They introduce a method for determining collective data prices based on privacy levels and employ the Shapley value from game theory for fair earnings distribution, which benefits both data providers and consumers by balancing privacy concerns and financial incentives.

\blue{\textit{\textbf{Lessons learned.} Incentive mechanisms are essential for motivating participants in FA, balancing individual rewards with collective goals while ensuring system efficiency. Game theory-based approaches, such as contract theory and multi-leader-follower games, effectively align participant actions with overall objectives, addressing challenges in privacy, scalability, and participation. Examples like CROWDFA and blockchain-integrated healthcare metaverses demonstrate how tailored incentive designs, such as additive secret sharing and privacy-preserving metrics, drive collaboration while maintaining privacy. Additionally, multi-task incentive mechanisms and coalition formation models for applications like HD map updates and data markets highlight the importance of fair resource allocation and financial rewards, often leveraging tools like the Shapley value to balance privacy and utility. These mechanisms showcase the critical role of game theory and innovative strategies in fostering effective and scalable FA systems.}}



\begin{figure*}[!]
\centering
\includegraphics[width=0.9\linewidth]{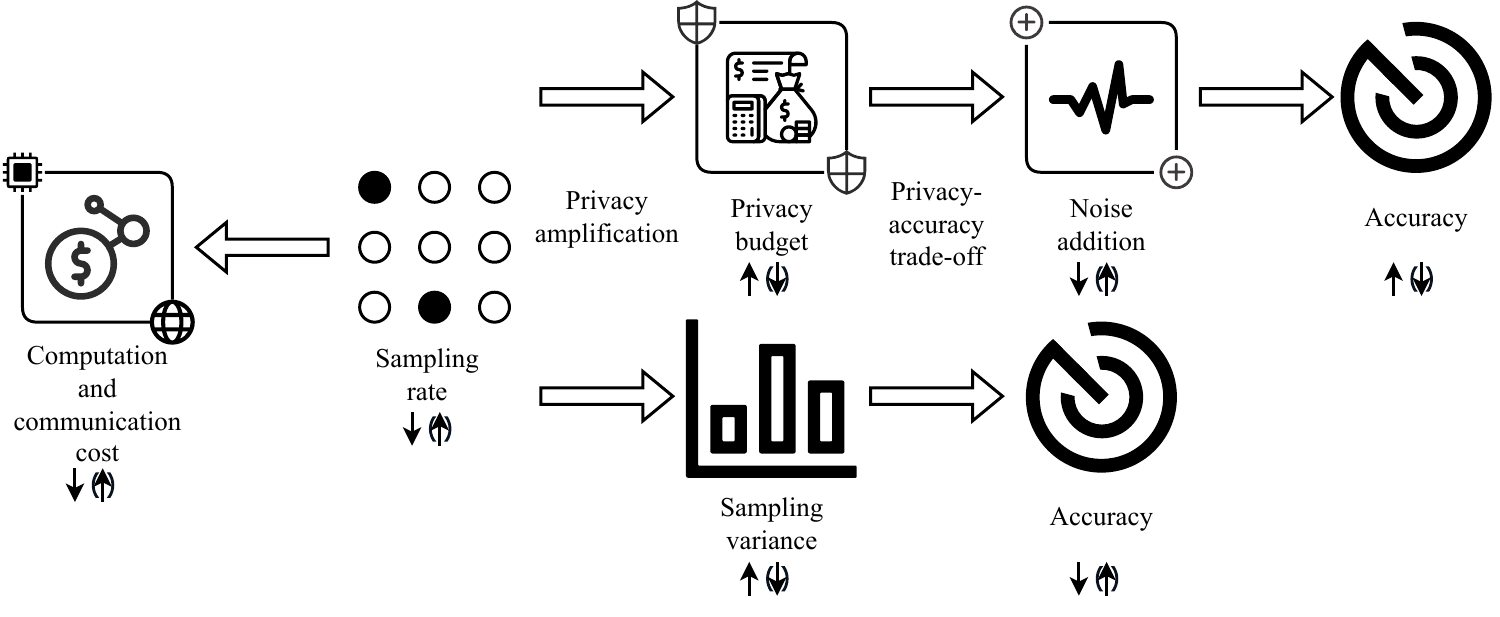}
\caption{Utilizing sampling in federated analysis reduces computation and communication costs by reducing the amount of data processed. Although it introduces result variance due to analyzing less data, it can amplify privacy budgets to improve result accuracy by reducing data perturbation.}
\label{fig_sampling}

\end{figure*}




\subsubsection{Sampling} Data sampling is the process of selecting a subset of data points from a larger dataset to analyze or draw conclusions about the whole, often to reduce computational cost while maintaining statistical representativeness \cite{noor2022simple,meister2023locally}. To aggregate insights across these distributed clients, FA often utilizes sampling techniques in terms of enhancing privacy and improving performance. There's also an optimization-wise problem for sampling techniques and the accuracy as shown in figure \ref{fig_sampling}. 1) When sampling is utilized, the total amount of data decreases so that the privacy budgets are amplified. The accuracy of results is improved since the federated data are less perturbed with more privacy budget. 2) While intuitively, less data analyzed leads to more variance in the results where sampling may cause accuracy reduction.

In early research within the domain of DP, the sampling technique is referenced and employed for privacy 
amplification\cite{li2012sampling}. Instead of examining the full data from all clients, using either data from a sample of clients or each client's sampled subset of data can enhance privacy protection in terms of amplifying the privacy bounds of a differentially private mechanism under various conditions. In other words, the utilization of sampling can improve the accuracy with the "amplified" privacy budgets. Most time, sampling is combined with other privatization techniques like noise injection to further protect data. Researchers in \cite{qin2016heavy} proposed the LDPMiner, a two-phase mechanism for heavy hitters tasks which applied sampled randomizer (combining sampling with privatization techniques like RAPPOR and Succinct Histogram) to each phase. They conducted the experiments on synthetic data and one of the results is the accuracy of targets from sampled randomizers outperforms that from naive randomizers under the same privacy budgets. It is worth noting that there's a built-in limitation of sampling that it always requires large enough total data to ensure its effectiveness, which is also indicated by another experimental result in \cite{qin2016heavy}. There is continuing work on LDPMiner, which proposed SVIM \cite{wang2018locally}, a protocol for searching frequent items under a set-valued LDP setting. They attempted new cooperated privatization techniques with sampling and found that some techniques benefit privacy amplification from sampling like Generalized Random Response while others do not such as Optimized Local Hashing. They explained the interesting phenomenon that the relation between reported and input values needs to satisfy a "many-to-one-wise" mapping for the benefit of privacy amplification from sampling, i.e., reported value after encryption "support" one input value instead of multiple values \cite{wang2018locally}. In addition, researchers utilize sampling in TrieHH \cite{zhu2020federated}, TrieHH++ \cite{cormode2022sample}, and OptPrefixTree \cite{chadha2023differentially} to choose a subset of clients in each iteration, where it is proved that sampling saves privacy budget.

However, sampling can improve the efficiency of the system attributed to the reduction in the amount of data processed while facing the concern of accuracy loss. In \cite{bater2020saqe}, authors proposed a private data federation system Secure Approximate Query Evaluator (SAQE) and denoted that sometimes sampling improves not only the efficiency but also the accuracy, \ie utilizing fewer data in the computation leads to improvement in accuracy. Given invariant privacy budgets, the sampling error decreases as the sampling rate grows, which means it requires more noise injection to maintain DP. Next, the total error summed by sampling error and error from noise is a convex function of the sampling rate for some tasks, meaning that sampling with the proper rate contributes to increasing accuracy. This utilization of sampling to increase the accuracy quantitatively of aggregation results by amplification of the privacy budget is uniquely used in the FA task. 
\blue{Utilization of the sampling technique in FA is highly associated with our taxonomy of \textit{data owner type}, where its utilization is different for cross-silo FA and cross-device FA. In cross-silo FA, sampling is conducted within the data points held by clients and obtains benefits of privacy amplification and reduction of overheads caused by cryptography. However, it requires the clients (silos) to be trusted to perform sampling correctly on their data. Meanwhile, sampling in cross-device FA is conducted as the server randomly samples the clients for participating in FA. Such sampling is the vanilla setting in most of the cross-device federated computation systems. However, it requires the central server to be trusted to some extent. It is challenging to maintain the secrecy of sampling in cross-device FA, \ie nobody knows anybody is sampled, which is an important prerequisite of privacy amplification.}





\blue{\textit{\textbf{Lessons learned.} Sampling techniques in FA are powerful tools for enhancing privacy, improving efficiency, and even increasing accuracy under certain conditions. By reducing the amount of data processed, sampling amplifies privacy budgets, allowing less perturbation and improved accuracy in DP mechanisms. However, the trade-off is a potential increase in variance due to fewer data points, highlighting the importance of balancing sampling rates to optimize accuracy and efficiency. Additionally, maintaining the secrecy of sampling in a one-data-per-person scenario—where no one knows who has been sampled—can be costly, making it less feasible for applications with strict privacy requirements.}}

\blue{In the previous sections, we have summarized many key factors of existing FA works and potential FA designs, including its general definition, taxonomy, challenges, and enabling techniques. In the following five sections, we introduce existing solutions on FA in detail. For reading convenience, we classify FA solutions based on their host data analytics tasks as mentioned in Section \ref{subsubsec_task}. Some sections will be further divided into subsections if the class of data analytics tasks can be further classified. Within each section/subsection, the existing FA solutions on  a particular class of data analytics tasks will be introduced ordered by their privatization methodology.}


\section{Statistical Metrics}\label{sec_statistical}

\begin{table*}
\caption{Summary of statistical metrics solutions.}
\label{table_statistical}
\centering
\begin{tabularx}{\linewidth}{lllX}
\hline
   \textbf{Reference}& \textbf{Data analytics task} &\textbf{Privacy} & \textbf{Note}\\
   \hline
   Cormode \etal \cite{cormode2021bit} & Mean computation & LDP & One-bit client upload\\
   Ding \etal \cite{ding2017collecting} & Mean computation & LDP & Handling continuous data collection\\
   Harmony \cite{nguyen2016collecting} & Mean computation & LDP & Laplace mechanism \\
   PrivKV \cite{ye2019privkv} & Mean computation & LDP & Means of key-value structure data \\
   Honeycrisp \cite{roth2019honeycrisp} & Mean computation & Cryptography\&CDP & Sparse vector theorem \\
   FEVA \cite{hu2021feva} & Mean computation & Hiding raw data & Federated Computation Builders architecture \\
   SecAgg \cite{bonawitz2017practical} & Mean computation & Cryptography & Utilization of two kinds of masks \\
   Bell \etal \cite{bell2020secure} & Mean computation & Cryptography & Interacting with only a small random part of clients for each client\\
   FastSecAgg \cite{kadhe2020fastsecagg} & Mean computation & Cryptography & Multi-secret sharing scheme based on fast Fourier transform\\
   Turbo \cite{so2021turbo} & Mean computation & Cryptography & Circular communication topology\\
   Zhao \etal \cite{zhao2022information} & Mean computation & Cryptography & Trusted third party\\
   LightSecAgg \cite{so2022lightsecagg} & Mean computation & Cryptography & a Secure and dropout-resilience secret sharing scheme\\
   Liu \etal \cite{liu2022efficient} & Mean computation & Cryptography & Lightweight and dropout-resilience secure aggregation\\
   Wei \etal \cite{wei2021user} & Gradient aggregation & LDP & Lightweight and dropout-resilience secure aggregation\\
   Kim \etal \cite{kim2021federated} & Gradient aggregation & LDP & Adjustable query sensitivity\\
   HFL-DP \cite{shi2021hfl} & Gradient aggregation & LDP & Server-edge-client architecture\\
   Geyer \etal \cite{geyer2017differentially} & Gradient aggregation & CDP & Hiding the participation of each participating client\\
   Mcmahan \etal \cite{mcmahan2017learning} & Gradient aggregation & CDP & Recurrent language models\\
  Hu \etal\cite{hu2020personalized} & Gradient aggregation & CDP & CDP-enhanced aggregation in personalized FL\\
   Aggarwal \etal \cite{aggarwal2010secure} & Medians and percentiles & Cryptography & Transforming the problem into a combinatorial circuit \\
   Tueno  \etal\cite{tueno2019secure} & Medians and percentiles & Hiding raw data &  \\
   Bohler \etal \cite{bohler2020secure} & Medians and percentiles & Cryptohraphy\& LDP &  Exponential Mechanism \\
   \cite{iutzeler2017distributed} & Medians and percentiles & Hiding raw data &  Model-based optimization approach in a federated way \\
   Dennis \etal\cite{dennis2021heterogeneity} & Clustering & Hiding raw data & Utilizing a hierarchical structure \\
   Zhou \etal\cite{zhou2022memory}  & Clustering & Hiding raw data & Applying kernel functions to transform data points into feature vectors \\
   UIFCA \cite{chung2022federated} & Clustering & Hiding raw data & Replacing the models used to be generative models that capture the distribution of one cluster\\
   Lubana \etal\cite{lubana2022orchestra}& Clustering & Hiding raw data & Uploading both the cluster model parameters and the local centroids to the server\\
   Servetnyk \etal\cite{servetnyk2020unsupervised}& Clustering & Hiding raw data & Uploading a vector representing the number of samples falling into each bin of the grid\\

   FedWalk \cite{pan2022fedwalk} & Graph metrics & LDP & FA version of a random walk\\
   
   LF-GDPR \cite{ye2020lf} & Graph metrics & LDP & Performing local computation by deriving the adjacency vector and the degree scaler each client\\
   
   Liu \etal\cite{liu2024edge}& Graph metrics & LDP & Protecting the information of vertex neighborhood within local graph data\\

\hline

\hline

\end{tabularx}  
\end{table*}
\blue{In this section, we focus on FA solutions on the first class of data analytics tasks: statistical metrics. We consider four variations of statistical metrics problems:  mean (Section \ref{subsec_statistical_mean}), median/percentile (Section \ref{subsec_statistical_median}), clustering center (Section \ref{subsec_statistical_clustering}), and metrics in graph data (Section \ref{subsec_graphmetric}), and introduce these solutions in the four subsections.}
\blue{Statistical metrics, like the means, variances, medians, and percentiles, as well as metrics in graph data, are simple but significant components in the field of data analytics. Deriving statistical metrics has various real-world applications, and also acts as intermediate steps of many complex data analytics algorithms. The demand for computing the statistical metrics of client data with privacy preservation motivates numerous FA studies. }

\subsection{Means}\label{subsec_statistical_mean}

The FA studies on calculating means usually consider that each client possesses a scalar value\footnote{The case of vector values can be formulated by executing multiple FA tasks on scaler values independently.}, \ie consider totally $n$ clients, where client $i$ possesses a scaler value $d_i$. The FA system derives the mean of their client-held scalers:
\begin{equation}
    \mu(d) := \frac{1}{n}\sum_{i=1}^{n} d_i.
\end{equation}

Mean computation is likely to be the most widely used data analytics task. It can be also transformed into other metrics. For example, sum computation can be conducted by multiplying the mean by $n$, and variance computation $V(d)$ can be derived as follows by executing the mean computation for the client data $d_i$ and square of client data $d_i^2$.
\begin{equation}
    V(d) = \mu(d^2) - \mu^2(d) = \frac{1}{n}\sum_{i=1}^{n} d_i^2 - \big(\frac{1}{n}\sum_{i=1}^{n} d_i\big)^2 .
\end{equation}

Many FA works are proposed for the sample analytics task of mean computation. In addition, some FA solutions tackle complex scenarios, but its mathematical model is essentially mean computation, like video analytics \cite{hu2021feva}. In the aggregation phase of FL, the FedAvg algorithm is computed on the client uploads. Essentially, the mean of gradient vectors from different clients is computed as the update on the global model. Many FL researchers try to improve the privacy and security of the aggregation phase, which falls into the category of FA-based mean computation \cite{bonawitz2017practical,bell2020secure,kadhe2020fastsecagg}.

Then, the FA mean computation solutions are introduced based on their privatization methodology, which is an effective taxonomy for FA mean computation tasks.

\textbf{Local differential privacy method(s).} As a strong and formal guarantee, LDP is applied in many FA-based mean computation solutions. LDP can be straightforwardly applied on mean computation by perturbing the client data $d_i$ before being uploaded. Based on the simple idea, researchers designed various solutions with their unique advantages.
Cormode and Markov \cite{cormode2021bit} propose an FA solution for the mean computation. In addition to satisfying the strong privacy criterion of LDP, their solution has a unique advantage, in that each client in their solution only needs to upload one bit to the server. Although it does not improve privacy preservation from a mathematical perspective, uploading only one bit is convincing for users with little mathematical backgrounds that the exposed information is minimized.
In \cite{ding2017collecting}, an LDP solution is proposed to privately estimate the mean and histogram of distributed data. It also proposes a discretization technique to handle the continuous collection of user data (\eg user data are collected daily).
Harmony \cite{nguyen2016collecting} provides an LDP solution for data analytics in mobile phone applications. It realizes mean estimation based on the Laplace mechanism, and frequency estimation based on the randomized response.
PrivKV and its variations \cite{ye2019privkv} extend the statistic tasks of mean estimation into the setting of data in the key-value structure. In that setting, each client holds a key-value pair; the data analyst expects to sort the keys of the clients and calculate the means of values with each same key. This paper then proposes an iterative structure to realize LDP with improved data utility.

LDP is also widely used for FA-empowered gradient aggregation in FL. 
In \cite{wei2021user}, user-level DP is satisfied by applying Gaussian noise on the gradient vectors to enforce LDP. The proposed approach also provides a theoretical FL convergence guarantee, and trade-off between convergence speed and computation overhead.
In \cite{kim2021federated}, another LDP-enhanced aggregation approach is proposed by applying a Gaussian mechanism to the gradients. \cite{kim2021federated} also provides privacy-utility-communication trade-off, adjustable query sensitivity for LDP, and an accounting scheme of LDP budgets.
LDP-FedSGD algorithm \cite{kim2021federated} considers privacy-preserving FL aggregation for IoT applications. It realizes LDP by perturbing the gradients with four LDP mechanisms. These mechanisms have different output structures and are optimal for different ranges of $\epsilon$ values.
In HFL-DP \cite{shi2021hfl}, FL under the server-edge-client architecture is considered. The employed Gaussian mechanism provides an LDP guarantee under different levels on client uploads and edge uploads.
Since the FL, and also the aggregation phases, take many rounds to complete, and the LDP budgets are cumulated by the composition theorem, na\"ively applying LDP in each round could result in quite a low privacy budget for each round. Therefore, many of the aforementioned works \cite{wei2021user,kim2021federated,shi2021hfl} propose additional privacy-utility trade-offs or privacy budgeting schemes to tackle the issue.

\textbf{Central differential privacy method(s).} 
Since LDP typically introduces significant noise on the uploads, which harms the accuracy of mean computation results, some researchers investigate privacy-preserving gradient aggregation approaches satisfying the CDP criterion.
In \cite{geyer2017differentially}, the authors consider applying CDP on the aggregated gradient, based on a trusted aggregator (server). Such a CDP-enforced gradient can hide the participation of each participating client so that the privacy of each client is preserved. Gaussian noise is adopted to enforce CDP in \cite{geyer2017differentially}. 
In \cite{mcmahan2017learning}, another privacy-preserving aggregation scheme where similar settings of trusted aggregator, client-level privacy, and Gaussian mechanism, are considered. It is particularly designed for recurrent language models.
The idea of CDP-enhanced aggregation is also employed in variations of FL like personalized FL \cite{hu2020personalized}, applying the Gaussian mechanism.

\begin{figure}[t]
\centering
\includegraphics[width=0.9\linewidth]{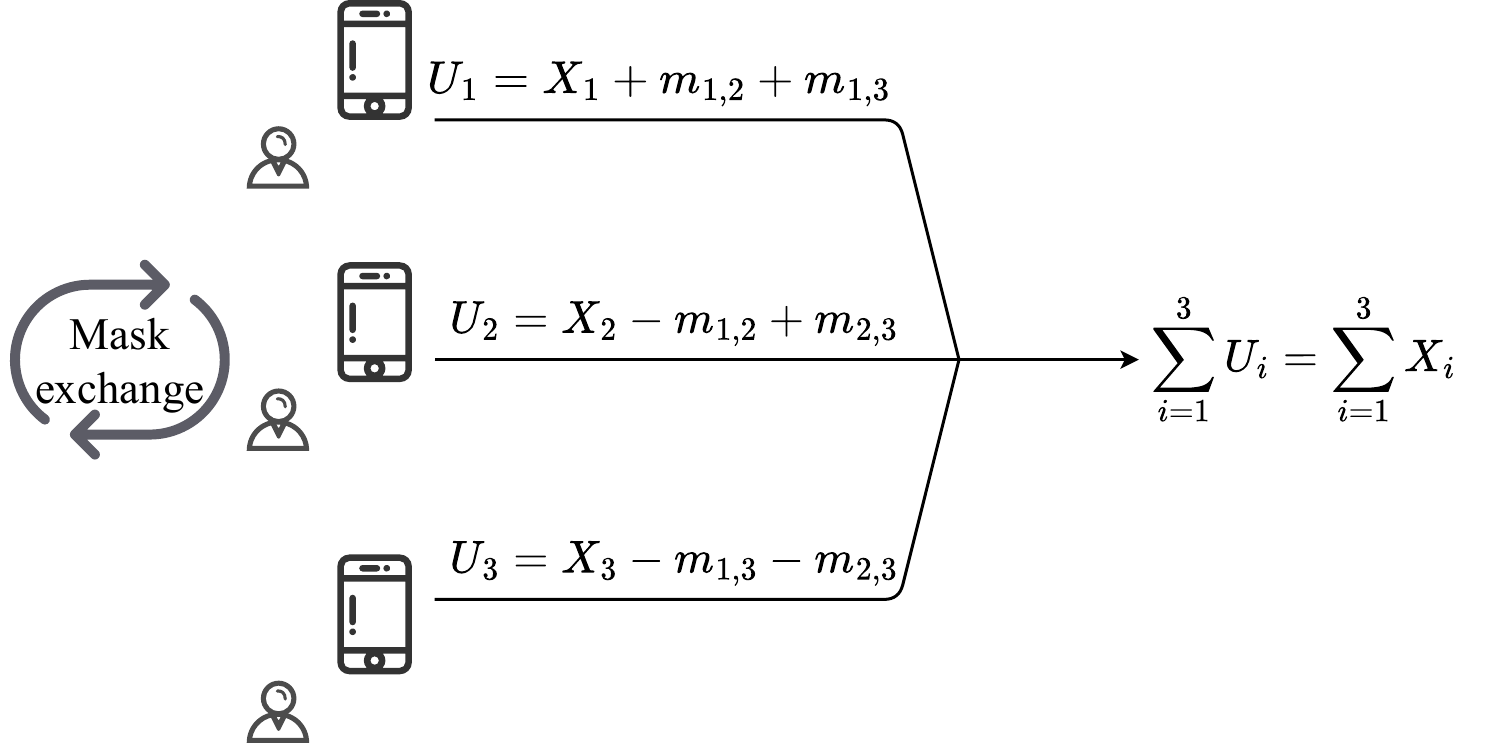}
\caption{A minimal running example of pairwise mask-based secure aggregation.}
\label{fig_mask_secureagg}
\end{figure}

\textbf{Cryptography method(s).} Many researchers also apply cryptography tools to protect privacy in mean computation, which is termed secure aggregation in the scenarios of FL gradient aggregation. With secure aggregation, the server can only learn the aggregated result (mean of gradient vectors), but cannot obtain any knowledge about the uploads of any single client.
Pairwise masking is the most lightweight and easy-to-implement technique to realize secure aggregation. In pairwise masking, each client communicates with some other clients, and a pairwise mask is generated for each pair of clients. The pairwise mask is added to the upload in one client in the pair, and its inverse is added to the other client in that pair. When an aggregator sums up the client uploads, the pairs of masks are eliminated, and the sum of the original data can be revealed. Fig. \ref{fig_mask_secureagg} provides an example of pairwise masking. In Fig. \ref{fig_mask_secureagg}, three clients, holding local upload $X_1, X_2$, and $X_3$ respectively, form three pairs and generate three pairwise masks. For example, mask $m_{1,2}$ is generated for clients 1 and 2, which is added to client 1's upload, and inversely added to client 2's upload. By summing up masked client uploads $U_1, U_2$, and $U_3$, the sum of $X_1,X_2$, and $X_3$ is revealed. To realize pairwise masking, the following works apply the novel \textbf{hybrid model} regarding its coordination model, so that masks can be exchanged via client-client interaction.

In \cite{bonawitz2017practical}, a pioneering secure aggregation scheme for FL SecAgg is proposed. It leverages the agreed random seed of client pairs to generate random masks on the uploads. The clients also apply self-masks on their uploads and leverage secret sharing to propagate the information of their self-masks. With the utilization of two kinds of masks, the FL system can preserve functionality and security with the existence of one-third of malicious clients and another one-third of dropout clients.
In \cite{bell2020secure}, the authors consider the heavy computation and communication loads of deriving and eliminating the masks and propose an improvement on SecAgg. In the new design, each client no longer needs to communicate with all other clients to share the masks. Instead, a client only needs to interact with a small random part of clients, which are logarithmic to the client size. The new design is still able to defend one-third of malicious clients and another one-third of dropout clients with high confidence, but not with 100\% confidence.
In \cite{kadhe2020fastsecagg}, FastSecAgg is proposed to improve the vanilla SecAgg protocol. It relies on a novel multi-secret sharing scheme based on fast Fourier transform, so that the per-client workload is reduced, but the guarantee on the portion of defended malicious/dropout clients is weakened.
In \cite{so2021turbo}, Turbo-Aggregate is proposed that employs the circular communication topology to reduce the communication overhead, but sacrifices privacy preservation as it only guarantees privacy in the average case, instead of the worst case considered in SecAgg.

Some other studies replace pairwise mask generation, which requires extensive client pair interactions, with non-pairwise masks. In non-pairwise mask generation, each client generates masks independently, and the server can eliminate the masks by learning the sum of the masks with a special protocol. In these non-pairwise mask schemes, the traditional \textbf{server-client architecture} is applied as the coordination model, which removes the extra requirement on clients to communicate with other clients.
In \cite{zhao2022information}, a trusted third party is employed to assist with the masking, unmasking, and dropout handling.
In LightSecAgg \cite{so2022lightsecagg}, the previous protocol is improved by removing the need for a trusted third party, based on a secure and dropout-resilience secret sharing scheme.
In \cite{liu2022efficient}, the homomorphic pseudorandom generator (HPRG) is leveraged as a non-pairwise mask generation scheme to achieve lightweight, dropout-resilience secure aggregation.

Privacy-preserving aggregation for FL, or privacy-preserving vector mean computation from the FA perspective, has gained numerous research interests in recent years. A thorough survey of these researches is in \cite{liu2022privacy}.

The cryptography tools are also utilized in FA-based mean computation in general settings. Honeycrisp \cite{roth2019honeycrisp} is a system to address large-scale count mean sketch tasks without a trusted server. A committee is randomly selected from users to generate HE keys. Client data is sent to the server utilizing additive HE, which allows the server to calculate the sum without accessing individual raw data. This sum is then decrypted by the committee. It utilizes the sparse vector theorem to manage the DP budget, enabling the committee to decrypt results and verify the aggregator's calculations before deciding on their disclosure. The system also includes a summation tree method to ensure robustness and prevent the aggregator from cheating, where each node is the sum of its children under HE and they are verified by clients to ensure integrity.

\textbf{Hiding raw data method(s).} In FEVA \cite{hu2021feva}, an FA-based video analytics solution is proposed. It requires the clients to derive features from the video frames to preserve privacy following the hiding raw data methodology, and performs scaling and averaging procedures on the server side to derive the analytics result, based on the Federated Computation Builders architecture.

\blue{\textit{\textbf{Lessons learned.} Mean computation, which is the most fundamental task in statistics, gains diverse research interests, and many FA-based mean computation tasks are proposed. These applications either treat mean computation as a stand-alone task or resolve mean computation procedure in the FL aggregation framework (often termed secure aggregation), as the FL aggregation procedure is essentially a vector mean computation task. Various privatization techniques are applied in FA mean computation solutions, including LDP, CDP, and cryptography, under the trade-offs among accuracy, privacy, and computation/communication overheads. Extra issues in mean computation, including special data types, coordination model, and dropout resilience, are also considered by researchers.}}

\subsection{Medians and percentiles}\label{subsec_statistical_median}

Median and percentile computation in the federated setting considers the clients each with many scaler data points $d_i$. $\mathcal{D}=\{d_i | i\}$ is the virtually aggregated data points from all clients. Consider a ranking function $\mathcal{R}(\mathcal{D},k)$ outputting the element in $\mathcal{D}$ exactly larger than $k$ portion of the elements in $\mathcal{D}$. The federated $k$-percentile computation lets the clients collaboratively derive the value of $\mathcal{R}(\mathcal{D},k)$ while preserving the privacy of data points $d_i$. The federated median computation is a special case of the federated $k$-percentile computation. The major challenge of federated median/percentile computation is that it requires different clients to compare their private data points to derive the ranking results.

As private value ranking is a classical problem setting in the field of MPC, some researchers design federated median/percentile solutions utilizing various cryptography tools.

\textbf{Cryptography method(s).} In \cite{aggarwal2010secure}, a private solution of median (and $k$-precentile) element over client local datasets is proposed. The solution is based on MPC and is realized by transforming the problem into a combinatorial circuit.
In \cite{tueno2019secure}, a federated median/percentile computation scheme is proposed following the server-client architecture. The clients provide privatized input to the computation, and the server conducts the computation-intensive workloads to derive the results without observing the raw data.

\textbf{Local differential privacy + Cryptography method(s).}
In \cite{bohler2020secure}, a median calculation scheme is proposed with both LDP and MPC. It can be extended to tasks like percentile computation. The LDP guarantee is realized by an exponential mechanism. The authors consider threat models of both semi-honest settings and adversarial settings.

\textbf{Hiding raw data method(s).}
Another federated median/percentile computation solution \cite{iutzeler2017distributed} relies on a model-based optimization approach in a federated way, without a formal privacy guarantee. In \cite{iutzeler2017distributed} deriving the quantile is formulated as an optimization problem. That optimization problem is then collaboratively computed by the clients with ADMM. No formal privacy guarantee is provided.

\blue{\textit{\textbf{Lessons Learned.} Computing medians and percentiles are different from other statistical metrics because the output is essentially an element of the input dataset. Therefore, the applications of DP mechanisms are limited, since adding random noise to the output is not desirable. As a result, DP applications on FA median/percentiles computation utilize the exponential mechanism, which adds noise to the output probabilities of the elements rather than the values of the elements. Cryptography tools that do not introduce randomness to the outputs also perform well in FA median/percentiles applications.}}

\subsection{Clustering}\label{subsec_statistical_clustering}

The clustering task aims at dividing data points (usually defined by fixed-length vectors) into groups, so that data points within a group are more similar to each other in some sense (\eg with lower vector distance), and data points in different groups are less similar. Clustering analysis is a major task in the field of exploratory data analytics and is an important technique in the fields of pattern recognition, sociology, data compression, and bioinformatics.

The federated clustering task extends the clustering problem into the federated data setting. In that setting, the clients, each possessing a subset of data points, want to collaboratively derive the global clustering results, \ie deriving what groups their data points belong to after the data points from different clients are virtually aggregated. The major challenge of federated clustering is that it needs to perform similarity measurements for data points from different clients while preserving data privacy.
Researchers have proposed federated clustering solutions following the FA paradigm. In these proposed algorithms, classical centralized clustering algorithms are adapted into the federated version, with the intermediate results serving as the FA insight. As for now, all existing FA-based clustering solutions only apply the idea of hiding raw data for privacy preservation. Designing FA-based clustering solutions with formal privacy preservation like DP and encryption is a valuable and promising research direction.

\textbf{Hiding raw data method(s).}
\cite{dennis2021heterogeneity} proposes a federated clustering solution. It utilizes a hierarchical structure, where data points are first clustered locally within local client data, and local centroids are then clustered by the server. The authors derive an interesting finding, that in contrast to other federated computation solutions, data heterogeneity is beneficial to the analytics results.
In \cite{zhou2022memory}, the authors consider kernel $k$-means, an effective clustering algorithm in capturing the nonlinear representation of the dataset using the nonlinear kernel functions, in the federated setting. To realize that, the clients first transform their data points into feature vectors by applying the kernel function.
UIFCA \cite{chung2022federated} considers the unsupervised clustering tasks (the authors take unlabelled image clustering tasks as an example). It follows the idea of the IFCA algorithm, which trains a model for each cluster, and assigns a data sample to a cluster with the smallest model loss. To modify IFCA, which was originally designed for personalized FA, UIFCA replaces the models used to be generative models that capture the distribution of one cluster.

In \cite{lubana2022orchestra}, a self-supervised learning-based federated clustering algorithm Orchestra is proposed. It takes a Sinkhorn-Knopp-based deep learning clustering algorithm \cite{genevay2019differentiable} which outputs equal-size clusters. The clients first cluster their local data to derive local centroids, and upload both the cluster model parameters and the local centroids to the server. In the server, the model parameters are aggregated by FL, and the local centroids are aggregated by FA via applying the Sinkhorn-Knopp algorithm again. Then, the updated model parameters and global centroids are sent back to the clients to start the next round, until a desirable clustering result is achieved.
In \cite{servetnyk2020unsupervised}, the FA technique is employed as a prepossessing step of a federated clustering algorithm. In that FA design, the data space is divided into a grid, and each client encodes their local data into a vector, representing the number of samples falling into each bin of the grid. Then, the vector is uploaded to the server as insight. The server aggregates the insights from all clients and derives a global view of data distribution. After that, the grid is updated following the method of the self-organizing map, and the updated map is sent to the clients, where the clients start a new round of bin checking.

\blue{\textit{\textbf{Lessons learned.} FA clustering tasks derive the clustering results (\eg centroids) from the data points held by different clients. Although clustering is an important statistics task, FA research on this topic is still somewhat primal. Existing solutions do not utilize advanced privatization mechanisms other than the fundamental idea of hiding raw data. These works convert the local data points into intermediate results of clustering (\eg local clustering centroids) and transmit these results to the server for aggregation, which exposes remarkable privacy risks.}}

\subsection{Graph metrics}\label{subsec_graphmetric}

The graph-structured data have yielded numerous promising graph analytics applications in the fields of social networks, multimedia, and AI4science. These applications require the utilization of novel graph metrics that are not present in conventional data structures. Leveraging FA to compute these graph metrics with privacy preservation can become a booster of graph analytics applications over edge data.

\textbf{Local differential privacy method(s).}
In \cite{ye2020lf}, a privacy-preserving graph metrics computation scheme named LF-GDPR is proposed. It supports various graph analytics tasks, including community detection and clustering coefficient estimation. Under the vertex-separated federated setting, each client performs local computation by deriving the adjacency vector and the degree scaler. Then, the insight is processed by LDP with careful privacy budget allocation. The derived metrics are aggregated on the server side to complete all kinds of graph analytics tasks utilizing dedicated aggregation algorithms proposed by the authors.
\cite{liu2024edge} focus on the triangle counting problem in the graph-separated federated setting. The authors proposed an enhanced privacy notion named edge relationship LDP which can protect the information of vertex neighborhood within local graph data. The insight processed by edge relationship LDP therefore will not reveal sensitive information about the existence of edges.

\textbf{Hiding raw data method(s).} FedWalk \cite{pan2022fedwalk} is an FA-based solution for graph embedding analytics. It derives expressive representations for the vertexes in a large graph. FedWalk considers the vertex-separated federated setting, where each client represents one vertex in the graph, and only possesses neighborhood information of the vertex. It performs the FA version of a random walk, a classical graph embedding learning algorithm.

\blue{\textit{\textbf{Lessons learned.} The graph data is a novel data structure with promising applications. The complex structure of a graph makes its privacy preservation quite different from conventional data, like scalar, vector, or discrete structures. Before applying particular privatization techniques, novel privacy criteria should be defined for the protection of edges, vertexes, or graphs, based on the particular data separation settings. After that, the corresponding FA graph analytics solutions are designed to optimize the performance of downstream graph analytics tasks.}}

\section{Frequency-related Applications}\label{sec_frequency}

\blue{In this section, we focus on the second class of data analytics tasks according to our taxonomy: frequency-related applications. Frequency-related applications refer to tasks that involve calculations/estimations on the frequency of particular patterns within the population of clients. After introducing the definitions of frequency-related applications as follows, we summarize some key issues, and then introduce the existing solutions in detail}

The tasks related to the frequency estimation on categorial data make up an important part of data analytics, and also its FA variations. In these tasks, the clients with indexes $i=1,...,n$ hold local data $d_i$. Denote $q$ as a queried data structure, for any $q$ and $d_i$, there exists a relationship of ``containing''. For simplicity, we borrow the symbols from set theory, so that $q\in d_i$ denotes the fact that ``$d_i$ contains $q$'', and $q\notin d_i$ denotes the fact that ``$d_i$ does not contain $q$''. For any $q$ and $d_i$, exactly one of $q\in d_i$ or $q\notin d_i$ must be true.

With the definition of a ``containing'' relationship, the frequency $f$ of any query data can be then defined as follows.

\begin{equation}
    f(q) = \frac{|\{i|q\in d_i \}|}{n},
\end{equation}
or 
\begin{equation}
    f(q) = \mathbb{P}(q\in d_i), {\rm~randomly~sample~} i {\rm~from~clients}
\end{equation}
can be used when the size of clients $n$ is not available.

With the definition of the frequency of queried data, multiple frequency-related tasks can be then defined. Here we list our taxonomy of these tasks.




\begin{itemize}
    \item \textbf{Frequency oracle.} The task of frequency oracle aims at constructing an abstraction (frequency oracle) of client data. Using the abstraction, a querier can arbitrarily derive estimations of any queried data without interacting with the clients. In some works, the frequency oracle problem is termed frequency estimation problem as the frequency oracle provides services of frequency estimation.
    \item \textbf{Frequent pattern mining.} The task of frequent pattern mining (FPM) aims at finding highly frequent queried data. There are two mainstream definitions of FPM. In the first definition, the system outputs all possible queried data $q$ where $f(q)$ is higher than a predefined threshold. In the second definition, the system outputs the top-$k$ queried data with the highest frequency. Based on the structures of client data $d_i$, queried data $q$, and definition of containing relationship, FPM is also named by its subtasks frequent item mining, frequent itemset mining, and frequent sequence mining.
    \item \textbf{Heavy hitter.} The term heavy hitters originated from the downstream tasks in web data analysis. The heavy hitter task is consistent with frequent item mining, where the containing relationship is defined to be the ``$\in$'' relationship in set theory.
\end{itemize}

The definitions of frequency-related tasks are highly connected, and many algorithms can tackle multiple tasks simultaneously. 

In the following part, we will first discuss some key issues in handling frequency-related tasks, and then introduce the existing FA solutions.


Data domain and data type are two key issues in the problem setting of any frequency-related task, which determine the difficulty of the task and introduce challenges in designing the solutions.
\begin{itemize}
    \item \textbf{Data domain.} The data domain (or data universe) is an important setting in frequency-related tasks. The data domain refers to the set of all possible queried data. The easiest setting is the \textit{small domain}, where the number of possible queried data is small enough so that the algorithm designer can simply enumerate all possible queried data, or design one-hot encoding of them. In the \textit{large domain} setting, the algorithm designer still knows all the possible queries, but can no longer apply one-hot encoding or similar techniques, due to the limit of computation and communication resources. In the most difficult \textit{infinite domain} setting, there are infinite possible queried data. An example of the infinite setting is that the queried data are all possible strings of characters with arbitrary lengths. 
    \item \textbf{Data type.} The frequency-related tasks can be categorized by the type of queried data (and also client data). The item type data is the default setting of frequency oracle and heavy hitters, where the possible queried data are indivisible distinct items. Complex types of queried data, like set, sequence, and graph of items, are investigated in frequent pattern mining problems. These problems are typically more difficult than the item-type problems, because they lead to exponentially increasing data domain, and require careful design to handle the relationship between different queried data.
\end{itemize}

We have introduced the key issues in FA applied for frequency-related applications. Then, we introduce the existing FA solutions, sorted by their utilized privatization methodologies.





\begin{table*}
\caption{Summery of FA-based solutions on frequency-related applications. (\textbf{FO}: frequency oracle; \textbf{HH}: heavy hitter/frequent item mining; \textbf{FIsM}: frequent itemset mining; \textbf{FSM}: frequent sequence mining)}
\label{table_frequency}
\centering
\begin{tabularx}{\textwidth}{llllX}
\hline
   \textbf{Reference}& \textbf{Task}& \textbf{Privacy} & \textbf{Data domain} & \textbf{Note}\\
\hline
RAPPOR \cite{erlingsson2014rappor} & FO \& HH  & LDP & Large   & Randomized response on modified bloom filter\\
Bassily \etal \cite{bassily2015local} & FO \& HH & LDP & Small & One-bit client upload\\
LDPMiner \cite{qin2016heavy} & FO \& HH \& FSM & LDP & Small &  Constant communication cost\\
PSFO \cite{wang2018locally} & FIsM & LDP & Small & Converting FIsM into HH via padding and sampling on itemsets\\
TreeHist \cite{bassily2020practical} & FO & LDP & Small & Near-optimal error\\
Bitstogram \cite{bassily2020practical} & FO \& HH & LDP & Small & Near-optimal error\\
Acharya \etal \cite{acharya2019hadamard} & FO & LDP & Small & Hadamard Response; No shared randomness\\
Acharya \etal \cite{acharya2019communication} & FO\& HH & LDP & Small & Hadamard Response; No shared randomness; One-bit client upload\\
FIML \cite{li2022frequent} & FIsM & LDP & Small & Top-$k$ setting of FIsM\\
OptPrefixTree \cite{chadha2023differentially} & HH \& FSM & LDP & Infinite & Handling infinite data domain with prefix tree\\
SFP \cite{apple2017learning} & FSM & LDP & Large & Uploading two count-mean-sketches to encode a sequence\\
FedFPM \cite{wang2022fedfpm} & HH \& FIsM \& FSM & LDP & Small & Unified framework for multiple FPM subtasks; One-bit client upload\\
TrieHH & HH \& FSM & CDP & Infinite & Building up the frequent strings by transmitting a prefix tree\\
TrieHH++ \cite{cormode2022sample} & HH \& FSM & CDP & Infinite & Randomized sample size\\
STAR \cite{davidson2022star} & HH & Cryptography & Infinite & Private threshold aggregation via secret sharing\\
Boneh \etal \cite{boneh2021lightweight} & FO \& HH \& FSM & Cryptography & Infinite & High security via incremental distributed point functions\\
B{\"o}hler \etal \cite{bohler2021secure} & HH & DDP & Small & Utilizing DDP for high data utility and formal DP guarantee\\
Bagdasaryan \etal \cite{bagdasaryan2021towards} & FO & DDP & Small & Utilizing DDP for high data utility and formal DP guarantee\\
FedWeb \cite{wang2024fedweb} & HH \& FIsM \& FSM & DDP & Small & Application in Web 3.0 scenario\\
\hline

\end{tabularx}  
\end{table*}



\textbf{Local differential privacy method(s).}
The majority of FA-based frequency oracle, heavy hitters, and frequent pattern mining solutions adopt LDP for their privacy guarantee, owing to its strong privacy preservation in the untrusted aggregator model. RAPPOR \cite{erlingsson2014rappor} is one of the earliest and most popular frequency oracle solutions with an LDP guarantee. It lets the clients encode their local data into bloom filters so that LDP can be enforced for arbitrary structure of local data. By aggregating the bloom filters, the frequency of the encoded elements can be estimated. \cite{bassily2015local} (succinct histograms) is another algorithm for frequency oracle and heavy hitter problems. With the help of shared randomness, the succinct histogram algorithm only requests clients to upload one-bit data. LDPMiner \cite{qin2016heavy} is another solution for heavy hitters. Compared to previous solutions, LDPMiner is the first one to resolve the set-value client data, where multiple $q$ can be contained by one $d_i$. PSFO \cite{wang2018locally} proposes its optimization over LDPMiner. While preserving the strengths of LDPMiner, PSFO enables frequent itemset mining tasks with its padding-and-sampling oracle. In \cite{bassily2020practical}, two algorithms, TreeHist and Bitstogram, are proposed, to resolve the tasks of frequency oracle, and heavy hitters. In the two algorithms, tailored data structures, shared randomness, and Hadamard transform are carefully designed and utilized. As a result, TreeHist and Bitstrogram manage to optimize the time and space complexity on both the server and client side. In \cite{acharya2019hadamard}, another frequency oracle algorithm based on Hadamard response is proposed that removes the need for shared randomness. In \cite{acharya2019communication}, an algorithm is proposed for frequency oracle and heavy hitters. In addition to removing the need for public randomness, \cite{acharya2019communication} only requires one-bit responses. FIML \cite{li2022frequent} tackles the top-$k$ setting of frequent itemset mining. It first builds a frequency oracle to find out top frequent items, then builds candidate itemsets, and queries their frequencies. OptPrefixTree \cite{chadha2023differentially} is a prefix tree-based heavy hitter algorithm. It leverages the prefix tree to handle the infinite data domain.

SFP \cite{apple2017learning} algorithm is proposed by Apple to tackle the difficult frequent sequence mining problem. It encodes the local client data with the count mean sketch so that LDP can be properly enforced. It partially tackles the large domain problem of domain element frequency estimation. It lets clients upload an extra count mean sketch of sequence fragments, where the short fragments can be easily enumerated and frequent short fragments can assemble longer sequences. FedFPM \cite{wang2022fedfpm} is proposed as a unified FA framework to tackle multiple FPM subproblems, including heavy hitters, frequent itemset mining, and frequent sequence mining. It follows a query-response scheme to estimate the frequency of the candidate patterns and derives Hoeffding's inequality-based bounds to filter the candidates. FedFPM can achieve a better data utility than existing solutions but requires sufficient participating clients. 

\textbf{Central differential privacy method(s).}
As LDP schemes request clients to add significant noise to the uploads, researchers propose CDP schemes in FA-based frequency-related tasks. The major advantage of CDP is that it requests clients to add smaller noise on the uploads, or even add no noise by utilizing the sampling mechanism. However, only enforcing CDP sacrifices the formal privacy guarantee, as the FA server is usually considered untrusted. TrieHH \cite{zhu2020federated} is an FA solution to digest string-typed heavy hitters. It iteratively builds up a prefix tree of characters with the interaction between the server and clients. Its major advantage is the ability to satisfy CDP without adding random noise to the outputs so that a better data utility of the mined results can be achieved. TrieHH++ \cite{cormode2022sample} is an expansion of TrieHH. It extends the applications from heavy hitters to quantile estimation and range query. It follows the basic ideas of TrieHH, that realizing CDP via sampling data without adding random noise to the outputs and generating a prefix tree to handle infinite data domain. In addition, it applies Poisson sampling, instead of fixed-size sampling, to hide the sample size.

\textbf{Cryptography method(s).}
Some works adopt cryptography tools instead of DP in FA-based frequency-related solutions. Compared to DP, cryptography tools allow deriving accurate results with maximal data utility. However, an adversary still may infer sensitive information from the accurate outputs. STAR \cite{davidson2022star} is a threshold aggregation (a variation of heavy hitters) solution for web data analysis. Its threshold aggregation enforces $k$-anonymity on the output data, and such functionality is realized by a secret sharing scheme. In \cite{boneh2021lightweight}, a private heavy hitter algorithm is proposed. The solution relies on the incremental distributed point function, a lightweight cryptography tool. It assumes that there exist two servers that collude with neither each other nor any client.

\textbf{Distributed differential privacy method(s).}
In \cite{bohler2021secure}, a private heavy hitter solution is proposed that utilizes DDP for its privacy preservation. DDP statistically satisfies CDP, and utilizes cryptography tools to prevent the server from learning any single client upload, so that local privacy can be achieved with CDP-level data utility.
In FedWeb \cite{wang2024fedweb}, another DDP-based solution is proposed. It can handle generic FPM tasks and proposes tailored designs to be applied in Web 3.0 applications.
In \cite{bagdasaryan2021towards}, an FA-based location heatmap generation solution is proposed, which can be regarded as a frequency oracle solution on geometric items. In \cite{bagdasaryan2021towards}, DDP, which combines the schemes of cryptography tools and DP, is applied to reduce the noise required for privacy preservation. It iteratively reduces the granularity of queried locations to better utilize the participating clients.

In Table \ref{table_frequency}, we summarize our surveyed FA-based solutions on frequency-related tasks and list their detailed tasks, privacy model, data domain setting, and extra features.

\blue{\textit{\textbf{Lessons learned.} FA applied in frequency-related applications focuses on estimating the frequency of any data pattern in the population of clients. Such frequency-related applications can be further classified in many aspects, including dedicated frequency tasks, domain of data, type of data, and type of pattern. These factors determine the difficulty and challenges of the tasks and affect the corresponding FA algorithmic design. Existing solutions utilize LDP, CDP, and cryptography tools for privatization. In addition, as the advanced DP mechanism, DDP is firstly utilized in frequency-related applications and mean computation, because the lightweight encryption tools in DDP are most effective when the insight aggregation phase can be formulated as averaging, where mean computation and frequency-related applications usually follow.}}

\begin{table*}
\caption{Summery of federated query solutions.}
\label{table_fedquery}
\centering
\begin{tabularx}{\textwidth}{lllXl}
\hline
   \textbf{Reference}& \textbf{Query type}& \textbf{Privacy} & \textbf{Cryptography technique} & \textbf{Maximal client}\\
\hline
    SMCQL \cite{bater2017smcql} & General & Cryptography & Garbled circuit & 2\\
    Conclave \cite{volgushev2019conclave} & General & Cryptography & Secret sharing & 3\\
    Senate \cite{poddar2021senate} & General & Cryptography (adversary model) & Garbled circuit & 16\\
    SAQE \cite{bater2020saqe} & General & DDP & Secret sharing & 2\\
    Orchard \cite{roth2020orchard} & General & DDP & HE & $>$1000\\
    Arboretum \cite{margolin2023arboretum} & General & DDP & HE & $>$1000\\
    Secure Yannakakis \cite{wang2021secure} & Free-connex join-aggregate query & Cryptography & Private set intersection (based on garbled circuit), secret sharing, \& oblivious extended permutation  & 2\\
    Zhang \etal \cite{zhang2022efficient} & Skyline query & Cryptography & Private set intersection (based on HE and garbled bloom filter) & 10\\
    Hu-fu \cite{tong2022hu} & Spatial query & Cryptography & Secret sharing & 10\\
    Mycelium \cite{roth2021mycelium} & Graph query & DDP & HE \& onion routing & $>$1000\\
\hline

\end{tabularx}  
\end{table*}

\begin{figure}[t]
\centering
\includegraphics[width=0.9\linewidth]{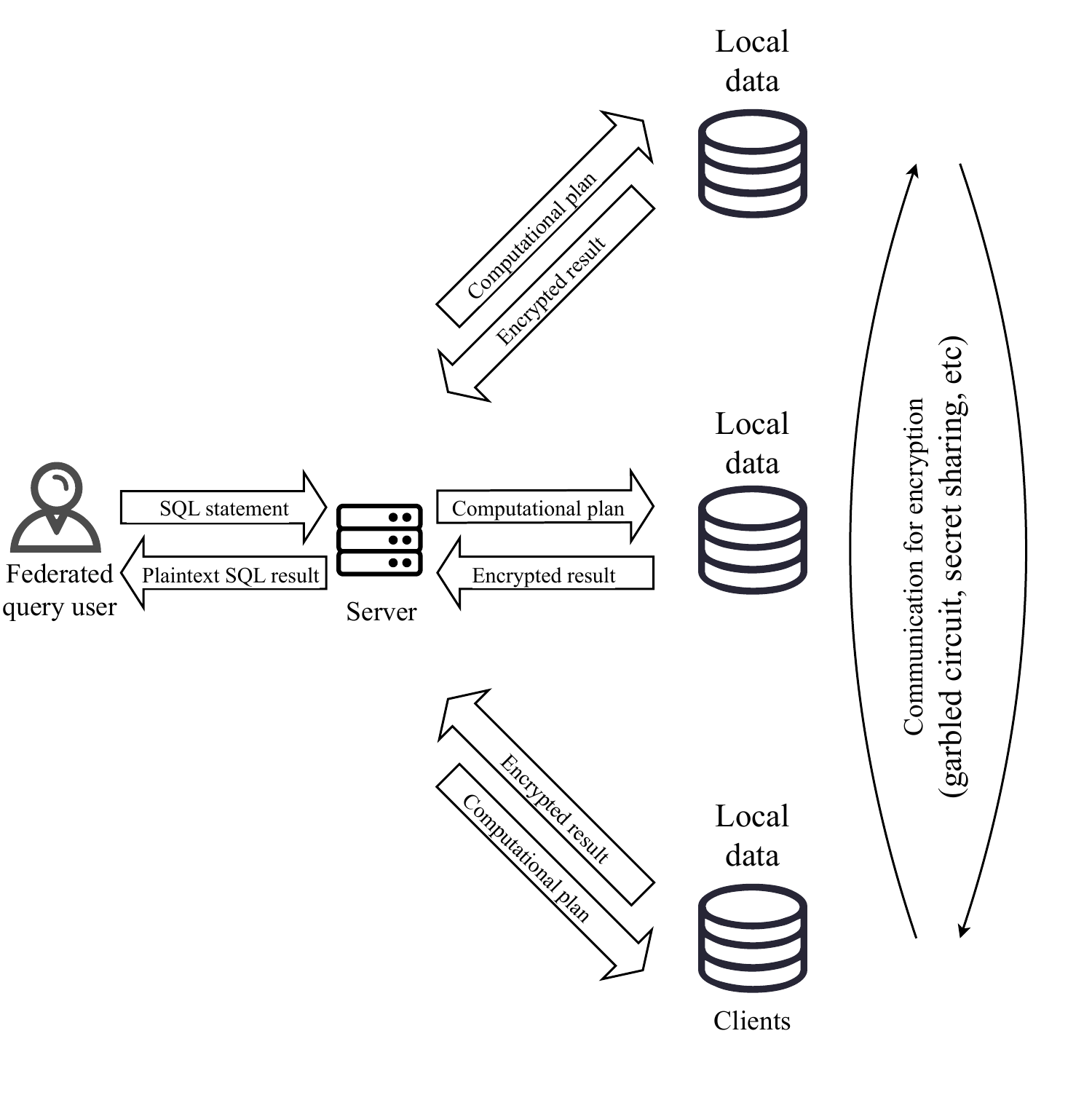}
\caption{Illustration of the server-client-based federated query architecture.}
\label{fig_fedquery_architecture}
\end{figure}

\section{Database Operations}\label{sec_database}

\blue{In this section, we focus on the third class of data analytics applications according to our taxonomy: database operations. The FA efforts in this section consider the novel idea of formulating data analytics into SQL queries.}

Federated query, also termed data federation SQL, targets enabling database operations (SQL queries) on the data separated in multiple clients. The result of federated query is expected to be the same as the case where data from clients is virtually gathered to form a centralized database, and the privacy of the client data is preserved. Privacy preservation of federated query solutions is usually achieved by various cryptography tools so that every participant cannot learn any information about the data of other clients unless it is necessary to be revealed by the query result. Federated query has the potential to be an important piece of FA research, as many simple data analytics tasks are equivalent to a single SQL query, and some complex data analytics tasks can be decomposed into a series of database operations.

As privacy-preserving federated query solutions rely on cryptography tools, especially MPC protocols, the heavy computation overhead of executing the cryptography tools becomes a significant burden to deploy these federated query solutions into real-world systems. Therefore, an important task of the federated query solutions is to optimize the computation time of executing the federated query and deriving the accurate result. To fulfill that goal, researchers propose various solutions to pursue desirable computation time performance, which utilizes various cryptography tools with different properties and carefully optimizes the computation graph of executing the query to reduce the computation cost.

Federated query solutions usually follow two kinds of architectures. The first architecture is inherited from classical MPC protocols like garbled circuits. In such a fully decentralized architecture, clients directly communicate with each other under the encrypted protocol. After the computation, all the clients can derive the decrypted query results. Another widely applied architecture is the server-client architecture, as is illustrated in Fig. \ref{fig_fedquery_architecture}. The server is in charge of receiving the SQL statement from the federated query user and composing the computation plan. The clients execute the computation plan to conduct local computation, where client-client communication is usually involved as is required by the encryption tools. The clients upload the encrypted results to the server. Then, the server derives the plaintext SQL result by decrypting the client uploads and returns the result to the federated query user.

Among the existing federated query solutions, the most ambitious ones are those supporting many types of queries (even the whole SQL standard) within one framework. These solutions are the most powerful FA solutions from the perspective of the number of supported data analytics tasks \cite{bater2017smcql,volgushev2019conclave,poddar2021senate,bater2020saqe,roth2020orchard,margolin2023arboretum}. However, these generate federated query solutions burdens many task-specific applications, and therefore perform bad regarding computation and communication efficiency. Some researchers propose specific federated query solutions as a compromise. Specific federated query solutions tackle one kind of query so that many tools other than general cryptography solutions can be applied \cite{wang2021secure,zhang2022efficient,tong2022hu,roth2021mycelium,roth2020orchard}. In other words, researchers only need to provide a cryptography version of a specific algorithm to enable a specific federated query. As a result, the specific federated query has the potential to achieve better performance in computation load and has wide application in scenarios where only a specific kind of query is required.

We have concluded the key issues and different types of federated query solutions. Then, we introduce the existing federated query solutions, sorted by the utilized privatization methodologies.

\textbf{Cryptography method(s).} Cryptography tools, like the garbled circuit and secret sharing, are the basis for privacy preservation in federated query solutions.
SMCQL \cite{bater2017smcql} is a pioneering federated query solution. It utilizes the garbled circuit to encrypt the whole query execution. The utilization of the garbled circuit makes SMCQL have the strong capacity to support the whole SQL standard but introduces significant computation overhead and only supports the computation over two clients. SMCQL also introduces a query planner to optimize the computation graph, so that some computation steps can be performed within one client and do not need MPC execution. Conclave \cite{volgushev2019conclave} is then proposed to reduce the heavy computation workload of SMCQL. It utilizes the respectively lightweight secret sharing schemes to replace garbled circuits so that the computation is faster; computation among three clients is supported; but some query (\eg window aggregate) is no longer supported. In addition, Conclave enables clients to optionally annotate some parts (columns) of their data to be non-private. The optional annotation reduces the scale of computation needed for MPC execution and further optimizes the computation time. Senate \cite{poddar2021senate} is another SQL executor over federated data. Compared to SMCQL and Conclave, it changes the semi-honest assumption into a stricter adversarial threat model. It proposes novel MPC decomposition and query planning techniques, letting some computations be executed on a subset of clients. 

Cryptography tool is also a necessity for specific federated query solutions. In \cite{wang2021secure}, a federated query solution is proposed to handle the free-connex join-aggregate query (a special kind of query). The authors provide a secure version of the Yannakakis algorithm over two clients, which is much more efficient in runtime compared to the garbled circuit. In \cite{zhang2022efficient}, the authors consider skyline queries over vertical data federations. Their algorithm design decomposes the execution into local computation and cross-client secure aggregation, where the secure part is reformulated into the private set computation problem, which can be efficiently handled by MPC solutions. Hu-fu \cite{tong2022hu} tackles SQL queries related to spatial data. The authors notice that various spatial queries, (including kNN, range counting, and range query) can be realized by only three secure operations (summation, comparison, and set union), while other distance-based operations can all be conducted without encryption. Hu-fu then designs effective dedicated algorithms to handle each of the three secure operators and therefore achieves good performance in computation load and gains capacity in supporting up to ten clients.

\textbf{Distributed differential privacy method(s).} Some researchers design advanced privacy preservation mechanisms for the federated query that combines cryptography tools and CDP, following the idea of DDP, so that an enhanced privacy guarantee can be achieved.
SAQE \cite{bater2020saqe} considers a new direction of optimizing federated query runtime. It removes the restriction that the SQL result must be accurate, and let the federated query derive approximate results. As the execution over a small sampled set of data is enough to derive the query result with a bounded accuracy guarantee, SAQE achieves much better performance by reducing the size of the data to be computed. In addition, the data sampling scheme of Sage realizes an additional DP guarantee of the approximate query result.
Orchard \cite{roth2020orchard} is a novel federated query solution that transforms queries into three ``zones" that computation on private individual data, aggregated but not noised data, and noised data, respectively. The results of the local computation in the first ``zone" are homomorphically encrypted and sent to the aggregator which performs summation. The aggregated results are decrypted, noised, and announced by a committee later that owns the key. The detailed procedure is the same as \cite{roth2019honeycrisp}. Orchard can efficiently answer queries at scale as long as there is additive aggregation in them such as k-means, logistic regression, perceptron, and PCA.
Arboretum \cite{margolin2023arboretum} is designed to efficiently answer a wide range of queries in large-scale FA setups with potentially billions of participants. Arboretum's key strengths include its ability to automatically optimize query plans and distribute computational tasks across participant devices. The system outperforms previous solutions by supporting new types of queries and matching the cost of existing systems that were hand-optimized for specific queries. Arboretum's approach significantly enhances the scalability and feasibility of executing complex queries in distributed environments while maintaining strong privacy guarantees.

The idea of DDP also applies to specific federated queries.
Mycelium \cite{roth2021mycelium} is a system that enables processing differentially private queries over large-scale distributed graphs, which are common in scenarios like disease or malware tracking. Mycelium achieves privacy by combining HE, a verifiable secret redistribution scheme (similar to Orchard \cite{roth2020orchard}), and a mixed network based on telescoping circuits. It can handle various queries relevant to medical research without compromising individual privacy or learning the graph's topology.

In Table \ref{table_fedquery}, we summarize our surveyed federated query solutions, regarding their supported query types, privacy models, cryptography techniques, and maximal supported clients.

\blue{\textit{\textbf{Lessons learned.} Federated query solutions are crucial for enabling privacy-preserving data analytics across distributed clients, but they face significant challenges in computation overhead due to the use of cryptographic tools like MPC. Optimizing computation time is essential, with researchers exploring various methods such as lightweight cryptographic techniques, query graph optimization, and approximations. While general solutions support a wide range of queries, they often struggle with scalability, whereas specific query solutions can achieve better performance by focusing on particular types of queries. Scalability and performance improvements highlight the importance of adapting federated query systems to the specific needs of large-scale environments and specialized queries.}}

\section{Assisting Federated Learning}\label{sec_assistfl}

\begin{table*}
\caption{Summary of assisting federated learning.}
\label{table_assistingFL}
\centering
\begin{tabularx}{\linewidth}{llX}
\hline
   \textbf{Reference}& \textbf{Data analytics task} & \textbf{Note}\\
   \hline
    FedACS \cite{wang2021fedacs} & Preprocessing: Client quality evaluation for FL  & Selecting clients with lower data heterogeneity based on their gradients\\
    FAVOR \cite{wang2020optimizing} & Preprocessing: Client quality evaluation for FL  & Using a reinforcement learning model to select clients based on their gradients to maximize validation accuracy \\
    Oort \cite{lai2021oort} & Preprocessing: Client quality evaluation for FL  &  Selecting clients with higher training losses\\
    Cho \etal \cite{cho2022towards} & Preprocessing: Client quality evaluation for FL  &  Prioritizing high-loss clients for participation \\
    CMFL \cite{luping2019cmfl} & Preprocessing: Client quality evaluation for FL &  Preventing gradient uploads of low-quality clients\\
    Sattler \etal \cite{sattler2020clustered} & Preprocessing: Personalize FL models via client clustering & Grouping clients with similar data distributions\\
    Briggs \etal \cite{briggs2020federated} & Preprocessing: Personalize FL models via client clustering  & Agglomeratively clustering clients after training a global model \\
    Liang \etal \cite{liang2023efficient} & Preprocessing: Personalize FL models via client clustering  & Decomposing and consolidating extreme clusters \\
    IFCA \cite{ghosh2020efficient} & Preprocessing: Personalize FL models via client clustering  & Assigning clients to clusters where they achieve the lowest local loss on distributed models\\ 
    COMET \cite{cho2023communication} & Preprocessing: Personalize FL models via client clustering  & Incorporating knowledge sharing with a regularization term during local model training \\
    FedSoft \cite{ruan2022fedsoft} & Preprocessing: Personalize FL models via client clustering  & Using proximal loss to align client training with multiple cluster models\\
    Oort \cite{lai2021oort} & Postprocessing: Model evaluation & Automatically adjusting the number of participating clients using confidence bounds derived from Hoeffding’s inequality\\
\hline

\end{tabularx}  
\end{table*}

\begin{figure*}[t]
\centering
\includegraphics[width=0.8\textwidth]{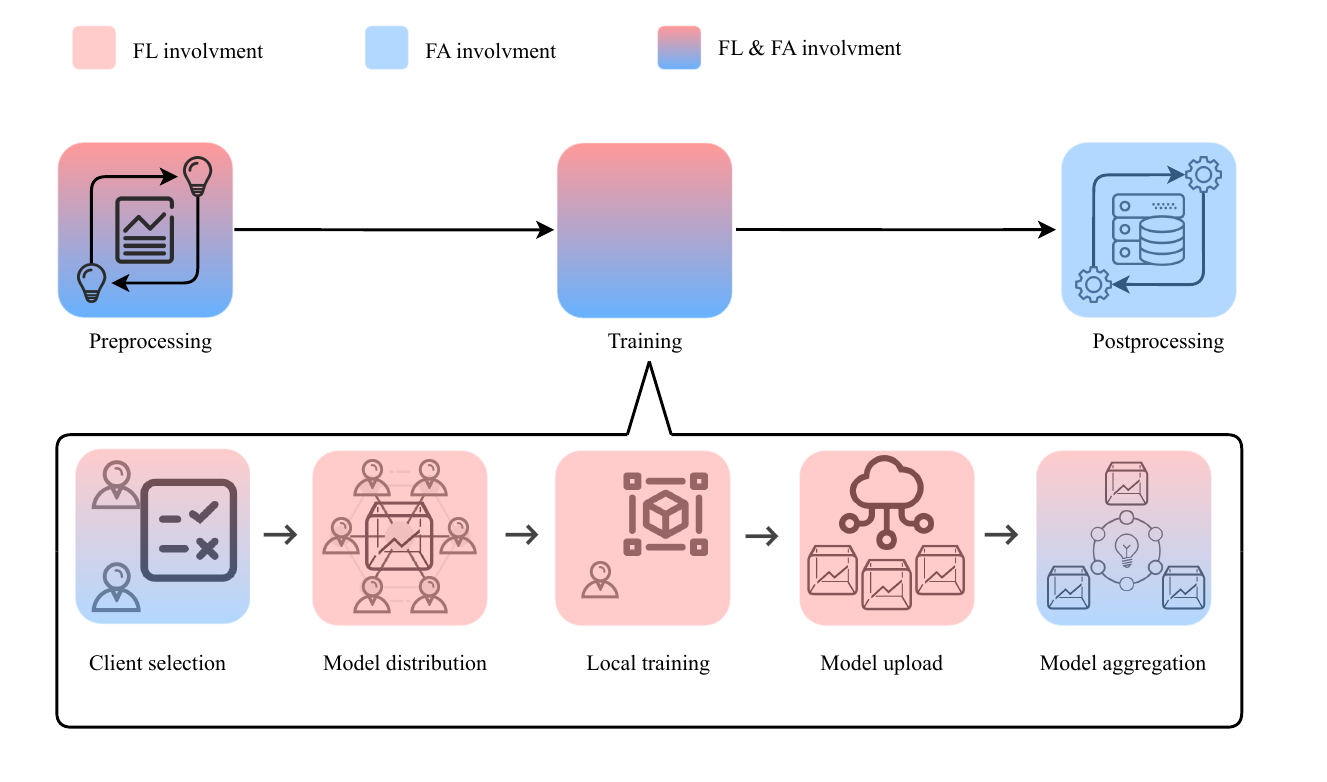}
\caption{A typical workflow of FL assisted by FA. FA is applied in the postprocessing phase of the FL pipeline. The preprocessing phase, as well as the client selection and model aggregation steps in the training phase, includes the involvement of both FL and FA.}
\label{fig_flworkflow}
\end{figure*}

\blue{In this section, we focus on the fourth class of data analytics applications following our taxonomy: assisting federated learning. We consider three scenarios where FL can gain benefit from FA-empowered analytics, and introduce them separately in the following part.}

FL has gained success in various data-incentive tasks where a neural network is trained for prediction/description applications. Many works are proposed to improve the FL performance regarding model accuracy, convergence rate, scalability, \textit{etc}. In these works, the data of the clients are usually analyzed to pursue optimization of the FL performance. However, as is restricted by the tenet of FL, the raw data should not be transmitted to the server while conducting such analysis. Therefore, in these works, the raw data of the clients are usually transformed into insensitive insights which help the system to perform decision-making in optimizing FL. Although not explicitly claimed, such an approach exactly follows the FA paradigm, and these solutions can be regarded as an FL instance optimized by another FA algorithm.

We categorize the solutions of FA-assisted FL based on when FA takes effect, as illustrated in Fig. \ref{fig_flworkflow}. We summarize the whole procedure into three phases: the preprocessing phase, where the clients are registered and evaluated by the server; the training phase, where the FL model is transmitted between server and clients, trained by the clients locally, and aggregated by the server, for multiple rounds; the postprocessing phase, where the performance of the derived FL model is then evaluated. 
Following the definition of the majority of FL research, FL usually includes the training phase and a part of the preprocessing phase, and the assisting FA schemes can be executed in the preprocessing phase, in the postprocessing phase, and sometimes in the training phase parallel with FL. Fig. \ref{fig_flworkflow} illustrates the involvement of FL and FA in the whole procedure with colors.

\blue{The FA schemes for FL assistance usually do not apply any privatization methodology other than ``hiding raw data''. Specifically, since many FA schemes recycle the client uploads in the host FL mechanisms as FA insights, they essentially inherit the privacy preservation of the FL schemes. For example, when the host FL schemes utilize LDP in the uploaded gradients, the assistance FA schemes also gain an LDP guarantee if they utilize the FL gradient as insights (it is common in FA-assisted FL).}

In the following parts, we will introduce some classes of FA schemes to assist FL. In Section \ref{subsec_assist_selection}, we investigate FA schemes that help FL to evaluate the data quality of clients, and further select proper participating clients, which executes on the preprocessing phase and client selection step of the training phase. In Section \ref{subsec_assist_personalize}, we investigate FA schemes that help FL model personalization via conducting clustering on the clients, which executes on the preprocessing phase and model aggregation step of the training phase. In Section \ref{subsec_assist_evaluation}, we investigate FA schemes to evaluate the performance of the FL model, which executes in the postprocessing phase. Table \ref{table_assistingFL} summarizes the FA schemes covered by this survey in assisting FL applications.

\subsection{Preprocessing: Client quality evaluation for FL}\label{subsec_assist_selection}

Being consistent with conventional model training, the quality of training data is a critical factor in influencing the outcome model performance in FL. When it comes to the scenario of federated data, the quality of data in different clients varies, providing different benefits (or even harms) to the FL model. The different quality of client data may originate from the data intrinsically, like different correctness of labels, and different scales of noise, or from the overall FL system, like different heterogeneity of data distribution compared to the global one (\textit{aka.} non-IID issue). Therefore, evaluating the quality of client data in the preprocessing phase becomes a profitable procedure. The results of client quality evaluation can be utilized for many FL system optimizations. In some FL research, the client quality evaluation is employed to improve the client selection of FL, rather than the random selection baseline. In other words, the uploaded gradients from different clients are processed by the server before aggregation, through reweighting or trimming, based on the client evaluation results.


\textbf{Hiding raw data method(s).}
FedACS \cite{wang2021fedacs} is an FA solution to measure the data skewness (severity of data heterogeneity) of the clients. Since data heterogeneity is a significant source of performance degrading of FL tasks, measuring data skewness is proven to be beneficial in assisting client selection for FL. In FedACS, the FL gradients of clients are utilized as FA insights, so that the computation and infrastructure of FL can be reused. The authors prove that the clients with local gradients close to the average gradient tend to have lower data skewness, and are more beneficial to the FL model. The server then formulates a dueling bandit to perform client selection, where clients with lower data skewness are more likely to be selected.
In FAVOR \cite{wang2020optimizing}, the authors formulate a reinforcement learning model running on the server side to perform client selection. In that reinforcement learning model, the states are defined to be a combination of the global model gradient and local gradients of all clients, where the latter is transmitted from the clients and serves as the FA insight. The actions are defined as all possible sets of participating clients in the incoming round. The rewards are defined by the validation accuracy of the model after the training. Such reinforcement learning formulation forces the reinforcement learning model to choose the participating clients that achieve higher validation accuracy.
In Oort \cite{lai2021oort}, the authors optimize the training and evaluation of FL models by wisely selecting participating clients. The optimization of training is based on an observation of the authors, that a client inducing higher training loss is more beneficial for the FL training. Therefore, in Oort, the clients upload their training loss (which is a scaler) to the server in each round they participate in, which works as the insight of FA. The server then utilizes a heuristic multi-armed bandit to perform client selection, where clients obtaining higher training losses in previous rounds will be more likely to be selected.
In \cite{cho2022towards}, the authors formally prove that selecting clients with higher training loss can accelerate the FL model convergence. Based on the theoretical result, the authors propose a power-of-choice client selection strategy, where a global model is sent to all available clients, the clients upload the training loss as the FA insight, and the server finally selects a set of high-loss clients as the participants in the following several rounds.
In CMFL \cite{luping2019cmfl}, the quality of clients is evaluated as the divergence between the local gradient and global gradient, based on the difference of signs throughout all dimensions of the gradient vector. After that, gradients from those low-quality clients are trimmed. Furthermore, by moving the evaluation procedure from the server to the clients, CMFL reduces the overall communication overhead by preventing the gradient uploads of low-quality clients.

\blue{\textit{\textbf{Lessons learned.} As the data of participating clients can greatly affect the performance of FL, evaluating the quality of client data becomes a significant applied domain in FA solutions in assisting FL. Existing solutions evaluate client data in many aspects, including the absolute quality and class distribution. These solutions usually recycle intermediate results of FL training to evaluate client data, such as the gradient and the loss. Such an approach has the benefits of computation/communication efficiency, and the inheritance of privacy preservation in FL.}}


\subsection{Preprocessing: Personalize FL models via client clustering}\label{subsec_assist_personalize}

The vanilla setting of FL considers an identical global model shared by the server and clients, which is applied in downstream tasks on all data. However, in practice, the client data are usually non-IID distributed, and the data held by different clients has quite different characteristics (distributions). As a result, it becomes difficult to let one model achieve good generalization on data from all clients, \ie the optimal model parameters for global data might be quite divergent from the optimal model parameters for data in particular clients. Personalized FL resolves such an issue by breaking the limit of one identical global model. In personalized FL, multiple models are trained by data from different clients and are applied to the data of those clients. With personalized FL, each client can obtain a personalized model tailored to its data characteristics, which achieves better performance when performing inference on its data \cite{tan2022towards}.

Among various approaches to personalized FL, clustering-based methods are widely selected for their high reasonability and strong performance. In clustering-based methods, clients are firstly clustered, where those with similar data characteristics are placed in the same cluster. Then, FL is fine-tuned based on the data within one cluster. With such approaches, the personalized models will be trained on sufficient data stored in several clients, and tailored for particular data characteristics. The clustering procedure requests the sharing of information about the client data. Similar to those vanilla federated clustering solutions introduced in Section \ref{subsec_statistical_clustering}, the FA schemes for client clustering also just apply the privatization of hiding raw data.


\textbf{Hiding raw data method(s).}
In \cite{sattler2020clustered}, the authors propose a pioneering clustering personalized FL solution named CFL. In CFL, the cluster assignment of clients is determined by a hierarchical clustering structure: the data distribution similarity of pairs of clients is calculated by the cosine similarity of local gradients; a bipartition of clients is derived by analyzing the similarity matrix; more clusters are derived by hierarchically executing the bipartition algorithm as long as it could improve the overall performance.
In \cite{briggs2020federated}, the authors propose another pioneering clustering personalized FL solution. It trains a global model with vanilla FL and then uses gradients from all the clients to perform clustering. The clustering is based on the cosine similarity between clients using an agglomerative clustering algorithm.
In \cite{liang2023efficient}, the authors propose a clustering personalized FL approach by extracting one layer of the gradient uploaded by the clients and clustering clients by calculating the cosine distance of the layer parameters. The authors also propose a decomposition and consolidation scheme to remove some extreme clusters and reassign the clients to other clusters.
In \cite{ghosh2020efficient}, the authors propose an Iterative Federated Clustering Algorithm (IFCA) to cluster the clients for personalized FL. In IFCA, the server first formulates one model for each of the clusters and distributes all the models to the participating clients. The clients calculate their local loss of all the models, and one client will be assigned to the cluster where the lowest loss is obtained on the corresponding model.
In \cite{cho2023communication}, the authors propose COMET to perform clustered co-distillation in FL. In COMET, there exists a public set of unlabeled data samples. The clients train their personalized model with its local data and then infer the public data to derive the soft decisions. The clients upload their soft decisions on the public data to the server, and the server performs $k$-means on that so that clients with similar soft decisions on the public data will be placed in the same cluster. After that, knowledge sharing is performed in model training. In detail, the clients then train their local model with their local data, while an extra regularization term is applied. The regularization term forces the model to obtain soft decisions similar to those of other clients in the same cluster on the public data.
In \cite{ruan2022fedsoft}, the authors consider the problem that the clients may find it hard to cluster into several distinct data distributions. They consider soft clustering, where the distribution of data in each client is considered as a mixture of several clusters. FedSoft is proposed to perform soft clustering by requesting the clients to train with a proximal loss, regulating the training to get close to a mixture of several cluster models. FedSoft system employs an FA subroutine to estimate the weight of each client with all the clustered. Each data sample of one client is evaluated by all the cluster model and is assigned to the cluster with the smallest loss. The weight of the client is then calculated as the portion of data samples belonging to all the clusters.

\blue{\textit{\textbf{Lessons learned.} Federated clustering is not only a typical statistical application of FA (discussed in Section \ref{subsec_statistical_clustering}) but also provides valuable optimization on FL. Based on the idea of clustered FL, \ie clients collaborate with other clients within the same cluster with higher similarity regarding client data. FA solutions for clustering FL derive clusters of clients based on various similarity measures of client data. To ensure privacy preservation, these methods usually reuse the gradient information in FL to represent a fingerprint of client data, so that the privacy level of FL can be preserved.}}

\subsection{Postprocessing: Model evaluation}\label{subsec_assist_evaluation}

While the mainstream of FL studies focuses on model training, \ie deriving a model that is expected to have high performance based on the federated data. Meanwhile, as the evaluation of the trained model also relies on the vast data, leveraging the federated data then becomes an effective approach. 

\textbf{Hiding raw data method(s).}
In the blog \cite{fa20} where the term FA is coined, the authors claim that the idea of FA is exactly motivated by the need to evaluate the trained FL model. In Oort \cite{lai2021oort}, the authors propose a system of federated evaluation. Oort automatically adjusts the number of participating clients, so that the accuracy of performance evaluation is guaranteed and the client effort is minimized. Such adjustment is achieved by the confidence bounds derived by Hoeffding's inequality.
In these FA-based model evaluation solutions, the transmitted computation model is usually the learning model, which is consistent with FL. However, the local computation part is conducted by performing model inference, and the derived insight is the performance metric of model inference. It is a simpler form of local computation compared to FL, only requiring the forward propagation phase of FL local computation, and dropping out the more computation-intensive backward propagation phase.

\blue{\textit{\textbf{Lessons learned.} Evaluating the FL model is the earliest formal application of FA, where the evaluation metrics are computed locally, and aggregated centrally so that the evaluation dataset can be protected. The aggregation of evaluation metrics is usually in the form of mean computation. However, extra issues}}


   

\section{Wireless Network Applications}
\label{sec_wireless}

\begin{table*}
\caption{Summary of wireless network application. }
\label{table_wirelessnetworks}
\centering
\begin{tabularx}{\linewidth}{llll}
\hline
   \textbf{Reference} & \textbf{Type} & \textbf{Wireless network application} & \textbf{Technical model}\\
    \hline
   Mulvey \etal \cite{mulvey2023cellular} & Wireless network model optimization & Anomaly detection& Clustering\\
   Wang \etal \cite{wang2020federated} & Wireless network model optimization & Energy consumption minimization & SVM\\
   Chen \etal \cite{chen2021federated} & Wireless network model optimization & Malicious entity detection & Customized model\\
   Xing \etal \cite{xing2023multi} & Assisting wireless network FL & Semantic communication & Customized model\\
   Wang \etal \cite{wang2020federateduav}& Assisting wireless network FL & UAV deployment planning & Customized model\\
   Zhao \etal \cite{zhao2022semi} & Assisting wireless network FL & Intrusion detection & Knowledge distillation\\
   Wang \etal \cite{wang2022content} & Assisting wireless network FL & Content popularity detection & Clustering\\
   \hline
   
\hline
\end{tabularx}  
\end{table*}

\blue{In this section, we focus on the fifth and last class of data analytics tasks in our taxonomy: wireless network applications.}

Wireless networks have become a key cornerstone of modern communication systems. Meanwhile, data analytics (machine learning) approaches are widely applied to optimize the performance of wireless network systems. These technologies have been utilized to handle critical tasks such as network slicing, caching management, anomaly detection, and semantic communication. Federated computation, including both FL and FA, is also applied by researchers to obtain the benefits of privacy preservation, edge computing utilization, and communication efficiency. In this section, we focus on the FA solutions applied for the critical optimizations on wireless network systems. 

\blue{Since the ``wireless network application'' is a relatively large domain, the specific data analytics tasks in FA solutions are diverse and request further classification. In this section, based on the essence of the host data analytics tasks, we further classify the wireless network applications into two types: wireless network model optimization, and assisting wireless network FL. The FA application in wireless networks is relatively primitive, and all existing FA solutions we surveyed do not apply advanced privatization methodology. Instead, they simply apply the methodology of hiding raw data via the federated computation paradigm.}

\subsection{Wireless network model optimization}

\blue{The complexity of wireless network forces their designers to include many model-based structures to give data-driven intelligence to the wireless network. Some of these models are not as complex as neural networks. Therefore, FA is applied in these tasks as their federated computation solutions to utilize privacy-preserving data. These model-based tasks include SVM \cite{wang2020federated}, clustering \cite{mulvey2023cellular,wang2022content}, and knowledge distillation \cite{zhao2022semi}.} 

\textbf{Hiding raw data method(s).}
In \cite{mulvey2023cellular}, an FA solution is proposed for anomaly detection in cellular network antenna tilt. It transforms the data regarding antenna electrical tilt into signature vectors and then performs federated clustering on these signature vectors, where the local clustering centroids serve as the FA insights. Then, anomaly items are then filtered out by measuring the distance between signature vectors and global clustering centroids.
In \cite{wang2020federated}, the problem of energy consumption minimization in mobile edge computing task offloading is considered. To handle the time-varying user task, a federated version of support vector machine (SVM) is proposed. In the federated SVM scheme, the covariance matrix between users is transmitted from the server to clients (base stations), and the SVM weights are uploaded from clients to the server and are then aggregated by the server. Eventually, the global SVM can derive the optimal solution of task offloading, with optimized communication and computation overheads.
In \cite{chen2021federated}, an FA scheme is utilized in mobile crowdsensing scenarios to discover malicious sensing tasks. In such a scheme, a machine learning module is equipped by each client (detection device). The clients predict whether a task is malicious, and uploads the prediction results to the server as the FA insight. The server aggregates the insights and derives the updated reputation score of the task proposers to the clients, which helps optimize the prediction results in the future. The global prediction results are also derived by aggregating the client predictions, with extra risk-aware computations.

\subsection{Assisting wireless network FL}

\blue{While many wireless network applications utilize non-complex models as introduced in the previous part, others apply neural networks to handle sophisticated tasks faced by wireless networks, and also FL when federated computation paradigm. Similar to those in Section \ref{sec_assistfl}, FA solutions are also applied to assist these wireless network FL models, which is another type of FA in wireless network applications}


\textbf{Hiding raw data method(s).}
In \cite{xing2023multi}, FL is employed to build up a semantic communication system, while FA is applied to assist it via generating the aggregation indicators. The aggregation indicator is generated by the FA local computation scheme, reflecting the suitability between the current channel condition and the local FL model. When the aggregation indicator is received by the server along with the FL model, it is transformed into the weight in weighted FL model aggregation.
In \cite{wang2020federateduav}, the FA-assisted FL scheme is deployed to control the UAV deployment in visible light communications scenarios. It needs to resolve the complex optimization problem jointly considering ``UAV deployment, user association, power efficiency, and predictions of the illumination distribution''. To predict the illumination distribution, a convolutional autoencoder is trained by FL methodology. The FA methodology is also applied to help with FL training. The FA scheme uploads the convolution kernels and bias to the server. By aggregating the convolution kernels, gap matrices can be calculated, which are essential for the evaluation and optimization of the UAV deployment plan.
In \cite{zhao2022semi}, an FL system is designed to conduct intrusion detection. To tackle the challenge of privacy risk of transmitting gradient, non-IID data, and high communication overhead, the proposed scheme exactly replaces the conventional FL insight derivation and aggregation schemes, by those in FA methodology. Such schemes, utilizing the idea of knowledge distillation, are also termed federated distillation from the FA perspective. In the proposed scheme, the locally generated labels on shared unlabelled data are sent to the server and then aggregated to form the global hard labels. Then, the server can use the shared data with global labels to train the global model.
In \cite{wang2022content}, an FL model is trained to perform content popularity prediction in Fog-RAN systems. The clustered FL scheme is further deployed based on the federated clustering solution. In the federated clustering scheme, the local features of the clients are computed and uploaded. The server computes the similarity between clients and splits up new clusters iteratively. Eventually, the final clustering results can be derived, and the personalization of the host FL scheme is improved.


\blue{We summarize the FA applications in wireless networks in Table \ref{table_wirelessnetworks}. \textit{Wireless network application} refers to their task from the perspective of wireless networks. \textit{Technical model} refers to the technical/mathematical task from the perspective of data analytics and artificial intelligence. It is marked ``customized model'' if its model formulation cannot be summarized into any well-known model.}

\blue{\textit{\textbf{Lessons learned.} FA applications on wireless network systems generally follow two types: optimizing some non-neural network models that request private data for functionality or assisting the FL models deployed in wireless network systems. FA helps wireless network systems in many tasks including anomaly detection and all kinds of planning and decision-making. The mathematical formulation of FA solutions may follow some classical problems in data analytics (\eg SVM and clustering) or include customized analytics tasks that fit the tailored need in wireless network systems.}}


\section{Open Issues and Future Directions}\label{sec_openissue}

\blue{In this section, we discuss some open issues and potential directions for future FA research, so that a bigger picture of future research will be presented. We present seven open issues that cover three directions to improve the research status of FA. In the first direction, FA extends its realm of applications, so that more data scenarios can be supported (Section \ref{subsec_openissue_scenario}), and eventually a unified FA framework can be realized to tackle all kinds of data scenarios (Section \ref{subsec_openissue_unified}). In the second direction, the privacy preservation of FA could be improved. Privacy preservation can be achieved in large-scale systems (Section \ref{subsec_openissue_privacyscale}), and different privacy criteria utilized in FA can be measured and compared (Section \ref{subsec_openissue_privacymeasure}). In the last direction, FA is desired to be more convenient for real-world deployment. FA systems are excepted to be proposed with considerations on many systematic factors (Section \ref{subsec_openissue_system}); novel technologies in the field of wireless communication are expected to optimize current FA workflow (Section \ref{subsec_openissue_wireless}); all kinds of resources in FA systems are expected to be properly managed (Section \ref{subsec_openissue_resource}). These open issues and future directions are promising to improve the value of FA to digest valuable information from big data analytics with the strongest respect to user privacy. }

\subsection{Applications to more complex data scenarios}\label{subsec_openissue_scenario}
The fundamental difference between FA and FL lies in the nature of their supporting tasks. The scope of data scenarios studied by existing FA remains small compared with the diversity of data science problems and models in the wild. Designing FA mechanisms to support non-trivial data types, including graph data \cite{9101863}, streaming data\cite{ren2022ldp}, key-value data\cite{wu2022poisoning}, multidimensional data \cite{8731512}, time series data, and important networked applications, including Internet telemetry\cite{9800915}, smart home \cite{fu2023comverse}, healthcare \cite{froelicher2021truly}, web 3.0 and other privacy-critical senarios\cite{8599856} all present interesting and open challenges. \blue{For example, current FA research in graph data focuses on deriving graph metrics from client data with private preservation. However, many real-world graph data are associated with attached data that is not in the form of graphs (\eg semantic web). How to perform fusion FA that performs analytics with fusion of structural and non-structural data, is a promising and challenging research direction.}

\subsection{A unified FA framework}\label{subsec_openissue_unified}
FL operates under a unified framework to embed similar computation procedures and insight structures when training various types of neural networks. Such a unified framework plays a vital role in boosting the wide studies and applications of FL. On the other hand, FA mechanisms designed so far are still highly task-specific. While it seems intuitive that different data analytics tasks are naturally different regarding their computation procedures, data structures, \textit{etc}, there already exists some efforts to provide a unified framework for a particular class of data science problem\cite{wang2022fedfpm}. Some FA solutions for database operations 
Attracted by the potential benefits of a unified framework, it remains a grand open problem to design a unified framework for broader classes of data analytics tasks or even universal data analytics tasks. 
Specifically, motivated by this, some open questions and interesting directions could be (i) What are the proper architecture, and interface design for a unified FA framework?  (ii) If we still adopt the current local insight upload and global aggregation scheme, how to design a general aggregator or general insight form that capable of incorporating diverse requirements? (iii) Witnessing the advancement in generative models, is it possible to use generative models at the server side, essentially, as a general aggregator, to further reduce the reliance on locally uploaded insights so that the communication cost can be reduced further and versatile privacy-preserving tasks can be supported.


\subsection{Privacy preservation at scale}\label{subsec_openissue_privacyscale}
As FA systems grow in scale with more data or users involved, ensuring privacy becomes more challenging. This challenge is exacerbated when analysts seek high-accuracy analytics. Existing privacy-preserving techniques exhibit distinct characteristics concerning privacy assurance, data utility, and scalability. For example, DP has long been criticized for offering a poor accuracy-privacy tradeoff, prohibiting its applications in domains that request accurate population-level profiling or analytics. Other cryptographic techniques, such as MPC, encounter difficulties in scaling to practical sizes for mobile applications. Consequently, FA mechanisms have to be carefully designed to strike a delicate balance between privacy, utility, and scalability. \blue{Some recent concepts regarding privacy preservation, like private cloud computing, are promising to enhance FA algorithmic design: it might be possible to decompose the FA algorithm into local computation part and private cloud computation part, where the local computation part conducts lightweight computations that extract insights to reduce communication overheads, and the computation-intensive schemes, such as cryptography-based privacy preservation, is conducted on the private cloud.}

\subsection{Formal measurement of privacy preservation}\label{subsec_openissue_privacymeasure}
\blue{Existing FA works utilize task-specific algorithms, and provide different privacy guarantees. These guarantees are under different threat models and assumptions and eventually provide different privacy preservation that are hard to trace and compare. Taking DP as an example, in DP-based FA solutions, in addition to conventional CDP/LDP separation, various DP definitions (\eg sequence DP, token DP) are designed with different focuses on privacy preservation. As a result, even for two FA schemes both providing DP guarantees with the same values of $\epsilon$ and $\delta$, their strength in privacy preservation may vary a lot. Therefore, it is crucial to provide a formal and comprehensive measure of privacy preservation, covering both DP, cryptography, and other privatization approaches. Such a formal measurement is beneficial for the development of FA platforms where FA are provided as a service, and also helpful for the ambition of a unified FA framework. Possible approaches include quantifying privacy preservation based on the performance of adversarial privacy attacks. However, such an approach lacks theoretical justification and is limited by attack methods and models.}

\subsection{System-efficient FA}\label{subsec_openissue_system}
Studies in the field of FA have primarily concentrated on algorithm design to enable the execution of various data analytics tasks in a federated manner. In contrast to the well-established problem hierarchy in FL, it is evident that a substantial portion of the FA domain remains unexplored, particularly from the systems' perspective. As a federated system, FA also faces practical challenges, such as heterogeneous computing power across devices, dynamic system sizes with devices joining and leaving, incentives and pricing issues, robust FA under adversaries, \textit{etc}. FA further introduces new interpretations of system measures, such as fairness, and novel system challenges, such as significant computing costs for certain privacy-preserving measures, and a limited privacy budget on the client side. As FA continues to evolve, addressing these system-level challenges will be crucial for its successful implementation and widespread adoption.

\subsection{Wireless communication for FA}\label{subsec_openissue_wireless}
Over-the-air computation (AirComp) leverages the superposition property of wireless channels to enable simultaneous transmission and aggregation of signals from multiple devices \cite{over-the-air,mao2023roar,hsu2023joint,yan2023device}. It has been applied to FL by aggregating model updates from numerous devices directly over the air. It not only enhances communication efficiency but also maintains the privacy of clients' local data, addressing key bottlenecks in traditional FL approaches, such as limited bandwidth and latency issues. However, the existing research on AirComp focuses on FL rather than FA. Since the model aggregation in FL usually are in a simpler form (FedAvg and its variations), FA aggregation usually requires a more complex computation procedure. How the idea of AirComp can be applied for complex aggregation schemes, and how FA could benefit from physical layer techniques, also present interesting and open challenges.

\subsection{Resource management in FA}\label{subsec_openissue_resource}

Essentially as an edge computing scheme, FA should naturally applied to all kinds of resource-limited scenarios. The deployment of FA is companied by all kinds of resource constraints: the local computation and insight aggregation phases require computation resources; the model and insight transmission require communication resources; the level of client data leakage consumes the novel privacy resource at the client side; even the participation of client can also be considered as a kind of resource to some extent. Significant research efforts have been made in resource management and resource saving in FL, including all these aspects of computation \cite{wang2022progfed}, communication \cite{luping2019cmfl}, privacy \cite{liu2024dpbalance}, and client participation \cite{wadu2021joint}. However, designing novel mechanisms for resource management and saving for FA, or adapting these FL mechanisms into FA applications, is a non-trivial unexplored topic. Many existing schemes for FL may highly rely on the FL architecture and are not applicable in FA applications. For example, gradient compression and quantization are widely used in communication-efficient FL but are unlikely to be useful in FA since the uploaded insights are not in the form of gradient. Therefore, designing resource management and saving schemes for FA is a critical issue for its wide deployment.


\section{Conclusion}\label{sec_conclusion}
Due to the exponential growth of edge data and the growing awareness of data privacy, privacy-preserving distributed data processing has attracted wide interest from both academia and industry. FA is an emerging collaborative data processing framework for descriptive data science tasks without centralizing the raw data. It brings significant benefits in privacy protection, communication reduction, and task coverage. 
Although FA has been widely studied in industry and academia, a systematic review of the existing efforts in FA has not been conducted yet. This survey fills the gap by first comprehensively reviewing the key concepts in FA, its relationship with similar techniques, and its key challenges. It presents a detailed taxonomy to categorize FA studies from both applications and system characteristics. Key enabling techniques are introduced with a specific focus on privatization, analytics, and optimization. A wide spectrum of FA tasks are then reviewed demonstrating the generality of FA applications. Finally, we discuss the open issues and future directions in FA, from the perspective of  application, privacy protection, framework, system optimization, and cross-layer design. This survey summarizes the existing efforts in FA and the huge intersection among FA and data science, privacy, distributed computing, wireless communications, and networking systems, and derives key lessons learned that represent our insights on this field. Overall, FA research is widely interdisciplinary, approachable, and contains many critical problems that would benefit greatly from the expertise of all related areas.

\bibliographystyle{IEEEtran}
\bibliography{ref}

\begin{IEEEbiography}[{\includegraphics[width=1in,height=1.25in,clip,keepaspectratio]{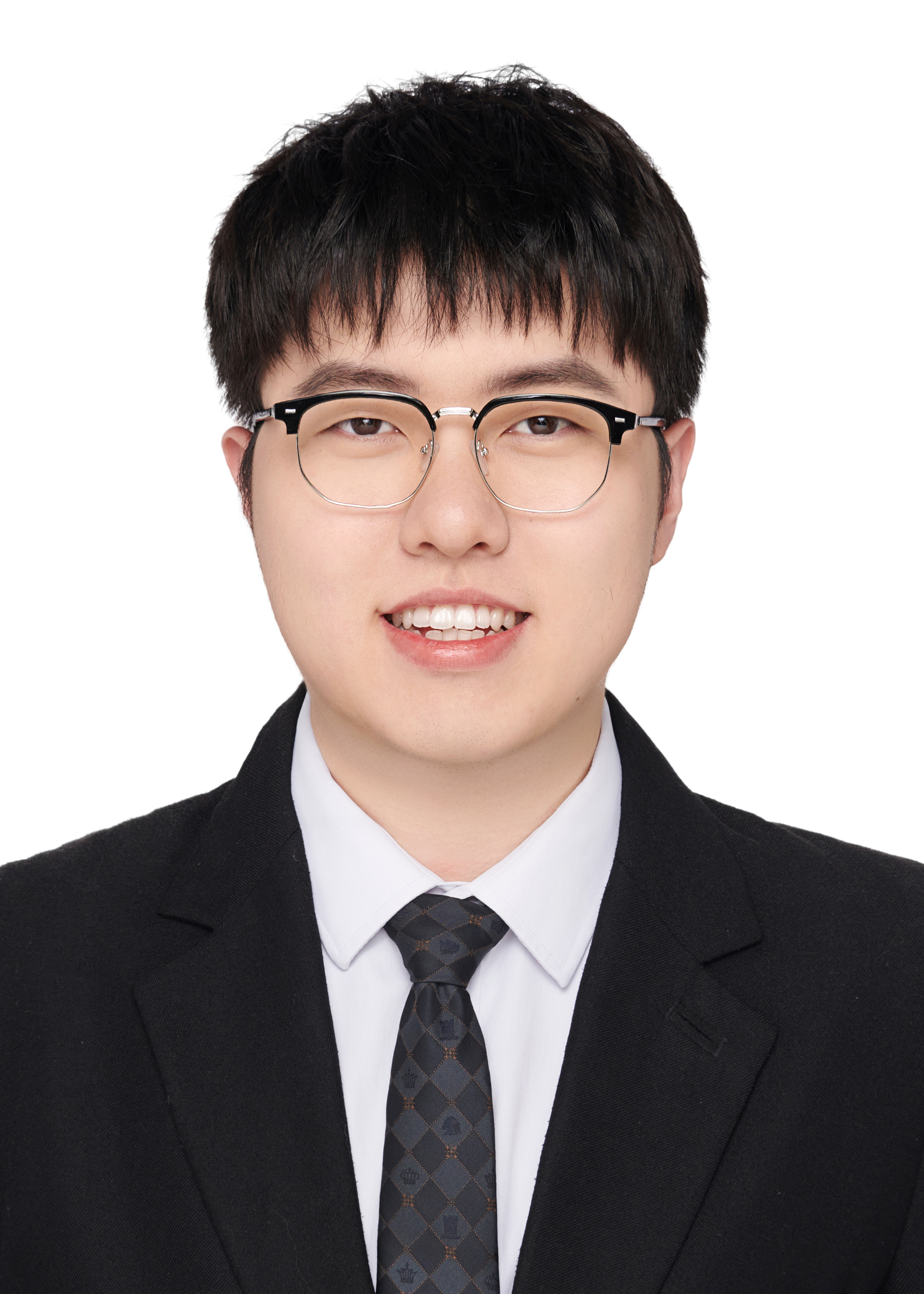}}]{Zibo Wang}
received the B.E. degree in Electrical and Computer Engineering from Shanghai Jiao Tong University in 2020. He is currently pursuing the Ph.D. degree in the same institute. His research interests include federated learning, federated analytics, and privacy computing.
\end{IEEEbiography}

\begin{IEEEbiography}[{\includegraphics[width=1in,height=1.25in,clip,keepaspectratio]{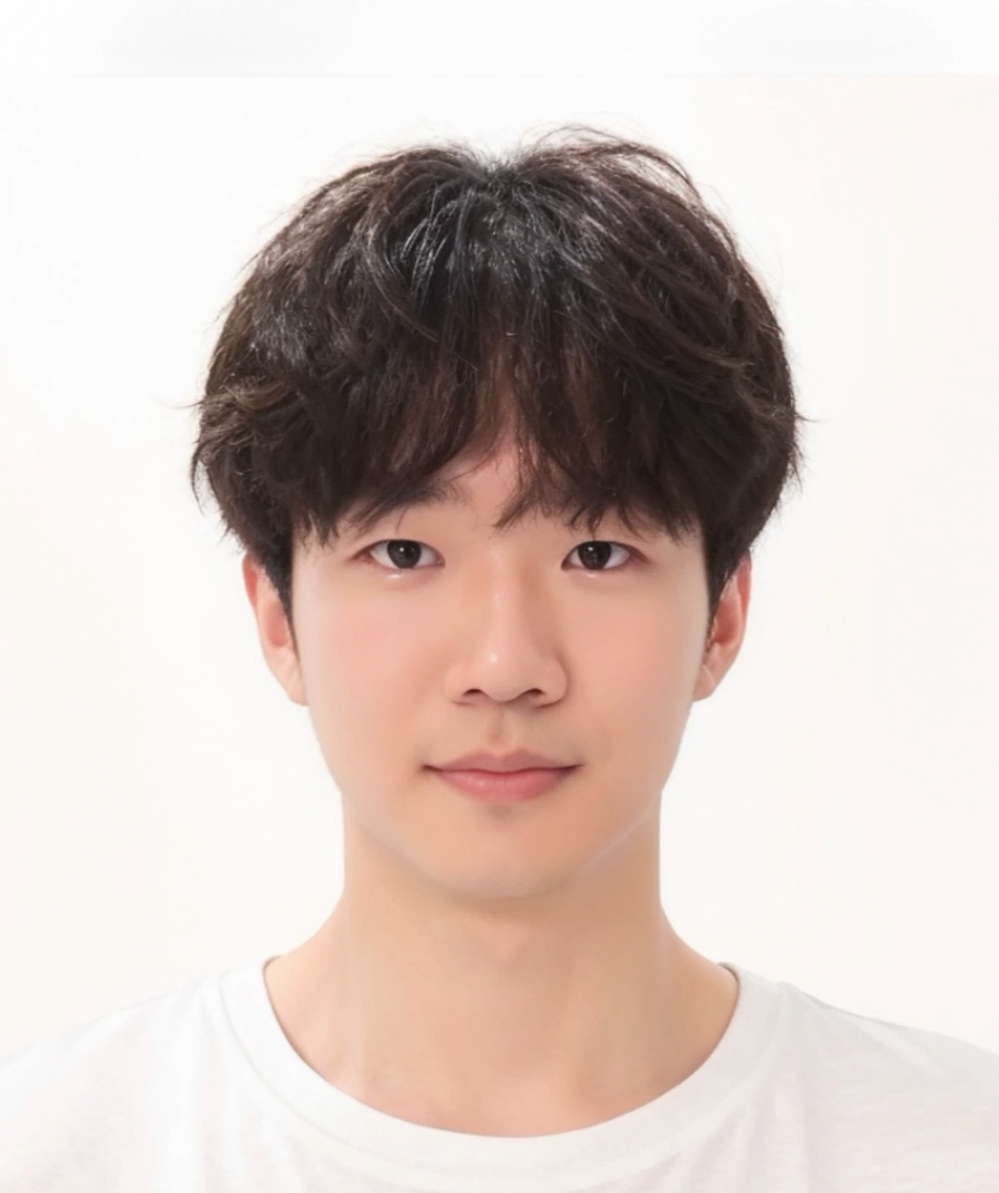}}]{Haichao Ji}
received the B.E. degree in Mechanical Engineering from Shanghai Jiao Tong University, China, in 2022, and M.S. degree in Applied Statistics from University of Michigan, Ann Arbor, US, in 2023. He is currently pursuing the Ph.D. degree in Shanghai Jiao Tong University. His research interests include federated analytics, privacy computing, and scheduling.
\end{IEEEbiography}

\begin{IEEEbiography}[{\includegraphics[width=1in,height=1.25in,clip,keepaspectratio]{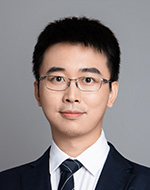}}]{Yifei Zhu}
(Member, IEEE)
is currently a tenure-track Associate Professor at the University of Michigan-Shanghai Jiao Tong University Joint Institute, Shanghai Jiao Tong University, China. He received his B.E. degree from Xi’an Jiaotong University, China, in 2012, his M.Phil. degree from The Hong Kong University of Science and Technology, China, in 2015, and his Ph.D. in Computer Science from Simon Fraser University, Canada, in 2020. His research interests include edge computing, multimedia networking, and distributed machine learning systems. He has published in top-tier venues such as ACM SIGCOMM, IEEE INFOCOM, and ACM Multimedia, among others. He is an Associate Editor of IEEE Internet of Things Journal, and has served as the area chair, and organization chair for several IEEE/ACM conferences such as ACM Multimedia.
\end{IEEEbiography}

\begin{IEEEbiography}[{\includegraphics[width=1in,height=1.25in,clip,keepaspectratio]{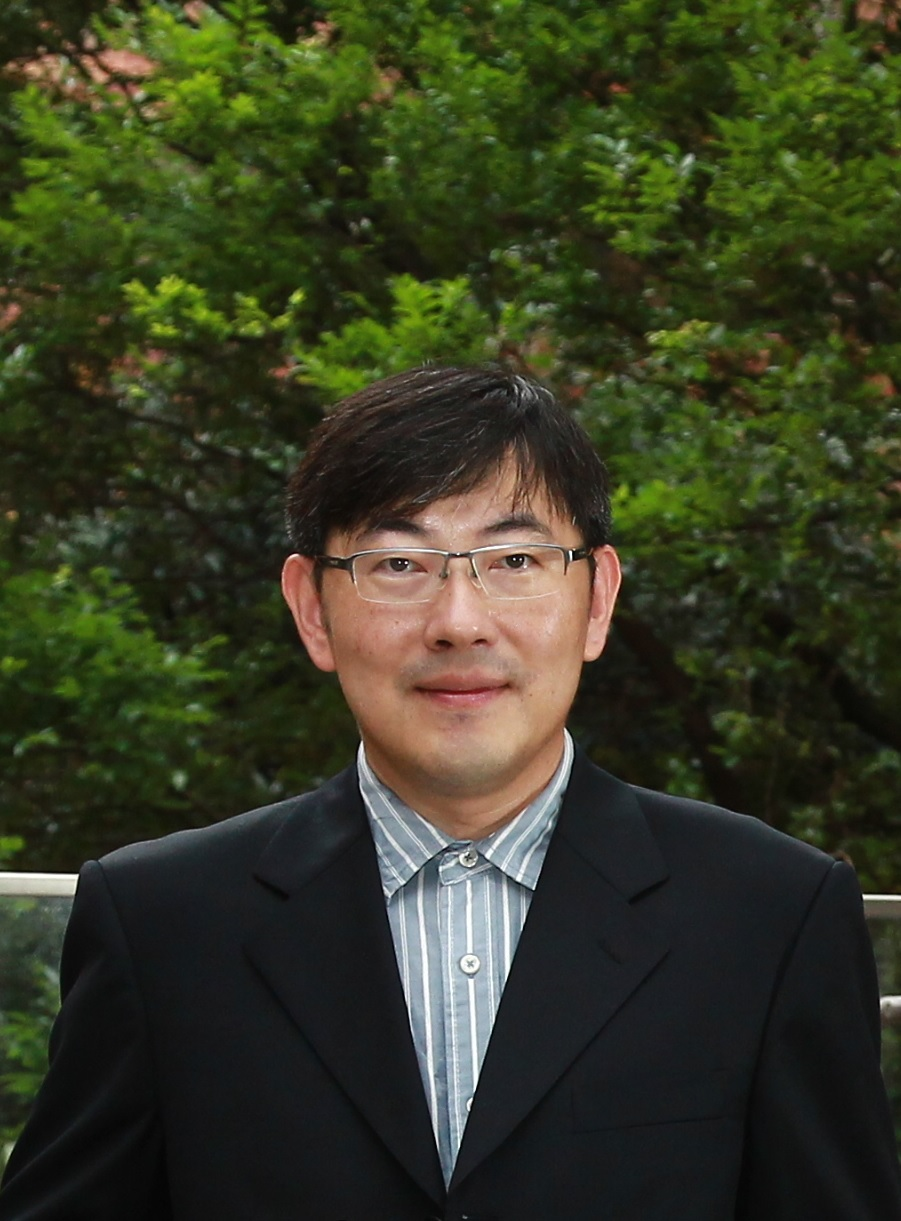}}]{Dan Wang}
(Senior Member, IEEE) receives his B.Sc., M.Sc., Ph.D. from Peking University, Case 
Western Reserve University and Simon Fraser University, all in 
Computer Science. His research falls in general computer networking and 
systems, where he published in ACM SIGCOMM, ACM SIGMETRICS and IEEE 
INFOCOM, and many others. He is the steering committee chair of IEEE/ACM 
IWQoS. He served as the TPC co-Chair of IEEE/ACM IWQoS 2020. His recent 
research focus on smart energy systems. He won the Best Paper Awards of 
ACM e-Energy 2018 and ACM Buildsys 2018. He has served as a TPC co-Chair 
of the ACM e-Energy 2020 and he will serve as General co-Chair of the ACM 
e-Energy 2022. He is a steering committee member of ACM e-Energy. He 
serves as a founding area editor of ACM SIGEnergy Energy Informatics 
Review. His research has been adopted by industry, e.g., Henderson, 
Huawei, and IBM. He won the Global Innovation Award, TechConnect, in 2017.
\end{IEEEbiography}

\begin{IEEEbiography}[{\includegraphics[width=1in,height=1.25in,clip,keepaspectratio]{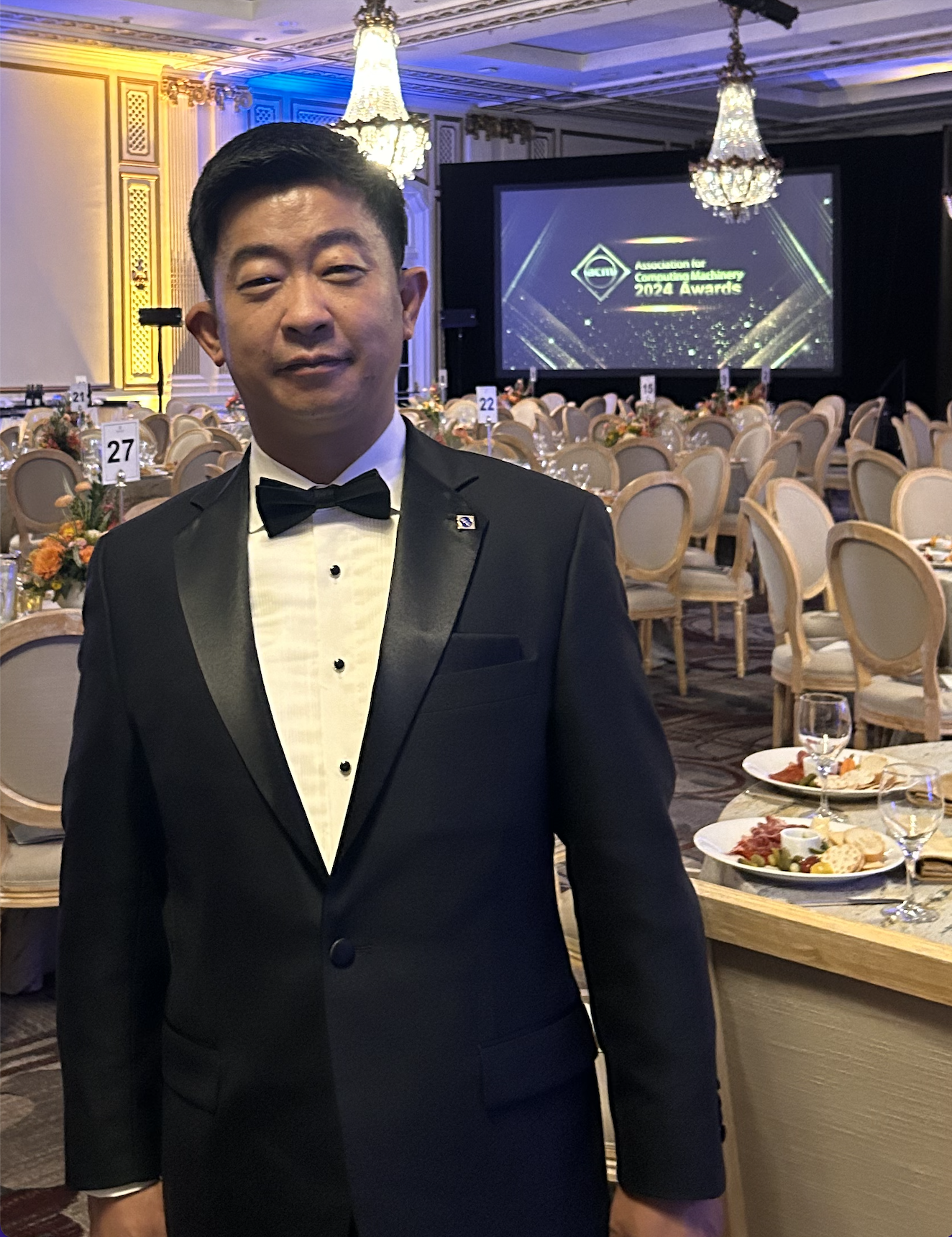}}]{Zhu Han}
(Fellow, IEEE) 
received the B.S. degree in electronic engineering from Tsinghua University, in 1997, and the M.S. and Ph.D. degrees in electrical and computer engineering from the University of Maryland, College Park, in 1999 and 2003, respectively. 

From 2000 to 2002, he was an R\&D Engineer of JDSU, Germantown, Maryland. From 2003 to 2006, he was a Research Associate at the University of Maryland. From 2006 to 2008, he was an assistant professor at Boise State University, Idaho. Currently, he is a John and Rebecca Moores Professor in the Electrical and Computer Engineering Department as well as in the Computer Science Department at the University of Houston, Texas. Dr. Han’s main research targets on the novel game-theory related concepts critical to enabling efficient and distributive use of wireless networks with limited resources. His other research interests include wireless resource allocation and management, wireless communications and networking, quantum computing, data science, smart grid, carbon neutralization, security and privacy.  Dr. Han received an NSF Career Award in 2010, the Fred W. Ellersick Prize of the IEEE Communication Society in 2011, the EURASIP Best Paper Award for the Journal on Advances in Signal Processing in 2015, IEEE Leonard G. Abraham Prize in the field of Communications Systems (best paper award in IEEE JSAC) in 2016, IEEE Vehicular Technology Society 2022 Best Land Transportation Paper Award, and several best paper awards in IEEE conferences. Dr. Han was an IEEE Communications Society Distinguished Lecturer from 2015 to 2018 and ACM Distinguished Speaker from 2022 to 2025, AAAS fellow since 2019, and ACM Fellow since 2024. Dr. Han is a 1\% highly cited researcher since 2017 according to Web of Science. Dr. Han is also the winner of the 2021 IEEE Kiyo Tomiyasu Award (an IEEE Field Award), for outstanding early to mid-career contributions to technologies holding the promise of innovative applications, with the following citation: ``for contributions to game theory and distributed management of autonomous communication networks."

\end{IEEEbiography}

\end{document}